# DOES THE PHOTON CARRY THE ARROW OF TIME? AN EXPERIMENTAL TEST

Darryl Leiter, Ph.D.
Interdisciplinary Studies Program, University of Virginia, Charlottesville, Virginia 22904

**Abstract**

If the photon has a negative parity under Wigner time reversal this generates a spontaneous CPT symmetry breaking effect that causes the photon to carry the quantum electrodynamic arrow of time (Leiter, D., 2009, 2010). In order to demonstrate the validity of this idea we show that a classic nonlinear optics experiment in the scientific literature, which involves a Michelson interferometer using combinations of ordinary mirrors and phase conjugate mirrors, contains experimental results which support the idea that the photon has a negative parity under Wigner time reversal.

**SECTION 1: INTRODUCTION**

It has been demonstrated (Leiter, D., 2009, 2010) that the quantum electrodynamic measurement process can be completed by inserting an operator symmetry of microscopic observer-participation called "Measurement Color" (MC) into Quantum Electrodynamics (QED).

The resultant Measurement Color Quantum Electrodynamics (MC-QED) formalism was shown to be a nonlocal quantum field theory which contains a time reversal violating description of the quantum electrodynamic measurement process which is independent of thermodynamic or cosmological assumptions. This occurred because the Measurement Color operator symmetry within MC-QED caused the photon operator to have a negative parity under Wigner time reversal. Then the requirement of a stable vacuum state generated a spontaneous CPT symmetry breaking effect which dynamically generated a quantum electrodynamic arrow of time in the Heisenberg operator equations of motion. This result differs from the case of QED which does not contain an intrinsic arrow of time since its photon operator has a positive parity under Wigner time reversal.

In this context we present analytical arguments which show that a well-known classic nonlinear optics experiment in the scientific literature, which uses combinations of ordinary mirrors and phase conjugate mirrors in a Michelson interferometer, has produced experimental results which represents strong evidence in favor of the MC-QED prediction that the photon has a negative parity under Wigner time reversal and that "the photon carries the quantum electrodynamic arrow of time".

## SECTION II: EXPERIMENT TO TEST FOR THE TIME PARITY OF THE PHOTON

In order to verify the correctness of the microscopic operator observer-participant paradigm underlying the structure of the MC-QED formalism it is necessary to demonstrate that experiments can be performed which can provide a test for its underlying prediction that the photon has a negative parity under Wigner time reversal.

The purpose of this paper is to demonstrate that the results of such an experimental test involving nonlinear optics in Michelson interferometers already exists in the literature and supports the predictions of the MC-QED formalism.

In order begin our analysis we will consider an experimental arrangement involving a Michelson interferometer in which a coherent optical laser beam sent thru a beam splitter to creates interference fringes due to multiple reflections in the vertical and horizontal arms of the apparatus.

Two different experimental scenarios are considered and their results are compared. The first scenario (M-M) is a Michelson interferometer involving the combination of two conventional mirrors while the second scenario (PCM-M) is a Michelson interferometer involving the combination of a Phase Conjugate mirror PCM and a conventional mirror M.

The schematic drawing of this experimental setup (Wolf, Mandel et, al 1987; Jacobs, et. al. 1987; Boyd, et. al. 1987) shown in figure 1 below is one in which one has the option of replacing one of the two conventional mirrors with a phase-conjugate mirror PCM.

A laser sends a coherent optical beam thru the interferometer and creates multiple interference fringes. Selective changes in the phase $\alpha$ of the internal beams can be generated by the use of a gas cell located in positions A, B, or C. The interference fringes can be recorded by the photo-detector, for both the M-M and the M-PCM configurations, and the results compared to the predictions of the QED and the MC-QED formalisms.

If the location of interference fringes recorded by the photo-detector are observed for the case A, where the phase shift $\alpha$ introduced by a gas cell located at position A in the figure induces a change the phase $\alpha$ of the incident waves on the PC mirror, the results which are obtained can experimentally distinguish between the predictions of the QED and the MC-QED formalisms.

In order to understand how this experiment has the potential to be able to distinguish between the QED and the MC-QED formalisms we will now discuss the underlying theoretical and experimental structure of it in more detail. In a series of three seminal papers (Wolf, Mandel et, al 1987; Jacobs, et. al. 1987; Boyd, et. al. 1987) published a detailed theoretical analysis, later supported by experimental observations, which demonstrated how experiments of this type could distinguish between the classical nature of the interference patterns produced by Michelson interferometers the M-M and the M-PCM configurations. More modern reviews of this type of experiment (Garuccio, 2007) have discussed the additional possibility of using it to test for quantum non-locality and anti-coherence effects in light.

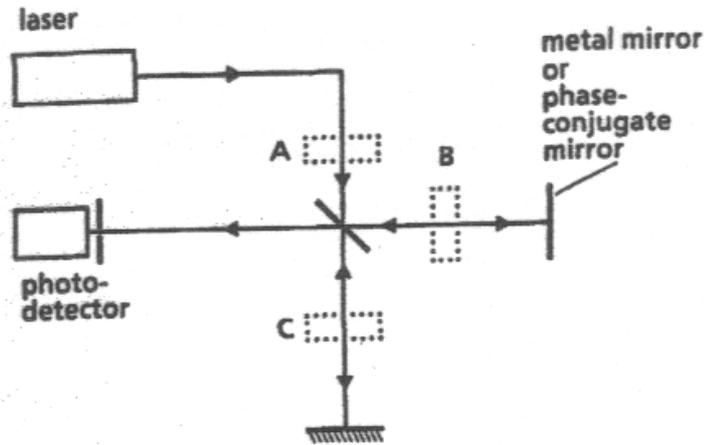

Figure 1. A modified version of a Michelson interferometer (Wolf, Mandel et, al 1987; Jacobs, et. al. 1987; Boyd, et. al. 1987) in which one has the option of replacing one of the two conventional mirrors with a phase-conjugate mirror PCM. A laser sends a coherent optical beam thru the interferometer and creates multiple interference fringes. Selective changes in the phase α of the internal beams can be generated by the use of a gas cell located in positions A, B, or C. The interference fringes are recorded by the photo-detector, for both the M-M and the M-PCM configurations, and the results compared to the predictions of the QED and the MC-QED formalisms.

In particular for the Michelson interferometer in the M-M configuration, which involved normally incident linearly polarized light with complex amplitude $|A|e^{i\alpha}$, it was shown that the locations of the bright maxima and dark minima of the time-averaged interference fringes were given respectively by:

Bright maxima of interference fringes
$$z = n(\lambda / 2) \qquad (n = 0,1,2,...)$$

Dark minimum of interference fringes
$$z = (n + 1/2)(\lambda / 2) \qquad (n = 0,1,2,...)$$

On the other hand for the Michelson interferometer in the M-PCM configuration, which involved normally incident linearly polarized light with complex amplitude $|A|e^{i\alpha}$ and where the complex amplitude reflectivity of the phase-conjugate mirror was assumed to be given by $\mu = |\mu|e^{i\phi}$ (here $|\mu| = 1$ and the phase shift $\phi$ was given in radians) it was shown that the locations of the bright maxima and dark minima of the time-averaged interference fringes were given respectively by:

Bright maxima of interference fringes
$$z = n(\lambda / 2) + (\lambda / 2) [(\phi / 2 - \alpha) / \pi ] \qquad (n = 0,1,2,...)$$

Dark minimum of interference fringes
$$z = (n + 1/2)(\lambda / 2) + (\lambda / 2) [(\phi / 2 - \alpha) / \pi ] \qquad (n = 0,1,2,...)$$

Hence in the context of (Wolf, Mandel et, al 1987; Jacobs, et. al. 1987; Boyd, et. al. 1987) it was demonstrated and later shown experimentally that the difference between the location of the maximum and minimum of the interference fringes for Michelson interferometers in the M-PCM and the M-M configurations was associated with a displacement of the interference pattern in radians given by

$$\Delta(\phi, \alpha) = [(\phi / 2) - \alpha] \text{ radians}$$

where $e^{i\phi}$ was the internal phase shift generated by the phase-conjugate mirror and $e^{i\alpha}$ was the phase of incident linearly polarized light

Next we point out that it can be shown (see Appendix I and II) that in MC-QED the connection between the coherent photon state vectors $|a(\alpha), \lambda\rangle$, and the corresponding coherent classical electromagnetic photon fields which they represent, is given by the expectation value of its negative Wigner time parity photon operator over the coherent states $|a(\alpha), \lambda\rangle$ in the formalism. Hence in the context of MC-QED a coherent photon state associated with propagation vector **k** and phase $\alpha$ is predicted to transform with a <u>negative time parity</u> into a coherent photon state associated with propagation vector -**k** and phase $-(\alpha + \pi)$ under Wigner time reversal. This is different from the case of QED, where a coherent photon state associated with propagation vector **k** and phase $\alpha$ is predicted to transform with a <u>positive time parity</u> into a coherent photon state associated with propagation vector -**k** and phase $-\alpha$ under Wigner time reversal.

In the context of this difference in the Wigner time reversal properties of coherent photon states in QED and MC-QED it is important to note that it has been shown (Chew, Habashi 1985) that the "healing effect" associated with removal of intermediate distortions of optical images generated by a phase conjugate mirror is physically equivalent, within a constant phase factor $\exp(i\phi)$ of modulus one, to the effects of time reversal. On this basis we conclude that both QED and MC-QED will predict the same phase conjugate mirror "healing effect" on the distortion of optical images since for QED the constant phase factor is $\phi = 0$ while for MC-QED the constant factor is $\phi = -\pi$.

Hence from the above experimental and theoretical discussion we find that the difference between the location of the maximum and minimum of the interference fringes for Michelson interferometers in the M-PCM and the M-M configurations will associated with a displacement of the interference pattern in radians is given respectively for QED and MC-QED by

$$\Delta(0, \alpha) = -[\alpha] \text{ radians} \qquad \text{QED}$$

$$\Delta(-\pi, \alpha) = -[(\pi/2) + \alpha] \text{ radians} \qquad \text{MC-QED}$$

In the limiting case where $\alpha = 0$ this corresponds to a displacement of the interference pattern in radians given by

$$\Delta(0, 0) = 0 \text{ radians} \qquad \text{QED}$$

$$\Delta(-\pi, 0) = -(\pi/2) \text{ radians} \qquad \text{MC-QED}$$

The difference between the QED and the MC-QED predictions is a reflection of the fact that the nonlocal photon operator acting within the quantum field theoretic structure of the MC-QED formalism has a negative parity under Wigner time reversal and hence carries the quantum electrodynamic arrow of time in the formalism. For this reason the apparatus discussed in figure 1 allows an experimental test to be performed to determine if the photon carries the arrow of time as predicted by the MC-QED formalism.

**SECTION III: DISCUSSION INTERPRETING THE RESULTS OF THE EXPERIMENT**

Results taken from the classic experiment performed by (Jacobs, et al. 1987) are shown in figure 2 below. While straight lines have been drawn through the data points for the M-PCM and M-M configurations in figure 2, it appears that nonlinear internal processes in the gas cell introduce oscillation errors into the data which break the predicted linearity for non-zero values of the phase shift. However these nonlinear gas cell oscillation errors will not affect the individual data points at phase shift equal to zero, since for these data points the gas cell is not active in the interferometer. For this reason only the data points at phase shift equal zero, for which the gas cell does not act in the interferometer, will be relevant in the analysis which follows.

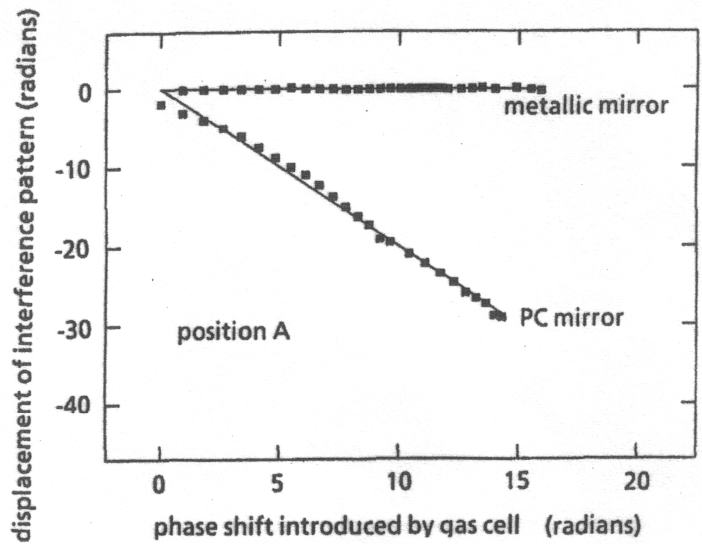

Figure 2. Measured displacement of the interference fringe pattern, for a Michelson interferometer for the M-M (metallic mirror) configuration and the M-PCM (PC mirror) configuration (taken from figure 3 of Jacobs, et al. 1987). The displacement of the interference pattern in radians is produced by interference between the signal and the phase-conjugate waves. It is plotted as a function of the phase shift $\alpha$ introduced by a gas cell whose location (in position A in figure 1 above) induces a change the phase $\alpha$ of the incident waves acting on the PC mirror. Note that the difference between the displacement of the interference pattern for the M-M (metallic mirror) and the M-PCM (PC mirror) configuration for the case of zero phase shift $\alpha = 0$ appears to be consistent with the MC-QED value $\Delta(-\pi, 0) = -(\pi/2) = -1.57$ Hence the results of this experiment appears to offer strong support in favor of MC-QED and its prediction that the photon carries the arrow of time.

In the graph the displacement of the interference pattern in radians, produced by interference between the signal and the phase-conjugate waves, is plotted as a function of the phase shift α introduced by a gas cell whose location, at position A as shown in figure 1, allows it to induce a change the phase α of the incident waves on the PC mirror.

*Note that the difference between the displacement of the interference pattern for the M-M (metallic mirror) and the M-PCM (PC mirror) configuration for the case of zero phase shift α = 0 appears to be consistent with the predicted MC-QED value of $\Delta(-\pi, 0) = -(\pi/2) = -1.57$ radians and not with the predicted QED value of $\Delta(0, 0) = 0$ radians.*

On this basis of these results, this experiment appears to offer strong support in favor of MC-QED and its prediction that the photon carries the arrow of time.

**SECTION III: CONCLUSIONS**

By incorporating an operator symmetry of microscopic observer-participation called "Measurement Color" into Quantum Electrodynamics (QED) the resultant Measurement Color Quantum Electrodynamics (MC-QED) contains a time reversal violating description of the quantum electrodynamic measurement process which is independent of thermodynamic or cosmological assumptions. This occurred because Measurement Color symmetry within MC-QED caused the photon operator in the formalism to have a negative parity under Wigner time reversal. This created a spontaneous CPT symmetry breaking effect which dynamically determined a causal quantum electrodynamic arrow of time in the formalism (Leiter, 2009, 2010).

On this basis it was shown (see Appendix I and II) that in MC-QED a coherent photon state, with propagation vector propagation vector **k** and phase (α), transformed under Wigner time reversal with negative time parity into a coherent photon state with propagation vector -**k** and phase $-(\alpha + \pi)$. This was to be compared to the case of a coherent photon state in QED, with a propagation vector k and phase (α), which was shown to transform under Wigner time reversal with a positive time parity into a coherent photon state with propagation vector -**k** and phase (–α) under Wigner time reversal.

Because of this difference between the Wigner time reversal symmetry of the photon in MC-QED and QED, we demonstrated that a Michelson interferometer experiment involving a combination of ordinary mirrors and phase conjugate mirrors could experimentally determine if the photon operator has a negative parity under Wigner time reversal and in this way test the MC-QED prediction that "the photon carries the arrow of time".

The experiment shown schematically in figure 1 involved measuring and comparing the displacement of the interference fringe pattern, for a Michelson interferometer for the M-M (metallic mirror) configuration and the M-PCM (PC mirror) configuration. In the context of this experiment it was shown that the displacement in radians $\Delta(\phi, \alpha)$ of the interference pattern in produced by interference between the signal and the phase-conjugate waves was given by $\Delta(\phi, \alpha) = [(\phi / 2) - \alpha]$, where $\phi$ was the internal phase shift generated by the phase-conjugate mirror and $\alpha$ was the phase of incident linearly polarized light.

In the limiting case of zero phase shift ($\alpha = 0$) the difference between the displacement of the interference pattern for the M-M (metallic mirror) configuration and the M-PCM (PC mirror) configuration predicted by MC-QED has a value of $\Delta(-\pi, 0) = -\pi/2$ radians which is distinctly different from the value of $\Delta(0, 0) = 0$ radians predicted by QED.

Because of the difference between the Wigner time reversal symmetry properties of the photon operator in QED and MC-QED we have shown that well-know classic Michelson interferometer experiment discussed in the literature, (Wolf, Mandel et, al 1987; Jacobs, et. al. 1987; Boyd, et. al. 1987) which involves combinations of ordinary mirrors and phase conjugate mirrors, appears to have experimentally demonstrated that the photon operator has a negative parity under Wigner time reversal and hence the MC-QED prediction that "the photon carries the quantum electrodynamic arrow of time". It is hoped that this paper will encourage experimental physicists to perform more modern versions of the Michelson interferometer experiment described in this paper, in order to further verify this experimental result within the context of the high accuracy of twenty-first century technology.

# APPENDIX I: PHOTON BARE STATE STRUCTURE IN THE MC-QED FORMALISM

It has been shown (Leiter, D. 2009, 2010, http://journalofcosmology.com/Contents.html) that the Measurement Color symmetric charge field photon Hamiltonian operator in the MC-QED formalism is given by

$$H_{ph} = \sum_{(j)} \{: \int dx^3 [ -1/2 (\partial_t A^{\mu(j)}_{(rad)} \partial_t A^{(j)}_{\mu\,(rad)}{}^{(obs)} + \nabla A^{\mu(j)}_{(rad)} \cdot \nabla A^{(j)}_{\mu\,(rad)}{}^{(obs)} )] :\}$$

where the symbols $: :$ denote operator normal ordering, ($j=1,2,\ldots,N \longrightarrow \infty$), and

$$A^{(j)}_{\mu\,(rad)}(x) = (\alpha_\mu(x) - A^{(j)}_{\mu\,(-)}(x))$$

$$A^{(j)}_{\mu\,(rad)}{}^{(obs)}(x) = \sum_{(k) \neq (j)} A^{(k)}_{\mu\,(rad)}(x) = A^{(j)}_{\mu\,(-)}(x)$$

$$\alpha_\mu(x) = \sum_{(j)} A^{(j)}_{\mu\,(-)}(x) / (N-1)$$

are linear functions of <u>the negative time parity operator</u> $A^{(j)}_{\mu\,(-)}(x) = \int dx^{4'} D_{(-)}(x-x') J^{(j)}_\mu(x')$ where $A^{(j)}_{\mu\,(-)}(x) = (A^{(j)}_{\mu\,(-)}(x))^\dagger$ which implies that $\Box^2 A^{(j)}_{\mu\,(-)}(x) = 0$ which from the above also implies that $\Box^2 A^{(j)}_{\mu\,(rad)}{}^{(obs)} = \Box^2 A^{(j)}_{\mu\,(rad)} = \Box^2 \alpha_\mu(x) = 0$.

The negative time parity operators $A^{(j)}_{\mu\,(rad)}$, $A^{(j)}_{\mu\,(rad)}{}^{(obs)}$ and $\alpha_\mu$ ($j=1,2,\ldots,N \to \infty$) can be respectively expanded as

$$A^{(j)}_\mu(x) = (\alpha_\mu(x) - A^{(j)}_{\mu\,(-)}(x)) = \int dk^3 / \sqrt{[2(2\pi)^3 k^2]} \{a^{(j)}_\mu(k)e^{-ik\cdot x} + a^{(j)}_\mu(k)^\dagger e^{ik\cdot x}\}$$

$$A^{(j)}_{\mu\,(rad)}{}^{(obs)}(x) = A^{(j)}_{\mu\,(-)}(x) = \int dk^3 / \sqrt{[2(2\pi)^3 k^2]} \{a^{(j)}_{\mu\,(-)}(k) e^{-ik\cdot x} + a^{(j)}_{\mu\,(-)}(k)^\dagger e^{ik\cdot x}\}$$

$$\alpha_\mu(x) = \sum_{(j)} A^{(j)}_{\mu\,(-)}(x) / (N-1) = \int dk^3 / \sqrt{[2(2\pi)^3 k^2]} \{\alpha_\mu(k)e^{-ik\cdot x} + \alpha_\mu(k)^\dagger e^{ik\cdot x}\}$$

where $k \cdot x = k_\nu x^\nu = \mathbf{k} \cdot \mathbf{x} + k_0 x^0$ and $\mathbf{k} = \mathbf{n}/\lambda$, $k_0 = \nu/c$

From the above we see that

$$a^{(j)}_{\mu\,(-)}(\mathbf{k}) = a^{(j)}_{\mu\,(rad)}{}^{(obs)}(\mathbf{k}), \quad \alpha_\mu(\mathbf{k}) = \sum_{(j)} a^{(j)}_{\mu\,(-)}(\mathbf{k})/(N-1) = \sum_{(j)} a^{(j)}_\mu(\mathbf{k})$$

$$a^{(j)}_\mu(\mathbf{k}) = \alpha_\mu(\mathbf{k}) - a^{(j)}_{\mu\,(-)}(\mathbf{k}) = \sum_{(j)} a^{(j)}_{\mu\,(-)}(\mathbf{k})/(N-1) - a^{(j)}_{\mu\,(-)}(\mathbf{k})$$

We now show that the time reversal violating Measurement Color symmetric operators $\alpha_\mu(\mathbf{k})$ and $\alpha_\mu(\mathbf{k})^\dagger$ act respectively as destruction and creation operators for Measurement Color symmetric charge field photon states in MC-QED.

We begin by substituting the above representations of $A_\mu^{(j)}(x)$, $A_\mu^{(j)}{}_{(obs)}(x)$, and $\alpha_\mu(x)$ into the above MC-QED commutation relations to find (j, m = 1,2, ... , N → ∞)

$$[a_\mu^{(j)}{}_{(-)}(\mathbf{k}), a_\nu^{(m)}{}_{(-)}{}^\dagger(\mathbf{k'})] = (1 - \delta^{jm})(-\eta_{\mu\nu}\lambda_0\,\delta^3(\mathbf{k}-\mathbf{k'}))$$

$$[\alpha_\mu(\mathbf{k}), a_\nu^{(j)}{}_{(-)}{}^\dagger(\mathbf{k'})] = -\eta_{\mu\nu}\lambda_0\,\delta^3(\mathbf{k}-\mathbf{k'})$$

$$[\alpha_\mu(\mathbf{k}), \alpha_\nu(\mathbf{k'})] = (-\eta_{\mu\nu}\lambda_0\,\delta^3(\mathbf{k}-\mathbf{k'}))(N/(N-1))$$

$$[\alpha_\mu(\mathbf{k}), a_\nu^{(j)}(\mathbf{k'})^\dagger] = 0$$

$$[a_\mu^{(j)}{}_{(-)}{}^\dagger(\mathbf{k}), a_\nu^{(m)}{}_{(-)}{}^\dagger(\mathbf{k'})] = 0$$

$$[a_\mu^{(j)}{}_{(-)}(\mathbf{k}), a_\nu^{(m)}{}_{(-)}(\mathbf{k'})] = 0$$

where $k_0 = \sqrt{(\mathbf{k}^2)} = \omega(\mathbf{k})$ and all other commutators vanish.

Next we substitute above representations of $A_\mu^{(j)}(x)$ and $A_\mu^{(j)}{}_{(obs)}(x)$ into the charge field photon hamiltonian $H_{ph}$ which gives the charge field photon Hamiltonian as\

$$H_{ph} = \sum_{(j)} \{: [-\int dk^3 / k_0 (\omega(\mathbf{k})\, a_\mu^{(j)}{}_{(-)}{}^\dagger(\mathbf{k})\, a^{\mu(j)}{}_{(-)}(\mathbf{k})] :\}$$

and normal ordering of operators inside of the symbols $\{: :\}$ has been taken.

In addition by inserting

$$a_\mu^{(j)}(\mathbf{k}) = \alpha_\mu(\mathbf{k}) - a_\mu^{(j)}{}_{(-)}(\mathbf{k}) \quad \text{and} \quad \alpha_\mu(\mathbf{k}) = \sum_{(j)} a_\mu^{(j)}{}_{(-)}(\mathbf{k})/(N-1) = \sum_{(j)} a_\mu^{(j)}(\mathbf{k})$$

into $H_{ph}$ the hermetian property of the photon hamiltonian $H_{ph} = H_{ph}{}^\dagger$ follows directly.

In this context if the bare MC-QED charge field photon vacuum state |0ph> is defined by

$$a_\mu^{(j)}{}_{(-)}(\mathbf{k})|\,0_{ph}> = 0 \quad (j = 1,2, ... , N \to \infty)$$

this implies that $H_{ph}\,|\,0_{ph}> = 0$ as required.

Now since $\alpha_\mu(\mathbf{k}) = \sum_{(j)} a_\mu^{(j)}{}_{(-)}(\mathbf{k}) /(N-1)$ the above definition of $|0_{ph}\rangle$ also implies that the bare charge field photon vacuum state also obeys

$$\alpha_\mu(\mathbf{k}) | 0_{ph} \rangle = 0$$

In this context the bare single charge-field photon in MC-QED can be defined as

$$|\lambda 1_\mu\rangle = \alpha_\mu(\mathbf{k1})^\dagger |0\rangle = (1/(N-1)) \sum_{(j)} a_\mu^{(j)}{}_{(-)}{}^\dagger (\mathbf{k1})^{(j)} |0\rangle$$

This can be seen by calculating

$$H_{ph} |\mathbf{k_1}\rangle = -\sum_{(j))} \int dk^3 / k_0 \, (\omega(\mathbf{k}) a_\nu^{(j)}(\mathbf{k})^\dagger a^{\nu(j)}{}_{(-)}(\mathbf{k}) \alpha_\mu(\mathbf{k1})^\dagger) |0\rangle$$

Then using the fact that

$$[\alpha_\mu(\mathbf{k}), a^{\nu(j)}{}_{(-)}{}^\dagger(\mathbf{k'})] = -\delta_{\mu\nu} k_0 \delta^3(\mathbf{k} - \mathbf{k'}) \quad \text{and} \quad \alpha_\mu(\mathbf{k1}) = \sum_{(J)} a_\mu^{(j)} \mathbf{k1})$$

we have

$$H_{ph} |\mathbf{k_1}\rangle = \sum_{(J)} \int dk^3 (\omega(\lambda) a_\nu^{(j)}(\lambda)^\dagger)(-\delta_\mu{}^\nu) \delta^3(\lambda - \lambda 1) |0\rangle$$

$$= \omega(\mathbf{k1}) \sum_{(J)} a_\mu^{(j)\dagger}(\mathbf{k1}) |0\rangle$$

$$= \omega(\mathbf{k1}) (\alpha_\mu(\mathbf{k1})^\dagger |0\rangle = \omega(\mathbf{k1}) |\mathbf{k1}\rangle \quad \text{as required}$$

Hence the N-bare charge-field photon states in MC-QED are defined as

$$|\mathbf{k}_{\alpha 1}, \mathbf{k}_{2\beta}, \mathbf{k}_{3\gamma}, \ldots \rangle = (1/N!)^{1/2} \alpha_\alpha(\mathbf{k1})^\dagger \alpha_\beta(\mathbf{k2})^\dagger \alpha_\gamma(\mathbf{k3})^\dagger \ldots |0\rangle$$

In a similar manner as that of the covariant form of QED, consistency with the expectation value of the operator form of Maxwell equations in the covariant form of MC-QED requires that an Indefinite Metric Hilbert space must be used.

In the context of an Indefinite Metric Hilbert space, the subset of physical bare charge field photon states in MC-QED contained within the above set of multiple charge field photon eigenstates of $H_{ph}$ are required to obey the Weak Subsidiary Condition $\lambda^\mu a_\mu^{(j)}(\mathbf{k}) |\psi\rangle = 0$ where

$$a_\mu^{(j)}(\mathbf{k}) = \alpha_\mu(\mathbf{k}) - a_\mu^{(j)}{}_{(-)}(\mathbf{k}) = \sum_{(j)} a_\mu^{(j)}{}_{(-)}(\mathbf{k})/(N-1) - a_\mu^{(j)}{}_{(-)}(\mathbf{k})$$

which requires them to contain equal numbers of timelike and longitudinal charge field photons. Since the Indefinite Metric Hilbert space implies that charge field photon states with an odd number of time-like charge field photons have an additional negative sign associate with their inner product, the combination of the Weak Subsidiary Condition and the Indefinite Metric Hilbert space together imply that the physical bare charge field photon states have a positive semi-definite norm and energy momentum expectation values similar to that of the QED formalism.

# APPENDIX II: COHERENT PHOTON STATES IN THE MC-QED FORMALISM

While the fermion current operators $J^{(k)\mu}(\mathbf{x},t)$ in MC-QED transform under Wigner time reversal as $T_w J^{(k)\mu}(\mathbf{x},t) T_w^{-1} = J^{(k)}_\mu(\mathbf{x},-t)$ similar to that of QED, the charge-field photon operators

$$A_\mu^{(j)}{}_{(rad)}{}^{(obs)}(x) = \sum_{(m) \neq (j)} A_\mu^{(m)}{}_{(rad)}(x) = A_\mu^{(j)}{}_{(-)}(x)$$

$$A_\mu^{(j)}{}_{(-)}(x) = \int dk^3 / \sqrt{[2(2\pi)^3 \mathbf{k}^2]} \{ a_\mu^{(j)}{}_{(-)}(\mathbf{k}) e^{-ik \cdot x} + a_\mu^{(j)}{}_{(-)}(\mathbf{k})^\dagger e^{ik \cdot x} \}$$

$$\alpha_\mu(x) = \sum_{(j)} A_\mu^{(j)}{}_{(-)}(x) /(N-1) = \int dk^3 / \sqrt{[2(2\pi)^3 \mathbf{k}^2]} \{ \alpha_\mu(\mathbf{k}) e^{-ik \cdot x} + \alpha_\mu(\mathbf{k})^\dagger e^{ik \cdot x} \}$$

where $k \cdot x = k_\nu x^\nu = \mathbf{k} \cdot \mathbf{x} + k_0 x^0$ and $\mathbf{k} = \mathbf{n}/\lambda$, $k_0 = \nu/c$, have a negative parity under Wigner Reversal operator $T_w$ as

$$T_w A_\mu^{(j)}{}_{(-)}(\mathbf{x},t) T_w^{-1} = -A^{\mu(j)}{}_{(-)}(\mathbf{x},-t)$$

$$T_w \alpha_\mu(\mathbf{x},t) T_w^{-1} = -\alpha^\mu(\mathbf{x},-t)$$

Hence this implies that under Wigner Time reversal $T_w$ the charge-field photon creation and annihilation operators in MC-QED also transform with a negative time parity respectively as

$$T_w a_\mu^{(j)}{}_{(-)}(\mathbf{k})^\dagger T_w^{-1} = -a^{\mu(j)}{}_{(-)}(-\mathbf{k})^\dagger \qquad T_w a_\mu^{(j)}{}_{(-)}(\mathbf{k}) T_w^{-1} = -a^{\mu(j)}{}_{(-)}(-\mathbf{k})$$

$$T_w \alpha^\mu(\mathbf{k})^\dagger T_w^{-1} = -\alpha^\mu(-\mathbf{k})^\dagger \qquad T_w \alpha^\mu(\mathbf{k}) T_w^{-1} = -\alpha^\mu(-\mathbf{k})$$

Now in MC-QED a coherent photon state $|a(\varphi), \mathbf{k}\rangle$ of frequency $\omega(\mathbf{k}) = \sqrt{(\mathbf{k}^2)} = k_0$ and complex phase $a(\alpha)$, is an eigenstate of the charge-field photon destruction operator $\alpha^\mu(\mathbf{k}) = \varepsilon^\mu \alpha(\mathbf{k})$ as $\alpha^\mu(\mathbf{k}) | a(\alpha), \mathbf{k}\rangle = \varepsilon^\mu a(\alpha) | a(\alpha), \mathbf{k}\rangle$ which is satisfied if

$$\alpha(\mathbf{k}) | a(\alpha), \mathbf{k}\rangle = a(\alpha) | a(\alpha), \mathbf{k}\rangle$$

Solving for a coherent photon state $| a(\alpha), \mathbf{k}\rangle$ with a mean photon number $\langle N \rangle$ yields

$$|a(\alpha), \mathbf{k}\rangle = \exp(-\langle N \rangle/2) \exp[a(\alpha) \alpha_\mu(\mathbf{k})^\dagger]|0\rangle$$

where $a(\alpha) = (\langle N \rangle)^{1/2} \exp(i\alpha)$. In terms of the n-photon state of frequency $\omega(\mathbf{k})$ is given by

$$|n, \lambda\rangle = (1/(n!))^{1/2} \alpha^\mu(\mathbf{k})^\dagger \alpha^\nu(\mathbf{k})^\dagger \alpha^\eta(\mathbf{k})^\dagger \ldots |0\rangle$$

the above expression for $|a(\alpha), \mathbf{k}\rangle$ can be written in the more explicit form

$$|a(\alpha), \mathbf{k}\rangle = \sum_{n=0,1,\ldots\infty} [\exp(-\langle N\rangle/2) (\langle N\rangle^n / n!)^{1/2} \exp(in\alpha) |n, \mathbf{k}\rangle]$$

By calculating $\langle a(\alpha), \mathbf{k} | a(\alpha), \mathbf{k}\rangle$ we see that the distribution of photons in the coherent state is obeys a Poisson statistical distribution $\Pi_n (\langle N\rangle) = [\exp(-\langle N\rangle) (\langle N\rangle^n / n!)]$ since

$$\langle a(\alpha), \mathbf{k} | a(\alpha), \mathbf{k}\rangle = \sum_{n=0,1,\ldots\infty} [\exp(-\langle N\rangle) (\langle N\rangle^n / n!)] = \sum_{n=0,1,\ldots\infty} \Pi_n (\langle N\rangle)$$

Since $\alpha^\mu(\mathbf{k}) = \varepsilon^\mu \alpha(\mathbf{k})$ and the bare charge field photon creation operators $\alpha(\mathbf{k})^\dagger$ have a negative parity under Wigner Time reversal $T_w$ given by $T_w \varepsilon^\mu \alpha(\mathbf{k})^\dagger T_w^{-1} = -\varepsilon_\mu \alpha(-\mathbf{k})^\dagger$, we find that Wigner time reversal $T_w$ acting on the coherent photon state $|a(\varphi), \mathbf{k}\rangle$ gives

$$|a(\alpha), \mathbf{k}\rangle_{Tw} = T_w |a(\alpha), \mathbf{k}\rangle = \sum_{n=0,1,\ldots\infty} [\exp(-\langle N\rangle/2) (\langle N\rangle^n / n!)^{1/2} \exp(-in\alpha) T_w |n, \mathbf{k}\rangle]$$

$$= \sum_{n=0,1,\ldots\infty} [\exp(-\langle N\rangle/2) (\langle N\rangle^n / n!)^{1/2} \exp(-in\alpha) (-1)^n |n, -\mathbf{k}\rangle]$$

$$= \sum_{n=0,1,\ldots\infty} [\exp(-\langle N\rangle/2) (\langle N\rangle^n / n!)^{1/2} \exp(-in(\alpha + \pi)) |n, -\mathbf{k}\rangle] = |a(-[\alpha + \pi]), -\mathbf{k}\rangle$$

Hence the negative time parity of the Wigner time reversed coherent photon state in MC-QED implies an observable difference between time reversed phase-conjugate coherent states in QED and MC-QED.

This is because for MC-QED

$$|a(\alpha), \mathbf{k}\rangle_{Tw} = T_w |a(\alpha), \mathbf{k}\rangle = |a(-[\alpha + \pi]), -\mathbf{k}\rangle$$

while for QED

$$|a(\alpha), \mathbf{k}\rangle_{Tw} = T_w |a(\alpha), \mathbf{k}\rangle = |a(-\alpha), -\mathbf{k}\rangle$$

Hence in QED a coherent photon state associated with wave vector $\mathbf{k}$ and phase $\alpha$ is predicted to transform into a coherent photon state associated with wave vector $-\mathbf{k}$ and phase $-(\alpha)$ under Wigner time reversal, while in MC-QED a coherent photon states associated with wave vector $\mathbf{k}$ and phase $\alpha$ is predicted to transform into a coherent photon state associated with wave vector $-\mathbf{k}$ and phase $-(\alpha + \pi)$ under Wigner time reversal.

Nonetheless the distribution of photons in the Wigner time reversed coherent state for both QED and MC-QED obey a Poisson statistical distribution since by direct calculations we find that for both theories

$$_{Tw}\langle a(\alpha), \mathbf{k} | a(\alpha), \mathbf{k}\rangle_{Tw} = \langle a(\alpha), \mathbf{k} | a(\alpha), \mathbf{k}\rangle = \sum_{n=0,1,\ldots\infty} \Pi_n (\langle N\rangle)$$

Since reflection from phase conjugate mirrors physically creates the effects of time reversal on coherent optical beam of photons, this difference between QED and MC-QED should be observable in the context of optical interferometer experiments involving combinations of ordinary mirrors and phase conjugate mirrors.

Even though MC-QED is a non-local quantum field theory, the formal similarity between the quantum field theoretic structure of MC-QED and QED implies that the connection between the coherent photon state vectors $|a(\varphi), \mathbf{k}\rangle$ and the corresponding coherent classical electromagnetic photon fields which they represent is given by the expectation value over the coherent state $|a(\alpha), \mathbf{k}\rangle$ of the observed radiation charge-field $A_\mu^{(k)}{}_{(rad)}^{(obs)}(x)$ as

$$\langle a(\alpha), \mathbf{k}| A_\mu^{(k)}{}_{(rad)}^{(obs)}(x) |a(\alpha), \mathbf{k}\rangle$$

where in the above we have

$$|a(\alpha), \mathbf{k}\rangle = \sum_{n=0,1,\ldots\infty} [\exp(-\langle N\rangle/2) \, (\langle N\rangle^n / n!)^{1/2} \, \exp(in\alpha) \, |n, \mathbf{k}\rangle]$$

$$|n, \mathbf{k}\rangle = (1/(n!))^{1/2} \, \alpha^\mu(\mathbf{k})^\dagger \alpha^\nu(\mathbf{k})^\dagger \alpha^\eta(\mathbf{k})^\dagger \ldots |0\rangle$$

$$A_\mu^{(j)}{}_{(rad)}^{(obs)}(x) = A_\mu^{(j)}{}_{(-)}(x) = \int dj^3 / \sqrt{[2(2\pi)^3 k^2]} \, \{a_\mu^{(j)}{}_{(-)}(\mathbf{k}) \, e^{-ik \cdot x} + a_\mu^{(j)}{}_{(-)}(\mathbf{k})^\dagger e^{ik \cdot x}\}$$

and $\quad k \cdot x = k_\nu x^\nu = \mathbf{k} \cdot \mathbf{x} + k_0 x^0 = (\underline{k} \cdot \underline{x} - \nu t), \quad \mathbf{k} = \mathbf{n}/\lambda, \, k_0 = \nu/c$

Now since the MC-QED commutation relations imply that

$$[a_\nu^{(j)}{}_{(-)}(\mathbf{k}'), \alpha_\mu(\mathbf{k})^\dagger] = [\alpha_\mu(\mathbf{k}), \alpha_\nu(\mathbf{k}')^\dagger] = -\eta_{\mu\nu}\lambda_0 \, \delta^3(\mathbf{k}-\mathbf{k}')$$

then the action of the $a_\nu^{(j)}{}_{(-)}(\mathbf{k}')$ operator on $|n, \mathbf{k}\rangle$ produces the same effect as the action of the $\alpha_\mu(\mathbf{k})$ operator on $|n, \mathbf{k}\rangle$.

Hence $\alpha_{(}\mathbf{k}_{)} \, | a(\alpha), \mathbf{k}\rangle = a(\alpha) \, |a(\alpha), \mathbf{k}\rangle$ implies that $a_\nu^{(j)}{}_{(-)}(\mathbf{k})| a(\alpha), \mathbf{k}\rangle = a(\alpha) \, |a(\alpha), \mathbf{k}\rangle$ where $a(\alpha) = (\langle N\rangle^{1/2} \exp(i\alpha)$,. From this we see that

$$\langle a(\alpha), \mathbf{k}| a_\mu^{(j)}{}_{(-)(j)} \, e^{-i\lambda \cdot x} |a(\alpha), \mathbf{k}\rangle = e^{-ik \cdot x} \langle a(\alpha), \mathbf{k}| a_\mu^{(j)}{}_{(-)(\lambda)} |a(\alpha), \mathbf{k}\rangle$$
$$= e^{-ik \cdot x} a(\alpha) \langle a(\alpha), \mathbf{k}|a(\alpha), \mathbf{k}\rangle$$
$$= C(\langle N\rangle) \, e^{-i(k \cdot x - \alpha)}$$

where

$$C(\langle N\rangle) = (\langle N\rangle^{1/2} \sum_{n=0,1,\ldots\infty} [\exp(-\langle N\rangle) \, (\langle N\rangle^n / n)]$$

Taking the hermetian conjugate of the above equations we also see that

$$\langle a(\alpha), \mathbf{k}| \, a_\mu^{(j)}{}_{(-)}(\mathbf{k})^\dagger \, e^{i k \cdot x} \, |a(\alpha), \mathbf{k}\rangle = C(\langle N\rangle) \, e^{i(\lambda \cdot x - \alpha)}$$

Hence for a coherent photon state $|a(\alpha), \mathbf{k}\rangle$ with a mean photon number $\langle N \rangle$

$$\langle a(\alpha), \mathbf{k}| \, A_\mu^{(j)}{}_{(rad)}{}^{(obs)}(x) \, |a(\alpha), \mathbf{k}\rangle = C(\langle N\rangle) \int dk^3 / \sqrt{[2(2\pi)^3 \mathbf{k}^2]} \, \{e^{-i(k \cdot x - \alpha)} + e^{i(k \cdot x - \alpha)}\}$$

where $\quad C(\langle N\rangle) = (\langle N\rangle^{1/2} \sum_{n=0,1,\ldots\infty} [\exp(-\langle N\rangle) \, (\langle N\rangle^n / n)$

Recalling for a Wigner Time reversed coherent photon state that

$$|a(\alpha), \mathbf{k}\rangle_{Tw} = T_w \, |a(\alpha), \mathbf{k}\rangle = |a(-[\alpha + \pi]), -\mathbf{k}\rangle$$

then it follows that

$$_{Tw}\langle a(\alpha), \mathbf{k}| \, A_\mu^{(j)}{}_{(rad)}{}^{(obs)}(x) \, |a(\alpha), \mathbf{k}\rangle_{Tw}$$
$$= \langle a(-[\alpha + \pi]), -\mathbf{k}| \, A_\mu^{(j)}{}_{(rad)}{}^{(obs)}(x) \, |a(-[\alpha + \pi]), -\mathbf{k}\rangle$$
$$= C(\langle N\rangle) \int dk^3 / \sqrt{[2(2\pi)^3 \mathbf{k}^2]} \, \{e^{-i(-\underline{k} \cdot \underline{x} - vt - [\alpha + \pi])} + e^{i(-\underline{k} \cdot \underline{x} - vt + [\alpha + \pi])}\}$$

where $C(\langle N\rangle) = (\langle N\rangle^{1/2} \sum_{n=0,1,\ldots\infty} [\exp(-\langle N\rangle) \, (\langle N\rangle^n / n)$

Hence in MC-QED under Wigner time reversal we see that

$$C(\langle N\rangle) \int dk^3 / \sqrt{[2(2\pi)^3 \mathbf{k}^2]} \, \{e^{-i(k \cdot x - \alpha)}\}$$
$$\longrightarrow C(\langle N\rangle) \int dk^3 / \sqrt{[2(2\pi)^3 \mathbf{k}^2]} \, \{e^{-i(-\underline{k} \cdot \underline{x} - vt + [\alpha + \pi])}\}$$

While in QED under Wigner time reversal we see that

$$C(\langle N\rangle) \int dk^3 / \sqrt{[2(2\pi)^3 \mathbf{k}^2]} \, \{e^{-i(k \cdot x - \alpha)}\}$$
$$\longrightarrow C(\langle N\rangle) \int dk^3 / \sqrt{[2(2\pi)^3 \mathbf{k}^2]} \, \{e^{-i(-\underline{k} \cdot \underline{x} - vt + \alpha])}\}$$

Hence in a similar manner as that of QED the connection between the coherent photon state vectors $|a(\alpha), \mathbf{k}\rangle$ and the corresponding coherent classical electromagnetic photon fields which they represent is given in MC-QED by the expectation value over the coherent state $|a(\alpha), \mathbf{k}\rangle$ of the observed radiation charge-field $A_\mu^{(k)}{}_{(rad)}^{(obs)}(x)$. However in contrast to QED, where a coherent photon state associated with propagation vector $\mathbf{k}$ and phase $\alpha$ is predicted to transform into a coherent photon state associated with propagation vector $-\mathbf{k}$ and phase $-\alpha$ under Wigner time reversal, in MC-QED a coherent photon states associated with propagation vector $\mathbf{k}$ and phase $\alpha$ is predicted to transform into a coherent photon state associated with propagation vector $-\mathbf{k}$ and phase $-(\alpha + \pi)$ under Wigner time reversal.

# THE PHOTON CARRIES THE ARROW OF TIME

Darryl Leiter, Ph.D.
Interdisciplinary Studies Program, University of Virginia Charlottesville, Virginia 22904

## Abstract

In order to describe the quantum electrodynamic measurement process in a relativistic observer-participant manner, an operator symmetry of "microscopic observer-participation" called Measurement Color (MC) is incorporated into the field theoretic structure of Quantum Electrodynamics (QED) in the Heisenberg Picture. It is found that the resultant Measurement Color Quantum Electrodynamics (MC-QED) contains a microscopic quantum electrodynamic arrow of time that emerges dynamically, independent of any thermodynamic or cosmological assumptions. This occurs because the Measurement Color symmetry within MC-QED implies that <u>the photon carries the arrow of time</u>. In this context the physical requirement of a stable vacuum state in MC-QED dynamically selects operator solutions containing a causal, retarded, quantum electrodynamic arrow of time, which causes a <u>spontaneous symmetry breaking of both T and CPT to occur</u>. Spontaneous CPT symmetry breaking is consistent with the observed CP symmetry invariance seen in quantum electrodynamic particle interactions since, in the CPT symmetry breaking context of the MC-QED formalism, CP invariance is not physically equivalent to T invariance. In this manner the existence of the microscopic arrow of time in MC-QED offers a quantum electrodynamic explanation for the existence of irreversible phenomena which complements that supplied by the statistical arguments in phase space associated with the Second Law of Thermodynamics. In this context further development of the MC-QED formalism may lead to a resolution of the problems associated with: a) the connection between the description of the microscopic and macroscopic "Arrows of Time" in the universe, b) the connection between the description of microscopic quantum objects and macroscopic classical objects, and (c) the search for a physical explanation of how macroscopic conscious observers emerge from the microscopic laws of quantum physics.

Key Words: Quantum Field Theory, Elementary Particles, Cosmology, Philosophy of Science, Consciousness Studies

**SECTION I.  INTRODUCTION**

In justifying the validity of the Copenhagen Interpretation of Quantum Theory (CI-QM), Niels Bohr emphasized that it was meaningless to ascribe a complete set of physical attributes to a microscopic quantum object prior to the act of quantum measurement being performed on it. Hence in the context of the CI-QM only the probability of an outcome of a quantum measurement could be predicted deterministically.

These probabilities represented quantum potentia, associated with the expectation values of various physical operators over the quantum wave function, whose structure and unitary time evolution was described by the Schrodinger equation. Hence it followed from the CI-QM that the physical nature of "Objective Reality", associated with the quantum actua generated from the quantum potentia by the quantum measurement process, could never be described in a deterministically predictable manner.

However this picture of the universe represented by the CI-QM remains problematic because of the logical asymmetry built into it which states that: a) large classical macroscopic systems associated with measuring instruments have local objective properties independent of their observation, while at the same time  b) microscopic quantum systems have non-local properties which do not have an objective existence independent of  the "act of observation", generated by their quantum measurement interaction with the large classical macroscopic measuring instruments.

This is paradoxical because macroscopic measuring instruments are made up of large numbers of atomic micro-systems and because of this fact the direct coupling of  these macro-aggregates of atomic micro-systems to nonlocal quantum micro-systems must occur in Nature. Hence the entire Universe, at both the macroscopic and microscopic level, is susceptible to the ghostly non-local quantum weirdness which lies at the heart of Quantum Theory.

Given this fact the puzzle is why we don't experience this macroscopic non-local quantum weirdness in our daily lives. Most physicists believe that there can be only one unified set of laws for the whole Universe and that in this context the quantum laws are more fundamental than the classical laws. In this picture the quantum laws should apply to everything, from atoms to everyday objects like tables and chairs and macroscopic conscious living beings. However this can lead to contradictory predictions when macroscopically objective systems are directly coupled to microscopic quantum systems in a superposition of states.

In the early days of quantum theory Bohr and Heisenberg debated about this problem in regard to the relationship of the CI-QM to the existence of conscious macroscopic observers. In the context of these discussions Heisenberg felt that the CI-QM was incomplete since it was unable to explain the existence of living conscious observers. Bohr replied that the CI-QM could be considered to be complete if the existence of physical systems and living conscious observers were considered to be complementary and not contradictory ways of looking at Nature. However Heisenberg was unsatisfied with Bohr's reply and remained convinced that the problem of the describing the existence macroscopic conscious observers implied that the CI-QM was incomplete.

Further progress toward a better understanding of the quantum measurement process in the context of the CI-QM was made by John Wheeler who pioneered the development of a new paradigm called "The Observer Participant Universe" (OPU). Within the context of the OPU macroscopic living conscious observers directly participate in the process of irreversibly actualizing the elementary quantum phenomena which make up the universe. Wheeler emphasized that in the context of the OPU "No elementary quantum phenomenon is a phenomenon until it is an irreversibly recorded phenomenon".

However the previously described logical asymmetry associated with the interpretation of the CI-QM remained in Wheeler's version of the Observer-Participant Universe, since the dynamical manner in which macroscopic living conscious observers irreversibly actualize microscopic elementary quantum phenomena was still unexplained. In order to address this problem the structure of this paper is as follows:

Section II presents a brief review of the structure of standard QED formalism in order to make it easier for the reader to better understand the issues discussed in the Introduction. The key points discussed in this brief review of QED will allow the reader will find it easier to understand the development of the observer-participant MC-QED formalism discussed in Section III

Section III uses the results of section II and generalizes them to in order to develop the MC-QED formalism (Leiter, D., 2009). This is done by requiring that the Abelian operator gauge symmetry of microscopic operator observer-participation called Measurement Color (Leiter, D., 1983, 1985, 1989) be incorporated into the operator equations of quantum electrodynamics in the Heisenberg picture. In this manner we show how a logical symmetry in regard to the quantum field theoretic definition of the "observer" and the "object" can be formally created at the microscopic level.

Section IV concludes by pointing out that the challenge of determining what is ultimately possible in physics will require the resolution of three fundamental issues : (1) the origin of the arrow of time in the universe; (2) the nature of objective existence in the context quantum reality, and (3) the spontaneous emergence of macroscopic conscious minds in the universe. It is then argued that in the context of the new paradigm of MC-QED the resolution of these three fundamental issues may be found within the paradigm of an observer-participant universe where the photon carries the Arrow of Time. It is then pointed out that the existence of the causal microscopic arrow of time in MC-QED represents a fundamentally quantum electrodynamic explanation for irreversible phenomena associated with the Second Law of Thermodynamics which complements the one supplied by the well-known statistical arguments in phase space

**SECTION II. THE STANDARD FORM OF QUANTUM ELECTRODYNAMICS**

The problem about the logical asymmetry between the object and observer discussed in the introduction makes itself felt directly within the context of the Copenhagen Interpretation of Quantum Electrodynamics (CI-QED).

In the CI-QED in the physical world is assumed to be arbitrarily divided into two complementary components: a) a microscopic quantum field theoretical world consisting of electrons positrons and photons , and b) a macroscopic classical world of macroscopic measuring instruments in which "macroscopic conscious observers" reside.

Clearly the validity of the Copenhagen Interpretation division of the world in CI-QED is limited to physical situations where, during the period of time between their preparation and detection, the microscopic quantum systems have no significant influence upon the classically described macroscopic measuring instruments.
However this cannot be always be the case since, as the physical size of an aggregate of microscopic quantum systems under consideration increases, a point must eventually be reached where the aggregate of quantum systems is neither small enough to have a negligible influence on the classically described measuring instruments, nor large enough to be able to be described in purely classical terms. For physical situations of this type, the basic assumption about the asymmetric nature of the classical/quantum interface which underlies the CI-QED is no longer valid. In order to resolve this paradox it is clear that a new formulation of the observer-participant quantum electrodynamic measurement paradigm is needed which is able to go beyond the limitations of the Copenhagen Interpretation. This new formulation of the quantum electrodynamic measurement process will be discussed in detail in Section III.

However in order to make it easier for the reader to better understand the manner in which this new paradigm of Measurement Color is inserted into the quantum electrodynamic operator formalism of QED in the Heisenberg Picture, Section II will be devoted to presenting a brief review of the structure of standard QED formalism. Then by using this discussion of standard QED as our logical baseline, we will find that the development of the observer-participant MC-QED formalism in Section III will follow in a natural easy to understand manner.

In Section II and Section III we will be using the metric signature (1,-1, -1,-1), and natural $(h/2\pi) = c = 1$ units. The relativistic notation, operator sign conventions, and operator calculation techniques, used below to generalize and extend the standard QED theory into the MC-QED theory, will be formally similar to those used in the book "Introduction to Relativistic Quantum Field Theory" by (Schweber, S., 1962).

We begin our brief review of the structure of the QED formalism in the Heisenberg picture starting with the standard charge-conjugation invariant QED action given by

$$I = \left\{ -\int dx^4 \left[ (1/4[\psi^\dagger \gamma^o, (-i\gamma^\mu \partial_\mu + m)\psi] + \text{Hermetian conjugate}) + (1/2 \partial_\mu A^\nu \partial^\mu A_\nu + J_\mu A^\mu) \right] \right\}$$

In the QED action written above, $\psi$ is the electron-positron field operator, $A^\mu$ is the electromagnetic field operator, and $J_\mu = -e[c\psi^\dagger \gamma^o, \gamma_\mu \psi]$ is the electromagnetic current operator. In the Heisenberg picture, by applying standard second quantization methods applied to the above action, we find that the QED Heisenberg operator equations of motion are given by

$$[(-ih/2\pi)\gamma^\mu \partial_\mu + mc + (e/c)\gamma^\mu A_\mu]\psi = 0 \quad \text{(Heisenberg equation for } \psi \text{ fermion operator)}$$

$$\Box^2 A_\mu = -e[c\psi^\dagger \gamma^o, \gamma_\mu \psi] \quad \text{(Heisenberg equation for the } A_\mu \text{ photon operator)}$$

For the convenience of scientific engineering readers who wish to check the dimensional consistency of the above QED operator equations of motion, we have written them using the (cgs) unit values of Planck's constant $(h/2\pi)$ and the speed of light c. However from this point on in order to maintain simplicity these and all other operator equations will be expressed in terms of the natural units where $(h/2\pi) = c = 1$.

In the indefinite metric Hilbert space of QED, the Heisenberg Picture State Vector $|\Psi\rangle$ is required to obey the Subsidiary Condition

$$\langle \Psi | (\partial_\mu A^\mu) | \Psi \rangle = 0$$

In addition the Heisenberg Picture operator equations of MC-QED are invariant under the Abelian Measurement Color gauge transformation

$$\psi'(x) = \psi(x) \exp(ie\Lambda(x))$$
$$A_\mu'(x) = A_\mu(x) + \partial_\mu \Lambda(x)$$

where $\Lambda(x)$ is a scalar field obeying $\Box^2 \Lambda(x) = 0$. Hence the current operator $J_\mu$ is conserved as $\partial^\mu J_\mu = 0$ which implies that the total charge operators $Q = \int dx^3 J_0$ commutes with the total Hamiltonian operator of the theory.

Following the standard procedures for the canonical quantization of fields applied to CED leads to the canonical equal-time commutation and anti-commutation relations in the QED formalism as

$$[A_\mu(x, t), \partial_t A_\nu(x', t)] = i\eta_{\mu\nu}\delta^3(x' - x)$$

$$\{\psi(x, t), \psi^\dagger(x', t)\} = \delta^3(x' - x)$$

$$[A_\mu(x, t), \psi(x', t)] = [A_\mu(x, t), \psi^\dagger(x', t)] = 0$$

where $\text{sig}(\eta_{\mu\nu}) = (1, -1, -1, -1)$, with other equal-time commutators and anti-commutators vanishing respectively.

In QED the most general operator solution to the operator equations of motion for $A_\mu$ is given by

$$A_\mu = A_\mu(+) + A_\mu^{(0)} = A_\mu(\text{ret, adv}) + A_\mu(\text{in, out})$$

where

$$A_\mu(+) = (A_\mu(\text{ret}) + A_\mu(\text{adv}))/2$$

and $A_\mu^{(0)}$ is the time symmetric free uncoupled photon operator.

It is useful to write $A_\mu$ in the equivalent form as

$$A_\mu = [(1+p)/2] A_\mu(\text{ret}) + [(1-p)/2] A_\mu(\text{adv}) + A_\mu(\text{rad, p})$$

where p is a c-number, and

$$A_\mu(\text{rad, p}) = A_\mu^{(0)} - p A_\mu(-)$$

With

$$A_\mu(-) = (A_\mu(\text{ret}) - A_\mu(\text{adv}))/2$$

The operator solution $A_\mu$ to the Maxwell operator equations are separately invariant under the "Radiation Flow Symmetry Operator" $T_p$ which changes the effects of "retarded fields into advanced fields by taking the value of the c-number p in the above equations and changing it into the c-number -p.

In addition the operator equations are separately invariant under the Wigner Time Reversal operation $T_W$ for which the combination of t → -t and complex conjugation occurs.

Hence the above QED formalism in the Heisenberg Picture is invariant under the generalized Time Reversal operator $T = T_W \times T_p$ which is the product of the Wigner Time Reversal Operator $T_t$ and the Radiation Flow Symmetry Operator $T_p$.

In addition the operator fields and their operator equations of motion of the QED formalism in the Heisenberg Picture in (1-a) and (1-b) above are also invariant under the action of the Charge Conjugation operator C, the Parity operator P. Thus it follows that QED is invariant under the CPT symmetry where $T = T_W \times T_p$.

In solving the above Wigner Time Reversal Operator $T_W$ invariant QED formalism to obtain the S-matrix one usually defines the "in-out" operator field solutions by imposing the "Asymptotic Condition" on the expectation values of the operator equations of motion. This is done first in the t → -∞) limit where $\psi(x, t \to -\infty) = \psi(in)$ as

$<A_\mu (x, t \to -\infty)> = <(\int dx^3 (J_\mu(x', t \to -\infty) / 4\pi |x-x'|) + A_\mu(in) >$     (kinematic condition)

$<\partial_t J_\mu (x, t \to -\infty)> = 0$     (dynamic stability condition)

where under these conditions the expectation value of the operator equations of motion become

$<(-i\gamma^\mu \partial_\mu + m - e\gamma^\mu A_\mu(x, t \to -\infty))\psi(in)> = 0$

$<\Box^2 A_\mu(in)> = -e <[\psi(in)\dagger\gamma^0, \gamma_\mu \psi(in)]>$

and then second in the limit as t → ∞) where $\psi(x, t \to \infty) = \psi(out)$ as

$<A_\mu (x, t \to +\infty)> = <(\int dx^3 (J_\mu(x', t \to +\infty) / 4\pi |x-x'|) + A_\mu(out) >$     (kinematic condition)

$<\partial_t J_\mu (x, t \to \infty)> = 0$     (dynamic stability condition)

where under these conditions the expectation value of the operator equations of motion become

$<(-i\gamma^\mu \partial_\mu + m - e\gamma^\mu A_\mu(x, t \to -\infty))\psi(out)> = 0$

$<\Box^2 A_\mu(out)> = -e <[\psi(out)\dagger\gamma^0, \gamma_\mu \psi(out)]>$

where in the above

$$A_\mu (\text{in, out}) = A(\text{rad}, p = \pm 1) = A_\mu^{(0)} - [\pm A_\mu(-)]$$

and hence

$$A_\mu(\text{out}) = A_\mu(\text{in}) + 2 A_\mu(-)$$

The kinematic components of the Asymptotic Conditions then respectively determine the values of the c-number p in the operator equations of motion to be either $p = 1$ or $p = -1$. However, because of the presence of the time symmetric free radiation field operators $A_\mu^{(0)}$ in QED, the dynamic stability components of the Asymptotic Conditions cannot physically distinguish between the $p = 1$ and $p = -1$ cases. To see this more clearly we note that under the Wigner Time Reversal Operator $T_w$ we have respectively that

$$T_w A_\mu(\text{in}) T_w^{-1} = A^\mu(\text{out})$$

$$T_w A_\mu(\text{ret}) T_w^{-1} = A^\mu(\text{adv})$$

Hence it follows that $A_\mu$ is invariant under $T_w$ since

$$T_w A_\mu T_w^{-1} = T_w [A_\mu(\text{ret}) + A_\mu(\text{in})] T_w^{-1} = [A_\mu(\text{adv}) + A_\mu(\text{out})] = A^\mu$$

In addition $A_\mu$ is invariant under since $T_p A_\mu T_p^{-1} = A_\mu$, Hence we see that QED is invariant under the generalized time reversal symmetry $T = T_w \times T_p$ as well as being separately invariant under both the $T_t$ and $T_p$ operations. Since QED is also invariant under the parity operation P and the Charge Conjugation operation C it follows that QED is also invariant under the CPT symmetry operation where $T = T_w \times T_p$. This is due to the presence of the time-symmetric the free photon field operator $A_\mu^{(0)}$ in QED which allows a stable vacuum state to exist in the context of the CPT invariance of the formalism.

Since QED is a local, relativistic, quantum field theory it obeys the CPT symmetry. The CPT symmetry is a fundamental property of all local, relativistic, quantum field theories, where T is the Wigner Time Reversal symmetry, P is the Parity Inversion symmetry, and C is the Charge Conjugation symmetry.

The fact that QED is CPT invariant implies that a CPT transformed version of the universe is an observable solution to the QED formalism. A CPT transformed version of the universe is one in which: a) positions of all objects are reflected by an imaginary plane mirror (parity inversion P); b) momenta of all objects are reversed (corresponding to time inversion T) and; c) all matter is replaced by antimatter (corresponding to charge conjugation C). The preservation of the CPT symmetry implies that every violation of the combined symmetry of two of its components (such as CP) must have a corresponding violation in the third component (such as T). Since QED and its Standard Model generalizations obey the CPT symmetry, in this context the observation CP symmetry in particle interactions is interpreted to be physically equivalent to the observation of T symmetry.

Because QED does not have a dynamically chosen microscopic arrow of time one must insert an arrow of time by hand. A causal retarded arrow of time can be imposed on the QED formalism in the Heisenberg picture by appealing to the Thermodynamic arrow of increasing entropy. This justifies the use of a low entropy boundary condition on the expectation values of the $A_\mu(in)$ operators in the far past of the Heisenberg Picture such that that all photons vanish for the $|\psi\rangle$ state vector at t → -∞ as

$$<\psi|A_\mu(in)|\psi> = 0.$$

<u>Hence in the context of the QED formalism in the Heisenberg Picture, the imposition of the Asymptotic Condition does not dynamically determine a Physical Arrow of Time and this implies that the Thermodynamic arrow of increasing entropy is the master time asymmetry in the universe</u> .

### III. MEASUREMENT COLOR QUANTUM ELECTRODYNAMICS

In this section will generalize and extend the results of section II and show that the requirement of a logical symmetry in regard to the definition of the "observer" and the "object" can be accomplished at the microscopic level by requiring that an Abelian operator gauge symmetry of microscopic operator observer-participation called Measurement Color (Leiter, D., 1983, 1985, 1989) be incorporated into the operator equations of quantum electrodynamics in the Heisenberg picture.

In this context we will show that the resultant formalism Measurement Color Quantum Electrodynamics (MC-QED) (Leiter, D. 2009) takes the form of a nonlocal, microscopically observer-participant quantum field theory, in which a microscopic electrodynamic arrow of time dynamically emerges independent of any external thermodynamic or cosmological assumptions.

We will also show that the MC-QED quantum electrodynamic arrow of time emerges dynamically because the microscopic observer-participant operator structure of the formalism implies that the local time-symmetric "free photon operator" is non-physical since it cannot be given a Measurement Color description. Instead it must be replaced by a Measurement Color Symmetric Total Coupled Radiation Charge-Field photon operator which is non-local and carries a negative time parity under the Wigner Time Reversal operator $T_W$.

However because of this difference between the QED and MC-QED photon operator structure, we will find that the physical requirement that a stable vacuum state exists dynamically constrains the MC-QED Heisenberg operator equations of motion to contain a causal retarded quantum electrodynamic arrow of time independent of external thermodynamic or cosmological assumptions.

In this manner the existence of the microscopic arrow of time in MC-QED will be shown to represent a fundamentally quantum electrodynamic explanation, for irreversible phenomena associated with the Second Law of Thermodynamics, which complements the one supplied by the well-known statistical arguments in phase space.

Measurement Color Quantum Electrodynamics (Leiter, D., 2009) is constructed by imposing an Abelian operator gauge symmetry of <u>microscopic operator observer-participation</u> called Measurement Color onto the operator equations of motion of standard Quantum Electrodynamics (QED) in the Heisenberg picture.

The Measurement Color symmetry is an Abelian operator labeling, associated with the integer indices k = 1,2, …, N in the limit as N --> ∞, which is imposed in an operational manner onto the <u>both</u> the electron-positron operators <u>and</u> the photon operators within the quantum field structure of the standard QED formalism. However since MC-QED is a theory of mutual quantum field theoretic observer-participation, its action principle must be constructed in a manner such that self-measurement interaction terms of the form $J_\mu^{(k)} A^{\mu(k)}$ (k=1,2,… , N→ -∞) are dynamically excluded from the formalism.

The MC-QED formalism which emerges operationally describes the microscopic observer-participant quantum electrodynamic process, between the electron-positron quantum operator fields $\psi^{(k)}$ and the charge field photon quantum operator fields $A_\mu^{(j)}$ (k ≠ j) which they interact with, in the Heisenberg Picture operator field equations.

Generalizing from the discussion in Section II about the QED formalism in the Heisenberg picture, it follows in this Measurement Color context that this can dynamically accomplished by constructing the charge-conjugation invariant MC-QED action principle given by

$$I = \{- \int dx^4 \ [\Sigma_{(k)} (1/4[\psi^{(k)\dagger} \gamma^o, (-i\gamma^\mu \partial_\mu + m)\psi^{(k)}] + \text{Hermitian conjugate}) + \Sigma_{(k)} \Sigma_{(j \neq k)} (1/2 \partial_\mu A^{\nu(k)} \partial^\mu A_\nu^{(j)} + J_\mu^{(k)} A^{\mu(j)})]\}$$

where (k, j =1,2,… , N→ -∞) and h / 2π = c = 1 natural units are being used. In the Heisenberg picture applying the standard second quantization methods taken to the above action for MC-QED we find that the MC-QED Heisenberg operator equations of motion are given by

$$(-i\gamma^\mu \partial_\mu + m - e\gamma^\mu A_\mu^{(k)}{}_{(obs)})\psi^{(k)} = 0 \qquad \text{(Heisenberg equation for } \psi^{(k)} \text{ fermion operator)}$$

$$A_\mu^{(k)}(\text{obs}) = \sum_{(j \neq k)} A_\mu^{(j)} \qquad \text{(electromagnetic operator field } A_\mu^{(k)}(\text{obs}) \text{ observed by } \psi^{(k)})$$

$$\Box^2 A_\mu^{(k)} = J_\mu^{(k)} = -e[\psi^{(k)\dagger}\gamma^o, \gamma_\mu \psi^{(k)}] \qquad \text{(Heisenberg equation for the } A_\mu^{(k)} \text{ operator)}$$

where the Measurement Color labels on the operator fields $\psi^{(k)}$, and $A_\mu^{(k)}$ range over (k= 1,2, , N --> ∞). In the context of an indefinite metric Hilbert space, the Subsidiary Condition

$$\langle \Psi | (\partial^\mu A_\mu^{(k)}) | \Psi \rangle = 0 \qquad (k =1,2, \ldots, N \to \infty))$$

must also be satisfied. Then the expectation value of the Heisenberg Picture operator equations of MC-QED are will be invariant under the Abelian Measurement Color gauge transformation (k =1,2, … , N → -∞)

$$\psi^{(k)'}(x) = \psi^{(k)}(x)\exp(ie\Lambda(x))$$

$$A_\mu^{(k)'}(x)_{(\text{obs})} = A_\mu^{(k)}(x)_{(\text{obs})} + \partial_\mu \Lambda(x)$$

where $\Lambda(x)$ is a scalar field obeying $\Box^2 \Lambda(x) = 0$

Hence the individual Measurement Color currents $J_\mu^{(k)}$ are conserved as $\partial^\mu J_\mu^{(k)} = 0$ (k =1,2, … , N → -∞) which implies that the individual Measurement Color charge operators

$$Q^{(k)} = \int dx^3 J_0^{(k)} \qquad (k =1,2, \ldots, N \to -\infty)$$

commute with the total Hamiltonian operator of the theory.

Following the standard procedures for the canonical quantization of fields applied to MC-CED leads to the canonical equal-time commutation and anti-commutation relations in the MC-QED formalism as

$$[A_\mu^{(k)}(x, t), \partial_t A_\nu^{(j)}(\text{obs})(x', t)] = i\eta_{\mu\nu}\delta^{kj}\delta^3(x' - x)$$

$$\{\psi^{(k)}(x, t), \psi^{(j)\dagger}(x', t)\} = \delta^{kj}\delta^3(x' - x) \qquad (k, j =1,2, \ldots, N \to \infty)$$

$$[A_\mu^{((k)}(x, t), \psi^{(j)}(x', t)] = [A_\mu^{((k)}(x, t), \psi^{(j)\dagger}(x', t)] = 0$$

with all other equal-time commutators and anti-commutators vanishing respectively,

$(k, j = 1, 2, \ldots, N \to \infty)$.

In this context the structure of the MC-QED operator equations of motion and the equal-time commutation and anti-commutation relations dynamically enforces a form of mutual operator observer-participation which dynamically excludes time-symmetric Measurement Color self-interaction terms of the form $e\gamma^\mu A_\mu^{(k)} \psi^{(k)}$ $(k = 1, 2, \ldots, N \to \infty)$ from the operator equations of motion.

In solving the Measurement Color Maxwell equations for the charge-fields $A_\mu^{(k)}$ within the context of the multi-field theoretic Measurement Color paradigm upon which MC-QED is based, "local time-symmetric free radiation field operators uncoupled from charges" $A_\mu^{(0)}$ must be excluded from the $A_\mu^{(k)}$ charge-field solutions since the $A_\mu^{(0)}$ fields cannot be defined in terms of Measurement Color charge-fields. This is in contrast to the case of QED where $A_\mu^{(0)}$ cannot be excluded from $A_\mu$ since Measurement Color does not play a role in its Maxwell field operator structure.

Hence in solving the Maxwell field operator equations for the electromagnetic field operators $A_\mu^{(k)}$ the MC-QED paradigm implies that a universal time-symmetric boundary condition, which mathematically excludes local time reversal invariant free uncoupled radiation field operators $A_\mu^{(0)}$ from contributing to the charge-field operators $A_\mu^{(k)}$ must be imposed on <u>each</u> of the $A_\mu^{(k)}$ operator solutions to the Maxwell operator equations. However since the operator boundary condition requirement which excludes $A_\mu^{(0)}$ operators is time-symmetric it does not by itself choose an arrow of time in the MC-QED formalism.

Hence within MC-QED "free uncoupled radiation field operators" $A_\mu^{(0)}$ are operationally excluded from MC-QED and in their place the physical effects of radiation are operationally described in a microscopic observer-participant manner by the measurement color symmetric, time anti-symmetric, total coupled radiation field operator"

$$A_\mu^{(TCRF)} = \sum_{(k)} A_\mu^{(k)}(-) \neq 0.$$

where

$$A_\mu^{(k)}(-) = 1/2 \int dx^{4'} (D_{(ret)}(x-x') - D_{(adv)}(x-x')) J^{(k)}(x')$$

In this context it follows that, in the operational observer-participant context of the MC-QED, that the Heisenberg operator field equations, the electron-positron operator fields $\psi^{(k)}$ $(k = 1, 2, \ldots N)$ "observe" the electromagnetic charge-field operator $A_\mu^{(k)}(obs)$ given by

$$A_\mu^{(k)}(obs) = \sum_{(k\neq j)} A_\mu^{(j)} = \sum_{(j\neq k)} A_\mu^{(j)}(+) + pA_\mu^{(TCRF)} \quad (k = 1, 2, \ldots, N \to \infty)$$

where

$$A_\mu^{(j)}(+) = 1/2 \int dx^{4'} (D_{(ret)}(x-x') + D_{(adv)}(x-x'))J^{(j)}(x'), \quad (j\neq k = 1, 2, \ldots N)$$

and

$A_\mu^{(TCRF)} = \sum_{(k)} A_\mu^{(k)}(-) \neq 0$ is the negative time parity total radiation charge-field operator
In the above the quantity p is a c-number constant whose value determines the amount of mixing between the time-symmetric and time anti-symmetric charge field operators which occur in the charge-field operator $A_\mu^{(k)}(obs)$.

For our purposes it is also useful to also write the $A_\mu^{(k)}(obs)$ charge-field operator in the equivalent form

$$A_\mu^{(k)}(obs) = [(1+p)/2] A_\mu^{(k)}(obs)(ret) + [(1-p)/2] A_\mu^{(k)}(obs)(adv) + A_\mu^{(k)}(obs)(rad, p)$$

where the $A_\mu^{(k)}(obs)(rad, p)$ are the negative time parity coupled charge-field photon "in and out" operators are defined as

$$A_\mu^{(k)}(obs)(rad, p) = p [A_\mu^{(TCRF)} - \sum_{(j\neq k)} A_\mu^{(j)}(-)] = p A_\mu^{(k)}(-)$$

Note that while the time-symmetric local free radiation field uncoupled to charges $A_\mu^{(0)}$ vanish in the MC-QED formalism, the time reversal violating nonlocal total coupled radiation field $A_\mu^{(TCRF)} = \sum_{(k)} A_\mu^{(k)}(-) \neq 0$ and its associated in-out fields do not vanish.

For this reason a consistent MC-QED quantum electrodynamic formalism is possible. Now in the <u>absence</u> of non-operational free radiation fields $A_\mu^{(0)}$, the <u>presence</u> of the negative time parity Total Coupled Radiation Field operator $A_\mu^{(TCRF)}$ in MC-QED implies that the MC-QED operator equations violate the following symmetries:

a) The "Radiation Flow Symmetry Operator" $T_p$, for which p → -p occurs, is violated in the operator equations (3) since they have a <u>negative parity</u> under the $T_p$ operation

b) The Wigner Time Reversal operator symmetry $T_W$, for which the combination of t → -t and complex conjugation occurs, is violated in the operator equations since by virtue of the presence of the Total Coupled Radiation Field operator $A_\mu^{(TCRF)}$

they have a <u>negative parity</u> under the $T_w$ operation

However, even though equations separately <u>violate</u> the $T_p$ and the $T_w$ symmetry, they are <u>invariant</u> under the generalized Time Reversal operator $T = T_w \times T_p$ which is the product of the Wigner Time Reversal Operator $T_w$ and the Radiation Flow Symmetry Operator $T_p$. Since the operator field equations of motion of the MC-QED formalism in the Heisenberg Picture are also invariant under the respective action of the Charge Conjugation operator C, and the Parity operator P, then even though it violates the $T_w$ time reversal symmetry, we find that MC-QED is CPT invariant where the T symmetry is generalized to become $T = T_w \times T_p$.

Now in the context of expectation values of the operator equations of motion in the Heisenberg Picture taken over the Heisenberg state vector $|\psi\rangle$, one can define the "in-out" operator field solutions to MC-QED by imposing time-symmetric same kind of Asymptotic Conditions as that which is done in the case of standard QED.

Hence the "In-Asymptotic Condition" is imposed in the limit as

$$\psi^{(k)}(x, t \to -\infty) = \psi^{(k)}(in) \qquad (k = 1, 2, \ldots, N \to \infty)$$

From which it follows that

*(In-kinematic condition)*

$$\langle A_\mu^{(k)}{}_{(obs)}(x, t \to -\infty)\rangle = \langle (\int dx^3 \, (J_\mu^{(k)}{}_{(obs)}(x', t \to -\infty) / 4\pi |x-x'|) + A^{(k)}{}_{(obs)}(in)\rangle$$

*(in-dynamic stability condition)*

$$\langle \partial_t J_\mu^{(k)}(x, t \to -\infty)\rangle = 0$$

Then in the limit as $t \to -\infty$ of the operator equations (1) become

$$\langle(-i\gamma^\mu \partial_\mu + m - e\gamma^\mu A_\mu^{(k)}{}_{(obs)}(x, t \to -\infty))\psi^{(k)}(in)\rangle = 0$$
$$\langle \Box^2 A_\mu^{(k)}(in)\rangle = \langle J_\mu^{(k)}(in)\rangle = -e \langle [\psi^{(k)}(in)^\dagger \gamma^o, \gamma_\mu \psi^{(k)}(in)]\rangle$$

In addition we also impose the "Out-Asymptotic Condition" as

$$\psi^{(k)}(x, t \to +\infty) = \psi^{(k)}(out) \qquad (k = 1, 2, \ldots, N \to \infty))$$

*(out-kinematic condition)*

$$< A_\mu^{(k)}(obs)(x, t \to +\infty)> = <(\int dx^3 (J_\mu^{(k)}(obs)(x', t \to +\infty) / 4\pi |x-x'|) + A^{(k)}(obs)(out)>$$

*(out-dynamic stability condition)*

$$<\partial_t J_\mu^{(k)}(x, t \to +\infty)> = 0$$

Then in the limit as $t \to +\infty$ of the operator equations (1) become

$$<(-i\gamma^\mu \partial_\mu + m - e\gamma^\mu A_\mu^{(k)}(obs)(x, t \to +\infty))\psi^{(k)}(out)> = 0$$
$$<\Box^2 A_\mu^{(k)}(out)> = <J_\mu^{(k)}(out)> = -e <[\psi^{(k)}(out)^\dagger \gamma^o, \gamma_\mu \psi^{(k)}(out)]>$$

Now by applying these Asymptotic Conditions to the MC-QED operator equations of motion it follows that a retarded quantum electrodynamic arrow of time emerges dynamically. This is because within the $A_{\mu\,(obs)}^{(k)}$ defined above we find that:

a) *The kinematic component of the Asymptotic Condition* formally determines two possible values for the c-number p which controls the arrow of time in the operator equations to be either p =1 or p = - 1, where

$$A_{\mu\,(obs)}^{(k)}(in, out)^{(k)} = A_{(obs)}^{(k)}(rad, p = \pm 1) = \pm A_\mu^{(k)}(-)$$

b) *The dynamic component of the Asymptotic Condition*, which is associated with the stability of the vacuum state, dynamically requires that the physical value of the c-number p which appears in the operator equations to be p=1 associated with a retarded, causal, quantum electrodynamic arrow of time.

We can see this more specifically by noting that for the case of p =1 the Heisenberg Picture operator equations of motion have the form

$$<(-i\gamma^\mu \partial_\mu + m - e\gamma^\mu A_\mu^{(k)}(obs))\psi^{(k)}> = 0$$
$$<A_\mu^{(k)}(obs)> = <\sum_{(j \neq k)} A_\mu^{(j)}(ret) + A^{(k)}(-)> \qquad (k, j = 1,2, \dots, N \to \infty))$$

The expectation value of the above operator equations physically describe the situation where charge field photons are *causally emitted and absorbed* between the $\psi^{(k)}$ and $\psi^{(j)}$ $k \neq j$ fermion operators, while *being spontaneously emitted into the vacuum* by the $\psi^{(k)}$ fermion operators, (k, j = 1,2, ... , N--> ∞)). For this reason these operator equations predict that electron-positron states can form bound states which spontaneously decay into charge field photons.

<u>Hence the p = 1 operator equations will satisfy the dynamic stability component of the Asymptotic Condition because they predict that a stable vacuum state exists.</u>

On the other hand for the case of p = -1 the Heisenberg Picture operator equations of motion have the form

$$\langle(-i\gamma^\mu\partial_\mu + m - e\gamma^\mu A_\mu^{(k)}{}_{(obs)})\psi^{(k)}\rangle = 0$$

$$\langle A_\mu^{(k)}{}_{(obs)}\rangle = \langle \sum_{(j \neq k)} A_\mu^{(j)}(adv) - A^{(k)}(-)\rangle \qquad (k, j = 1, 2, \ldots, N \to \infty))$$

On the other hand the expectation value of these operator equations physically describe the situation where charge field photons are *acausally absorbed and emitted* between the $\psi^{(k)}$ and $\psi^{(j)}$ $k \neq j$ fermion operators, while *being spontaneously absorbed from the vacuum* by the $\psi^{(k}$ fermion operators, $(k, j = 1, 2, , N \to \infty))$.
For this reason these operator equations predict that electron-positron states will be spontaneously excited from the vacuum.

<u>Hence the p = -1 operator equations cannot satisfy the dynamic stability component of the Asymptotic Condition because they predict that a stable vacuum state cannot exist.</u>

Hence in MC-QED the action of the nonlocal negative time parity Total Coupled Radiation Field $A_\mu^{(TCRF)} = \sum_{(k)} A_\mu^{(k)}(-) \neq 0$, in conjunction with the time-symmetric Asymptotic Condition on the operator field equations, implies that the requirement of a stable vacuum state in the MC-QED formalism dynamically determines the choice of a retarded quantum electrodynamic Physical Arrow of Time associated with the p =1 operator equations in, independent of requiring that any Thermodynamic or Cosmological boundary conditions (Zeh, D., 2007) be imposed on the MC-QED formalism.

This is because, in contradistinction to the case of QED, the existence of a causal arrow of time in MC-QED does <u>not</u> require the boundary condition $\langle A^{(k)}_{\mu\,(obs)}(in)^{(k)}\rangle = 0$ associated with the low entropy assumption that the contribution of all photons, which occur in the expectation value of the Heisenberg state vector, must vanish as time goes to minus infinity as

In the context of the multi-field-operator theoretic Measurement Color paradigm upon which MC-QED is based, a microscopic, causal, electrodynamic arrow of time exists in the universe, (independent of any additional external thermodynamic or cosmological assumptions), because the dynamic role of the free photon operator (which is absent in MC-QED since it cannot be given a Measurement Color description) is replaced by the measurement color symmetric, negative time parity Total Coupled Radiation Charge-Field Photon operator in the MC-QED formalism.

This result can be understood as being due to the phenomenon of <u>spontaneous symmetry breaking with respect to time reversal invariance</u> which occurs in terms of the following logical sequence in the MC-QED formalism:

a) In the parameterized solutions to the time-symmetric MC-QED operator equations of motion, the Measurement Color symmetry in MC-QED ruled out the existence the local time-symmetric photon operator in favor of the negative time parity total coupled radiation charge-field photon operator;

b) Application of standard time-symmetric asymptotic conditions to these parameterized time-symmetric solutions to the MC-QED operator equations of motion selected out the time-symmetric pair of operator causal and acausal operator charge-field solutions which were parameterized respectively by the c-numbers $p = 1$ and $p = -1$;

c) Because of the presence of the negative time parity total, coupled radiation charge-field photon operator within the time symmetric $p = \pm 1$ pair of parameterized solutions to the MC-QED operator equations of motion, the physical requirement of a stable vacuum state spontaneously broke this time-reversal symmetry by dynamically selecting the $p = 1$ solution with a causal, retarded, quantum electrodynamic arrow of time.

On a more general level since MC-QED is a relativistic quantum field theory then CPT symmetry is conserved by the operator solutions to the operator equations of motion. However the T symmetry in MC-QED is generalized to become the product of the Wigner Time Reversal $T_w$ and Radiation Flow Reversal $T_p$ operators as $T = T_w \times T_p$.

Then in this context the physical requirement of a stable vacuum state in MC-QED spontaneously breaks the T and the CPT symmetry by dynamically selecting the operator solution containing a causal, retarded, quantum electrodynamic arrow of time.

Since MC-QED is a non-local quantum field theory in which the photon carries the arrow of time, the requirement of a stable vacuum state spontaneously breaks the CPT symmetry and leads to solutions which are CP invariant but not T invariant. The spontaneous breakdown of CPT symmetry in MC-QED implies that the CPT transformation cannot turn our universe into its "mirror image". This occurs because the photon carries the arrow of time in MC-QED which implies that time in the universe can only run forward in a causal sense and not backward. For MC-QED and its Standard Model generalizations, C, P, and CP symmetry is preserved but CPT symmetry is spontaneously violated. Hence it follows that in the context of the MC-QED formalism, the observed invariance of CP in quantum electrodynamics is not physically equivalent to T invariance.

Having used the MC-QED formalism to resolve the apparent asymmetry in the description of the microscopic and macroscopic "Arrows of Time" in the universe, we can next apply it to the problem of the asymmetry between microscopic quantum objects and macroscopic classical objects inherent in the laws of quantum physics.

We begin by first noting that the origin this problem lies within the nature of Copenhagen Interpretation of QED. This is because within QED macroscopic bodies, associated with macroscopic measuring instruments and macroscopic conscious observers, are assumed to obey a strict form of "Macroscopic Realism", on a complementary classical level of physics external to the microscopic quantum electrodynamic system. Macroscopic bodies that satisfy the strict form of Macroscopic Realism are assumed have the property that they are at all times in a macroscopically distinct state which can be observed without affecting their subsequent behavior.

However this concept of strict Macroscopic Realism is not valid for the case of MC-QED because its Measurement Color symmetry implies that the photon operator carries the arrow of time. This fact has a profound effect on the nature of the time evolution of the state vector in the Schrodinger Picture of the MC-QED formalism.

In particular it has been shown (Leiter, D,. 2009) that this causes the Hamiltonian operator in the Schrodinger Picture of MC-QED to contain a time reversal violating quantum evolution component and a time reversal violating retarded quantum measurement interaction component. The time reversal violating quantum measurement interaction part of the Hamiltonian operator contains components which have causal retarded light travel times, connected to the values of the physical sizes and/or spatial separations associated with the physical aggregate of Measurement Color symmetric fermionic states into which the fermionic sector of state vector is expanded.

For the retarded light travel time intervals in between the preparation and the measurement, the expectation values of the time-reversal violating retarded quantum measurement interaction operator will be negligible compared to the expectation values of the time reversal violating quantum evolution operator and the net effect generates the "quantum potentia" of what may occur.

On the other hand for the retarded light travel time intervals corresponding to the preparation and/or the measurement, the expectation values of the time-reversal violating retarded quantum measurement interaction operator will be dominant compared to the expectation values of the time reversal violating quantum evolution operator and the net effect causes the "quantum potentia" to be converted into the "quantum actua" of observer-participant measurement events.

In this context it can be shown (Leiter, D., 2009) that for a sufficiently large aggregate of atomic systems, described by the by the bare state component of MC-QED Hamiltonian and assumed to exist in an "environment" associated with the retarded quantum measurement interaction component of the Hamiltonian, the net effect of the quantum measurement interaction in MC-QED will generate time reversal violating decoherence effects on the reduced density matrix in a manner which can give large aggregates of atomic systems apparently classical properties.

Hence, in contradistinction the Copenhagen Interpretation of QED with its strict form of "Macroscopic Realism", it follows that MC-QED obeys a dynamic form of Macroscopic Realism in which the classical level of physics emerges dynamically in the context of local intrinsically time reversal violating quantum decoherence effects which can project out individual states since they are generated by the time reversal violating quantum measurement interaction in the formalism.

This is in contrast to the time reversal symmetric case of QED where the local quantum decoherence (Schlosshauer, M., 2007) effects only appear to be irreversible. This occurs in the time symmetric description of decoherence in QED because a local observer does not have access to the entire wave function and, while interference effects appear to be eliminated, individual states have not been projected out.

Hence we conclude that the resolution of the problem of the asymmetry between microscopic quantum objects and macroscopic classical objects inherent in the laws of quantum physics can be found in the MC-QED formalism, because the intrinsically time reversal violating quantum decoherence effects inherent within it imply that MC-QED does not require an independent external complementary classical level of physics obeying strict Macroscopic Realism in order to obtain a physical interpretation.

## IV. CONCLUSIONS

In this paper it has been shown that order to describe the quantum electrodynamic measurement process in a relativistic, observer-participant manner, an Abelian operator symmetry of "microscopic observer-participation" called Measurement Color (MC) was incorporated into the field theoretic structure of the Quantum Electrodynamics (QED).

The resultant formalism, called Measurement Color Quantum Electrodynamics (MC-QED), was constructed by defining a particle-field Measurement Color operator labeling symmetry associated with the integer indices (k = 1,2, ... , N ≥ 2, and imposing this labeling symmetry onto both the electron-positron operators $\psi^{(k)}$ and their electromagnetic charge-field operators $A_\mu^{(k)}$, in an mutual observer-participant manner which dynamically excluded time-symmetric self interactions from the formalism.

Within the multi-charge-field operator theoretic structure upon which the paradigm of MC-QED was based, "local time-symmetric free photon field operators uncoupled from charges" could not be operationally defined in terms of the Measurement Color charge-fields in the MC-QED formalism and hence were dynamically excluded from the formalism.

Mathematically this required that universal time-symmetric boundary conditions had to be imposed on <u>each</u> of the $A_\mu^{(k)}$ operator solutions to the $N \geq 2$ Maxwell operator equations, which prevented local the time reversal invariant free uncoupled photon operator $A_\mu^{(0)}$ from contributing to the $N \geq 2$ charge-field operators $A_\mu^{(k)}$. In this context the physical effects of radiation in MC-CED were generated in the operator equations of motion by a Total Coupled Radiation Charge-Field operator $A_\mu^{(TCRF)} \neq 0$, which obeyed an operator field equation $\Box^2 A_\mu^{(TCRF)} = 0$ with $\partial^\mu A_\mu^{(TCRF)} = 0$ similar to that obeyed by the $A_\mu^{(0)}$.

However $A_\mu^{(TCRF)}$ was found to be fundamentally different from $A_\mu^{(0}$ since by virtue of being Measurement Color symmetric it was non-locally coupled to the sum of the all the currents $\Sigma_{(k)} J^{(k)}(x')$ k = 1,2, …, N (N $\geq$ 2) in a manner which gave it <u>a negative parity under Wigner reversal</u> $T_W$.

In this context the MC-QED formalism and its "in-out" charge-field structure was found to be invariant under the generalized Time Reversal operator $T = T_W \times T_p$ (where $T_W$ is the Wigner time reversal operator and $T_p$ is the radiation flow reversal operator) while at the same time having a negative time parity for b<u>oth the $T_W$ and</u> the $T_p$ symmetry operations taken separately.

Then by applying same time-symmetric Asymptotic Conditions to MC-QED as is done in standard QED, it was shown that a causal, retarded electrodynamic arrow of time emerged dynamically from the stability conditions within the formalism independent of any Thermodynamic or Cosmological assumptions.
In this manner the new paradigm of Measurement Color upon which MC-QED was based implied that the arrow of time in the universe was quantum electrodynamic in origin. This result can be understood in a more general manner as follows::

    a) MC-QED is a <u>nonlocal</u>, relativistic quantum field theory whose operator solutions obey CPT symmetry where $T = T_W \times T_p$, and the Total Coupled Radiation Charge-Field photon operator $A_\mu^{(TCRF)}$ is non-local and has a negative parity under both $T_W$, and $T_p$ symmetries;

    b) Because the photon operator in MC-QED has a negative time parity under both the $T_W$, and $T_p$ symmetries, the physical requirement of a stable vacuum state in <u>dynamically requires</u> the charge-field operator solutions in MC-QED to contain a causal, retarded, classical electrodynamic arrow of time;

    c) For this reason the Measurement Color symmetry within the nonlocal quantum field theoretic structure of MC-QED dynamically leads to its CPT symmetry being spontaneously broken, and this is what causes the photon to carry the arrow of time.

The spontaneous breakdown of CPT symmetry in MC-QED implies that the CPT transformation cannot turn our universe into its "mirror image". This occurs because the photon carries the arrow of time in MC-QED which implies that time in the universe can only run forward in a causal sense and not backward. For MC-QED and its Standard Model generalizations, C, P , and CP symmetry is preserved but CPT symmetry is spontaneously violated. <u>Hence the observed invariance of CP in particle interactions is not physically equivalent to T invariance in the context of the MC-QED formalism</u>.

Hence using only time-symmetric boundary conditions, within the context of the Measurement Color paradigm underlying the MC-QED formalism, a dynamic explanation for the existence of a microscopic quantum electrodynamic arrow of time has been found in terms of the spontaneous symmetry breaking of both the T and the CPT symmetry in the formalism, independent of any Thermodynamic arguments.

In this manner the existence of the causal microscopic arrow of time in MC-QED represents a fundamentally quantum electrodynamic explanation for irreversible phenomena associated with the Second Law of Thermodynamics which complements the one supplied by the well-known statistical arguments in phase space (Zeh, D., 2007). Hence from the point of view of MC-QED, the Thermodynamic arrow of increasing entropy is not the source of the master time asymmetry in the universe.

This is because the MC-QED formalism implies the dynamic existence of a causal radiation arrow in the universe which automatically implies that the entropy associated with spontaneous emission of a cloud of photons from a aggregate of fermions will always increase. <u>Hence the dynamic radiation arrow of time inherent in the MC-QED formalism implies the Second Law of Thermodynamics in the fundamental form which states that the heat associated with radiation is an irreversible process which will spontaneously flow from hot to cold and not the other way around.</u>

In addition, since the microscopic observer-participant paradigm of Measurement Color with its dynamically generated microscopic dynamic arrow of time is a general concept, its application can be applied to quantum gauge field theories which are more general than Quantum Electrodynamics. Hence Measurement Color generalizations of higher symmetry quantum gauge particle field theories associated with the Standard Model and Grand Unified Models should be attainable, within which the gauge bosons as well as the photon would carry the Arrow of Time.

In future papers on MC-QED we will demonstrate in more detail how, in addition to being able to the explain the origin of the arrow of time, MC-QED can explain the existence of macroscopic objective reality in a quantum field theoretic context, as well as being able to offer two possible approaches to explain the apparently spontaneous emergence of macroscopic conscious minds in the universe from the microscopic laws of quantum physics.

The first approach is a global one which can be found by noting the fact that MC-QED describes the universe in terms of myriads of microscopic, time reversal violating, observer-participant quantum field theoretic interactions which span both the classical and the quantum world. On the other hand living, macroscopic conscious observers also appear to have physical properties which simultaneously span both the classical and the quantum world.

Because of this similarity it follows that the MC-QED formalism has the capability of being able to explain how macroscopic conscious observer-participant entities emerge in a microscopic observer-participant universe. Since this occurs a Measurement Color quantum field theoretic manner, it implies that a global quantum holographic description of consciousness may exist which connects the "minds of macroscopic conscious observers" to the "mind of the universe" as a whole.

The second approach is a local one which can be found by extending the Measurement Color paradigm into the recently developed quantum field theoretic domain of consciousness research called Quantum Brain Dynamics QBD, (Jibu, M., and Yasue, K., 1995) , (Vitiello, G., 2001 ). Since MC-QED is a quantum field theoretic formalism which contains both the effects of quantization and dissipation, it may be possible that the ideas underlying QBD can be consistently generalized into a (MC-QBD) formalism. In this way it may be possible to find a local cybernetic description of how macroscopic conscious observer-participant entities emerge in a microscopic observer-participant universe.

Acknowledgments: The author wishes to thank Rudy Schild at the Harvard Center for Astrophysics, Carl Gibson at the University of California San Diego, and Rhawn Joseph for useful comments and suggestions about the organization of the paper.

# QUANTUM REALITY IN A UNIVERSE WHERE THE PHOTON CARRIES THE ARROW OF TIME

Darryl Leiter, Ph.D.
Interdisciplinary Studies Program, University of Virginia Charlottesville, Virginia 22904

## ABSTRACT

We show how the new observer-participant paradigm of Measurement Color Quantum Electrodynamics (MC-QED) discussed earlier (Leiter, D., Journal of Cosmology, 2009, Vol 3, pp 478-500) can resolve the fundamental problem of the asymmetry between microscopic quantum objects and macroscopic classical objects inherent in the laws of quantum physics. Since spontaneous CPT violation implies that the photon carries the arrow of time in MC-QED, the total Hamiltonian operator in the Schrodinger Picture contains quantum potentia and quantum measurement interaction operator components which are time reversal violating. The quantum measurement interaction operator component contains causal retarded light travel times that are related to the physical sizes and/or spatial separations associated with the physical aggregate of Measurement Color symmetric fermionic states into which the fermionic sector of the state vector is expanded.

For light travel time intervals *in between* the preparation *and* the measurement, the expectation values of the time-reversal violating retarded quantum measurement interaction operator will be negligible compared to the expectation values of the time reversal violating quantum evolution operator, and the net effect generates the "quantum potentia" of what may occur. On the other hand for light travel time intervals *corresponding to* the preparation *and/or* the measurement, the expectation values of the time-reversal violating retarded quantum measurement interaction operator will be dominant compared to the expectation values of the time reversal violating quantum evolution operator, and the net effect causes the "quantum potentia" to be converted into the "quantum actua" of observer-participant measurement events.

For sufficiently large aggregates of atomic "systems" described by the bare state component of the total Hamiltonian, which are assumed to exist in an "environment" associated with the quantum measurement interaction component of the total Hamiltonian, the net effect of the quantum measurement interaction generates time reversal violating decoherence-dissipation effects on the reduced density matrix in a manner which can give large aggregates of atomic systems apparently classical properties. In this context MC-QED obeys a "dynamic form of Macroscopic Realism" in which the classical level of physics emerges dynamically in the context of local intrinsically time reversal violating quantum decoherence-dissipation effects. Because of the intrinsic time reversal violating quantum decoherence-dissipation effects generated by its time reversal violating photon structure, MC-QED does not require an independent external complementary classical level of physics obeying strict Macroscopic Realism in order to obtain a physical interpretation. Hence a resolution of the fundamental problem of the asymmetry between microscopic quantum objects and macroscopic classical objects inherent in the laws of quantum physics can be found in the MC-QED formalism. This offers the possibility of new insights into the emergence of macroscopic conscious observers in an observer-participant universe where the photon carries the arrow of time.

## I. INTRODUCTION

The Copenhagen Interpretation of Quantum Mechanics (CI-QM) contains an inherent logical asymmetry between object and observer which leads to contradictions. This problem is associated with the circular reasoning associated with the two assumptions in CI-QM that: a) "conscious observers" are associated with macroscopic measuring systems which have local objective properties and, b) "microscopic quantum systems" have non-local properties which do not have an objective existence independent of the irreversible "act of observation" generated by its interaction with the macroscopic measuring instruments associated with "conscious observers". Because of this
problem the CI-QM leads to contradictory predictions when macroscopically objective systems become directly coupled to microscopically non-objective quantum systems in a superposition of states. This problem occurs because "conscious observers" and their "macroscopic measuring instruments" are made up of large numbers of quantum micro-systems.  This problem cannot be avoided since the direct coupling of macro-aggregates of quantum systems to nonlocal micro-quantum systems must occur in Nature.

In an attempt to better understand the full implications of the quantum measurement process described by the CI-QM John Wheeler pioneered the development of the "Observer Participant Universe" (OPU). Within the OPU macroscopic conscious observers directly participate in the process of irreversibly actualizing the elementary quantum phenomena which make up the universe. Wheeler emphasized this point in his well known dictum that  "No elementary quantum phenomenon is a phenomenon until it is an irreversibly recorded phenomenon". However the logical asymmetry between the observer and the object associated with the CI-QM remained in Wheeler's Observer-Participant Universe, since the dynamical manner in which macroscopic living conscious observers irreversibly actualize microscopic elementary quantum phenomena was still unexplained.

In order to solve this problem a new observer-participant paradigm of the quantum measurement process is needed which generalizes Wheeler's concept of the observer-participant universe into a microscopic quantum operator form that is symmetric in regard to the definition of the "observer" and the "object". In a recent paper (Leiter, D., Journal of Cosmology, 2009, Vol 3, pages 478-500)  it was shown that this new paradigm could be found by incorporating an Abelian operator gauge symmetry of microscopic operator observer-participation called "Measurement Color" into the operator equations of Quantum Electrodynamics in the Heisenberg picture. This was shown to require that the Measurement Color labeling symmetry be imposed onto the quantum field theoretic structure of <u>both</u> the electron-positron operators <u>and</u> the electromagnetic field operators in the QED formalism. The resultant formalism, called Measurement Color Quantum Electrodynamics (MC-QED), took the form of a non-local quantum field theory which described the quantum measurement process in terms of myriads of microscopic electron-positron quantum operator fields undergoing <u>mutual</u> microscopic observer-participant quantum measurement processes mediated by the charge-field photon quantum operator fields through which they interact.

Since the time-symmetric free photon operator could not be given a Measurement Color description within the microscopic observer-participant operator symmetry in MC-QED, it was automatically excluded from the formalism, Instead the photon operator in MC-QED was given by the nonlocal Measurement Color Symmetric "Total Coupled Radiation" charge-field photon operator, which carried a negative time parity under the Wigner Time Reversal operator. In this context, applying the same time-symmetric Asymptotic Conditions to the non-local MC-QED operator equations of motion as was done for the local operator equations of motion in standard QED, the physical requirement of a stable vacuum state in MC-QED dynamically required that the MC-QED Heisenberg operator equations must contain a causal retarded quantum electrodynamic arrow of time.

It was shown that this surprising result could be better understood in a broader context by noting that, within the nonlocal quantum field theoretic structure of the MC-QED formalism, the physical requirement of a stable vacuum state generated a spontaneous symmetry breaking of both the T and the CPT symmetry. Spontaneous symmetry breaking of the T and the CPT symmetry occurred in MC-QED because the nonlocal photon operator acting within it has a negative parity under Wigner time reversal. In this manner the requirement of a stable vacuum state dynamically selected the operator solutions to the MC-QED formalism that contain a causal, retarded, quantum electrodynamic arrow of time, independent of any external thermodynamic or cosmological assumptions (Zeh, D., 2007).

Spontaneous CPT breaking in MC-QED implies that the photon carries the arrow of time. In this paper we will show that this fact implies that MC-QED contains both the Von Neumann Type 1 and Von Neumann Type 2 of time evolution of the state vector. For this reason we will find that MC-QED contains its own microscopic observer-participant description of the quantum measurement process, independent of the use of the Copenhagen Interpretation or the Everett "Many Worlds Interpretation". It is for this reason that the paradigm of MC-QED can be used to solve the problem of macroscopic quantum reality.

The origin of the problem of macroscopic quantum reality lies in the nature of Copenhagen Interpretation of QED. This is because within the QED formalism macroscopic bodies, associated with macroscopic measuring instruments and macroscopic conscious observers, are assumed to obey a strict form of "Macroscopic Realism" on a complementary classical level of physics external to the microscopic quantum electrodynamic system. Macroscopic bodies that satisfy the strict form of Macroscopic Realism are assumed have the property that they are at all times in a macroscopically distinct state which can be observed without affecting their subsequent behavior.

In this paper we will show that strict Macroscopic Realism is not valid for MC-QED. This is because its Measurement Color symmetry implies that the photon operator carries the arrow of time. This fact will be shown to have a profound effect on the nature of the time evolution of the state vector in the Schrodinger Picture of the MC-QED formalism, since it causes the total Hamiltonian operator acting on the state vector in the Schrodinger Picture of MC-QED to become a differential-delay equation containing time reversal violating quantum evolution and quantum measurement interaction components.

The time reversal violating quantum measurement interaction part of the Hamiltonian operator will be shown to contain causal retarded light travel times, which are connected to the values of the physical sizes and/or spatial separations associated with the physical aggregate of Measurement Color symmetric fermionic states into which the fermionic sector of state vector is expanded. For the retarded light travel time intervals in between the preparation and the measurement, the expectation values of the time-reversal violating retarded quantum measurement interaction operator will be negligible compared to the expectation values of the quantum evolution operator which generates the "quantum potentia" of what may occur. On the other hand for retarded light travel time intervals corresponding to the preparation and/or the measurement, the expectation values of the time-reversal violating retarded quantum measurement interaction operator will be dominant compared to the expectation values of the quantum evolution operator and this will cause the "quantum potentia" to be converted into the "quantum actua" of observer-participant measurement events.

In this context we will show that for a sufficiently large aggregate of atomic systems, described by the by the bare state component of MC-QED Hamiltonian and assumed to exist in an "environment" associated with the time reversal violating quantum measurement interaction component of the total Hamiltonian operator, the effects of the quantum measurement interaction will *generate time reversal violating decoherence and dissipation effects* on the reduced density matrix in a manner which will give these large aggregates of atomic systems apparently classical properties. This *dynamic form of Macroscopic Realism* within the MC-QED formalism is in stark contrast to the Copenhagen Interpretation of QED with its *strict form of Macroscopic Realism*.

Since MC-QED obeys a dynamic form of Macroscopic Realism, the classical level of physics emerges dynamically in the context of local intrinsically time reversal violating quantum decoherence effects which project out individual states since they are generated by the time reversal violating quantum measurement interaction in the formalism. Hence MC-QED does not require an independent external complementary classical level of physics obeying strict Macroscopic Realism in order to obtain a physical interpretation. This is in contrast to the time reversal symmetric case of QED where the local quantum decoherence effects only have the appearance of being irreversible because a local observer does not have access to the entire wave function and, while interference effects appear to be eliminated, individual states have not been projected out.

Hence while MC-QED uses standard canonical quantization methods in the development of its observer-participant time reversal violating microscopic operator equations of motion in the Heisenberg Picture, it does not require an independent external complementary classical level of physics in order to obtain a physical interpretation of the quantum measurement process. For this reason the Copenhagen Interpretation division of the world does not play a role in the MC-QED formalism. Hence the MC-QED formalism represents a more general observer-participant approach to quantum electrodynamics in which a consistent description of quantum electrodynamic measurement processes at both the microscopic and macroscopic levels can be obtained. Because of this we will find that the paradigm of Measurement Color in Quantum Electrodynamics represents a new observer-participant quantum field theoretic language in which both microscopic and macroscopic forms of quantum de-coherence and dissipation effects may be studied in a relativistically unitary, time reversal violating quantum electrodynamic context.

The structure of this paper is as follows: Section II discusses the structure of the microscopic, time reversal violating, observer-participant quantum measurement process in the MC-QED formalism. Section III discusses how the time reversal violating quantum measurement process in MC-QED acts to convert the "quantum potentia" of potential events into the "quantum actua" of actual events. Section IV discusses how the time reversal violating, observer-participant quantum measurement process in MC-QED leads to a well-defined dynamic description of the transition from the quantum to the classical level. Finally Section V concludes with discussions about the possible extension of the new paradigm of MC-QED into the broader realms of physical phenomena associated with emergence of macroscopic conscious observers in an observer-participant universe where the photon carries the arrow of time.
For the convenience of readers with higher levels of expertise who want more technical information about the arguments underlying the general discussions given in Sections II - IV of this paper, more detailed analysis in support of these sections is given in Appendices I –VI.

## II. THE OBSERVER-PARTICIPANT QUANTUM MEASUREMENT PROCESS

In this section we continue the development of the time reversal violating Measurement Color Quantum Electrodynamics (MC-QED) formalism published earlier (Leiter, D., Journal of Cosmology, 2009, Vol 3, pages 478-500). For the convenience of those readers who are not familiar with this paper, a brief summary of the MC-QED formalism has been given in Appendix I.

Previously we have shown how MC-QED can resolve the apparent asymmetry in the description of the microscopic and macroscopic "Arrows of Time" in the universe. In this section we will show how MC-QED can resolve the problem of the asymmetry between microscopic quantum objects and macroscopic classical objects inherent in the laws of quantum physics.

We begin our discussion by noting that the origin of this problem lies within the nature of Copenhagen Interpretation itself. This occurs in the Copenhagen Interpretation of QED because within it macroscopic bodies are assumed to obey a strict form of "Macroscopic Realism" on a complementary classical level of physics external to the microscopic quantum electrodynamic system. Macroscopic bodies that satisfy this strict form of Macroscopic Realism must have the property that they are at all times in a macroscopically distinct state which can be observed without affecting their subsequent behavior.

Since the Measurement Color symmetry in MC-QED implies that the photon operator carries the arrow of time this has a profound effect on the nature of the time evolution of the state vector in the Schrodinger Picture of the MC-QED formalism such that the assumption of Macroscopic Realism is not required in the MC-QED formalism. The fact that the photon carries the arrow of time causes the Hamiltonian operator in the Schrodinger Picture of MC-QED to contain quantum evolution and quantum measurement interaction components which are both time reversal violating. This causes the time reversal violating quantum measurement interaction part of the Hamiltonian operator to contain components which have causal retarded light travel times, which are connected to the values of the physical sizes and/or spatial separations associated with the physical aggregate of Measurement Color symmetric fermionic states into which the fermionic sector of state vector is expanded.

For the retarded light travel time intervals in between the preparation and the measurement, the expectation values of the time-reversal violating retarded quantum measurement interaction operator will be negligible compared to the expectation values of the time reversal violating quantum evolution operator and the net effect generates the "quantum potentia" of what may occur. On the other hand for the retarded light travel time intervals corresponding to the preparation and/or the measurement, the expectation values of the time-reversal violating retarded quantum measurement interaction operator will be dominant compared to the expectation values of the time reversal violating quantum evolution operator and the net effect causes the "quantum potentia" to be converted into the "quantum actua" of observer-participant measurement events.

Let us now discuss the above comments in more technical detail. We begin by recalling that in MC-QED the reason why the Photon carries the arrow of time is because of the effects of spontaneous CPT symmetry breaking inherent within the formalism. In this context the expectation value of the electron-positron operator equations of motion in the Heisenberg Picture, whose Wigner time reversal violating structure is determined by the Asymptotic Condition requirement that a stable vacuum state exists, are

($k = 1, 2, \ldots, N$   ($2 \leq N \to \infty$)

$$\langle (-i\hbar \gamma^\mu \partial_\mu + m - e/c \gamma^\mu A_\mu^{(k)}(obs)) \psi^{(k)} \rangle = 0$$

where

$$\langle A_\mu^{(k)}(obs) \rangle = \langle \sum_{(j \neq k = 1\ldots, N)} A_\mu^{(j)}(ret) + A_\mu^{(k)}(-) \rangle,$$

The expectation value of the <u>time reversal violating MC-QED operator equations of motion</u> describe the Universe as being made up of an infinitely large number of countable ($2 \leq N \to \infty$) microscopic observer-participant quantum measurement interactions, in the context of which:

  a) charge field photons are causally being emitted and absorbed between the $\psi^{(k)}$ and $\psi^{(j)}$ fermions ($k \neq j = 1, 2, \ldots, N$), and
  b) charge field photons are spontaneously being emitted into the vacuum by each of the $\psi^{(k)}$ fermions $k = 1, 2, \ldots, N$),

In the Heisenberg picture the Total Heisenberg State Vector $|\psi_H\rangle$ obeys $\partial_t |\psi_H\rangle = 0$, and the Total Heisenberg Hamiltonian Operator $H = H^\dagger$ obeys $dH/dt = 0$. Since they are both conserved in time then it follows that both $|\psi_H\rangle$ and $H = H^\dagger$ are time reversal invariant. In the Heisenberg picture the
MC-QED Hamiltonian operator $H$ can be written as

$$H = H^\dagger = [H_0 + V_{qp} + V_{ret-qa}]$$

where $H_0$ is the bare fermion and bare hamiltonian operator given by

$$H_0 = H_f + H_{ph}$$

Inside of $H_0$ we have that:

$H_f$ is the bare electron-positron Hamiltonian operator given by

$$H_f = \sum_{(k)} \{ : \int dx^3 [\psi^{(k)\dagger} (\alpha \cdot p + \beta m - e\varphi^{(k)}(ext)) \psi^{(k)}] + J^{\mu(k)} A_\mu^{(k)}{}_{Breit}{}^{(obs)} : \}$$

(where (k =1,2, ... , N→ ∞,  the symbols **: :** represent the use of normal ordering, $A_\mu^{(k)}{}_{Breit}{}^{(obs)} = \sum_{(j) \neq (k)} \int dx^3 J_\mu^{(j)}(x',t) / 4\pi |x-x'|$ represents the Breit vector potential operator, and the external potential $\varphi^{(k)}{}_{(ext)})$ represents the lowest order Coulombic effects of baryonic nuclei).

and $H_{ph}$ is the MC-QED charge-field photon hamiltonian operator given by

$$H_{ph} = -1/2 \int dx^3 [\sum_{(k)} \{:[(\partial_t A_{\mu\,(rad)}^{(k)} \partial_t A^{\mu}{}_{(rad)}^{(k)(obs)} + \nabla A_{\mu\,(rad)}^{(k)} \cdot \nabla A^{\mu}{}_{(rad)}^{(k)(obs)})]:\}$$

where

$$A_\mu^{(k)}{}_{(rad)} = (\alpha_\mu - A_\mu^{(k)}{}_{(-)})$$

$$\alpha_\mu = \sum_{(j)} A_\mu^{(j)}{}_{(-)} / (N-1)$$

$$A_\mu^{(k)}{}_{(rad)}{}^{(obs)} = \sum_{(j) \neq (k)} A_\mu^{(j)}{}_{(rad)} = A_\mu^{(k)}{}_{(-)}$$

and $A_\mu^{(k)}{}_{(-)}$ is the non-local, negative time parity, Heisenberg picture operator

$$A_\mu^{(k)}{}_{(-)} = \int dx'^4 \, (D_{(-)}(x-x')) \, J_\mu^{(k)}(x') \qquad (k = 1, 2, \ldots, N \to \infty)$$

In this context the <u>time reversal violating</u> "quantum potentia interaction operator" $V_{qp}$ is given by

$$V_{qp} = \sum_{(k)} : \int dx^3 \, J_\mu^{(k)} A^\mu{}_{(rad)}^{(k)(obs)} :$$

where

$$A_\mu^{(k)}{}_{(rad)}{}^{(obs)} = \sum_{(j) \neq (k)} A_\mu^{(j)}{}_{(rad)} = A_\mu^{(k)}{}_{(-)}$$

and the <u>time reversal violating</u> "retarded *quantum actua* interaction operator" $V_{ret-qa}$ is given by

$$V_{ret-qa} = :\{\int dx^3 \sum_{(k)} [J_\mu^{(k)} (A^{\mu(k)}{}_{(ret)}{}^{(obs)} - A_\mu^{(k)}{}_{Breit}{}^{(obs)})]$$
$$-1/2 [\partial_t A^\mu{}_{(ret)}^{(k)} \partial_t A_{\mu\,(ret)}^{(k)(obs)} + \partial_t A^\mu{}_{(ret)}^{(k)} \partial_t A_\mu^{(k)}{}_{(rad)}^{(obs)}$$
$$+ \partial_t A^\mu{}_{(rad)}^{(k)} \partial_t A_{\mu\,(ret)}^{(k)(obs)} + \nabla A^\mu{}_{(ret)}^{(k)} \cdot \nabla A_\mu^{(k)}{}_{(ret)}^{(obs)}$$
$$+ \nabla A^\mu{}_{(ret)}^{(k)} \cdot \nabla A_\mu^{(k)}{}_{(rad)}^{(obs)} + \nabla A^\mu{}_{(rad)}^{(k)} \cdot \nabla A_{\mu\,(ret)}^{(k)(obs)}]\}:$$

where inside of the expression for the quantum measurement interaction operator $(V_{ret\text{-}qa})$

$$(A^{\mu(k)}{}_{(ret)}{}^{(obs)} - A_\mu{}^{(k)}{}_{Breit}{}^{(obs)}) = \sum_{(j) \neq (k)} (A_\mu{}^{(j)}{}_{(ret)} - A_\mu{}^{(j)}{}_{Breit})$$

Since the above quantum measurement interaction operator $V_{ret\text{-}qa}$ involves $(k \neq j = 1, 2, \ldots, N)$ retarded electromagnetic operators, then the "quantum actua states" that it selects out of the available "quantum potentia states" generated by $V_{qp}$, will be restricted to those which contain $(k \neq j = 1, 2, \ldots, N)$ measurement color labels. The total Hamiltonian $H = H^\dagger = H_S$ is conserved in time as

$$idH/dt = i\partial H/dt + [H, H] = i\partial H/dt = 0$$

Hence it follows that the total Hamiltonian operator H is invariant under the action of the Wigner Time Reversal operator $T_w$ in the MC-QED theory as

$$[H, T_w] = [(H_0 + V_{qp} + V_{ret\text{-}qa}), T_w] = 0$$

which implies that

$$[H_0, T_t] = -[V_{qp} + V_{ret\text{-}qa}, T_t]$$

However because of the presence of nonlocal charge-field charge field photon operators with negative parity under Wigner time reversal $T_w$, the operator $[V_{qp} + V_{ret\text{-}qa}]$ violates Wigner time reversal Invariance as

$$[V_{qp} + V_{ret\text{-}qa}, T_w] \neq 0$$

Hence from the above equations this implies that $H_0$ violates Wigner time reversal as well

$$[H_0, T_t] \neq 0$$

Now since the MC-QED Heisenberg density matrix operator $\rho_H = |\psi_H\rangle\langle\psi_H|$ obeys $\rho_H = (\rho_H)^2$ and $Tr(\rho_H) = 1$), then the expectation value the H operator given by $\langle\psi_H|H|\psi_H\rangle = Tr(\rho_H H)$ evolves in a time reversal invariant unitary manner.

However in this context the expectation value of the operator $[V_{qp}(t) + V_{ret\text{-}qa}(t)]$ given by

$$\langle\psi_H | [V_{qp}(t) + V_{ret\text{-}qa}(t)] | \psi_H\rangle = \text{Tr}(\rho_H [V_{qp}(t) + V_{ret\text{-}qa}(t)])$$

violates Wigner time reversal, and hence follows that the time evolution of the expectation value of the operator $H_0(t)$

$$\langle\psi_H | H_0(t) | \psi_H\rangle = \text{Tr}(\rho_H H_0(t))$$

also violates Wigner time reversal.

However this internal time reversal violating process still preserves global unitarity, since it dynamically preserves the value of the total energy associated with the quantity $\langle\psi_H | H | \psi_H\rangle = \text{Tr}(\rho_H H)$.

## III. CONVERSION OF "QUANTUM POTENTIA" INTO "QUANTUM ACTUA"

Since the above MC-QED time reversal violating property is valid in the Heisenberg Picture, it must also be true in both the Schrodinger Picture and the Interaction Picture as well. To see this more explicitly we transform the total Heisenberg picture state vector $|\psi_H\rangle$ into the total Schrodinger picture state vector $|\psi_S(t)\rangle$ by the unitary time transformation $\exp[-iH(t-t_o)]$

$$|\psi_S(t)\rangle = \exp[H(t-t_o)/i\hbar] |\psi_H\rangle$$

From which it follows that

$$i\hbar \partial_t |\psi_S\rangle = H_S |\psi_S\rangle$$

where

$$H = H_S = (H_0)_S + (V_{qp})_S + (V_{ret\text{-}qa})_S$$

$$(V_{qp})_S = \sum_{(k)} \{ :(\int dx^3 \, J_\mu^{(k)} A^{\mu\,(k)(obs)}_{(rad)})_S : \}$$

and

$$(V_{ret\text{-}qa})_S = \{ : \int dx^3 \sum_{(k)} [J_\mu^{(k)} (A^{\mu(k)\,(obs)}_{(ret)} - A_\mu^{(k)\,(obs)}_{Breit})]$$
$$-1/2\, [\partial_t A^{\mu\,(k)}_{(ret)} \partial_t A_{\mu\,(ret)}^{(k)\,(obs)} + \partial_t A^{\mu\,(k)}_{(ret)} \partial_t A_{\mu\,(rad)}^{(k)\,(obs)}$$
$$+ \partial_t A^{\mu\,(k)}_{(rad)} \partial_t A_{\mu\,(ret)}^{(k)\,(obs)} + \nabla A^{\mu\,(k)}_{(ret)} \cdot \nabla A_\mu^{(k)\,(obs)}_{(ret)}$$
$$+ \nabla A^{\mu\,(k)}_{(ret)} \cdot \nabla A_\mu^{(k)\,(obs)}_{(rad)} + \nabla A^{\mu\,(k)}_{(rad)} \cdot \nabla A_{\mu\,(ret)}^{(k)\,(obs)}]_S : \}$$

and inside of the expression for the quantum measurement interaction operator $(V_{ret-qa})_S$ we have

$$(A^{\mu(k)}{}_{(ret)}{}^{(obs)} - A_\mu{}^{(k)}{}_{Breit}{}^{(obs)})_S = \sum_{(j) \neq (k)} (A_\mu{}^{(j)}{}_{ret} - A_\mu{}^{(j)}{}_{Breit})_S$$

$$(A_\mu{}^{(k)}{}_{(rad)}{}^{(obs)})_S = (\sum_{(j) \neq (k)} A_\mu{}^{(j)}{}_{(rad)})_S = (A_\mu{}^{(k)}{}_{(-)})_S$$

$$(A_\mu{}^{(k)}{}_{(rad)})_S = (\alpha_\mu - A_\mu{}^{(k)}{}_{(-)})_S$$

Hence in the same manner as occurred in the Heisenberg Picture, it follows in the Schrodinger Picture the relativistic, time reversal violating property of

$$Tr(\rho_S[V_{qp}(t) + V_{ret-qa}(t)]_S) = \langle\psi_S(t)| [V_{qp} + V_{ret-qa}]_S |\psi_S(t)\rangle$$

dynamically causes the value of

$$Tr(\rho_S[H_0]_S) = \langle\psi_S(t)| [H_0]_S |\psi_S(t)\rangle$$

to change in a time reversal violating manner which is unitary since it preserves the value of the total energy associated with $Tr(\rho_S H_S) = \langle\psi_S(t)| | [H]_S |\psi_S(t)\rangle$. The dynamic source of this time reversal violating property can be seen more explicitly by examining the form of the non-local Schrodinger state vector potential operators $(A^{\mu(j)}{}_{(ret)})_S$, $(A^{\mu(j)}{}_{(adv)})_S$, and $(A^{\mu(j)}{}_{(-)})_S$ given respectively by

$$(A^{\mu(j)}{}_{(ret)})_S = \left(\int dx'^3 \exp(-iHR/\hbar c) J^{\mu(j)}(\mathbf{x}') \exp(iHR/\hbar c) / 4\pi R\right)_S$$

$$(A^{\mu(j)}{}_{(adv)})_S = \left(\int dx'^3 \exp(iHR/\hbar c) J^{\mu(j)}(\mathbf{x}') \exp(-iHR/\hbar c) / 4\pi R\right)_S$$

and

$$(A^{\mu(j)}{}_{(-)})_S = [(A^{\mu(j)}{}_{(ret)})_S - (A^{\mu(j)}{}_{(adv)})_S ] / 2$$

where $R = |\mathbf{x} - \mathbf{x}'|$.

Since linear combinations of these nonlocal time reversal violating Schrodinger operators and their derivatives appear inside of the $(V_{ret-qa})_S$ operator component of $H_S$ in the Schrodinger Picture equation of motion for the MC-QED state vector, they influence time evolution of the MC-QED state vector. In this context since

$$|\psi_S(t)\rangle = \exp(-i/\hbar H_S t)|\psi_S(0)\rangle$$

then it follows that

$$\exp(+iHR/hc)\,|\psi_S(t)\rangle = \exp(H_S[t-R/c]/i\hbar)|\psi_S(0)\rangle = |\psi_S(t-R/c)\rangle$$

$$\exp(-iHR/hc)\,|\psi_S(t)\rangle = \exp(H_S[t+R/c]/i\hbar)|\psi_S(0)\rangle = |\psi_S(t+R/c)\rangle$$

Hence we see that the effects of the non-local Schrodinger state vector potential operators $(A^{\mu(j)}{}_{(ret)})_S$,

and $(A^{\mu(j)}{}_{(-)})_S$ inside of the Hamiltonian equation of motion for the Schrodinger Picture state vector $|\psi_S(t)\rangle$, convert it into a retarded time-irreversible integro-differential-delay equation for the Schrodinger state vector $|\psi_S(t)\rangle$ which contains nonlocal-in-time volume integrals over the state vector quantities $|\psi_S(t-R/c)\rangle$ and $(|\psi_S(t-R/c)\rangle - |\psi_S(t-R/c)\rangle)/2$.

Since the Schrodinger Hamiltonian operator $H_S$ in MC-QED contains components which are nonlocal in time, this prevents locally defined eigenstates $|E_S\rangle$ of the total Schrodinger Picture Hamiltonian $H_S$ operator from being able to be defined at finite time t = to.

However since the bare state hamiltonian operator $H_0$ is local-in-time then, as shown in Appendices III and IV, locally defined bare states $|E_o\rangle = |E_{fermion}\rangle|E_{photon}\rangle$ can be defined for finite times t = to and the total MC-QED state vector in the Schrodinger Picture $|\psi_S(t)\rangle$ can be expanded into these bare local-in-time microscopic observer-participant eigenstates $|E_0\rangle$ of the local-in-time bare hamiltonian operator $H_0$ as

$$|\psi_S(t)\rangle = \sum_{(Eo)} C_{(Eo)} |E_o\rangle$$

where

$$H_o\,|E_o\rangle = E_o\,|E_o\rangle$$

Then from the Schrodinger state vector equation of motion

$$i\hbar\partial_t|\psi_S(t)\rangle = H_S\,|\psi_S(t)\rangle$$

we find that the time dependent coefficients $C_{(E_o)}(t)$, of the observer-participant bare states $|E_o\rangle$ into which $|\psi_S(t)\rangle$ was expanded, obeys a time irreversible, retarded, integro-differential equation.

Hence in the context of the expectation value associated with states $|E_o\rangle$ given by

$$\mathrm{Tr}(\rho_S[H_0]_S) = \langle\psi_S(t)|[H_0]_S|\psi_S(t)\rangle = \sum_{(E_o)} E_o |C_{(E_o)}(t)|^2$$

the time irreversible, retarded, integro-differential equation obeyed by the $C_{(E_o)}(t)$ physically represents the fact that there exist characteristic light travel time intervals $\Delta E/h < (t-t_o) < (t - R/c)$, in between the creation of "quantum potentia" by the action of $V_{qp}(t)$ and their conversion into "quantum actua" by the time irreversible action of $V_{ret-qa}(t)$. In the context of these characteristic time intervals, the $|C_{(E_o)}(t)|^2$ can be interpreted as the relative probabilities, associated with *observer-participant quantum-potentia*, of potential events being converted into actual events associated with *observer-participant quantum-actua*.

We now transform back from the Schrodinger Picture in MC-QED to the Interaction Picture in MC=QED as

$$|\psi_I(t)\rangle = \exp[i(H_0)_S(t-t_o)]|\psi_S(t)\rangle$$

Then we can write

$$|\psi_I(t)\rangle = U(t, t_o)|\psi_H\rangle$$

where $U(t-t_o)$ is the unitary operator given by

$$U(t-t_o) = \exp[i(H_0)_S(t-t_o)]\exp[-iH(t-t_o)]$$

and $(H_0)_S$ and $H=H_S$ are constant in time. (The more technical oriented reader can look at Appendix II for a more detailed discussion of the Interaction Picture in the MC-QED formalism)

In this context it follows from the Schrodinger Picture state vector equation

$$i\partial_t|\psi_S(t)\rangle = H_S|\psi_S(t)\rangle$$

that the Interaction Picture state vector equation which $|\psi_I(t)\rangle$ obeys is given by

$$i\partial_t|\psi_I(t)\rangle = [V_{qp}(t) + V_{ret-qa}(t)]_I \ |\psi_I(t)\rangle$$

This can be formally integrated as

$$|\psi_I(t)\rangle = |\psi_I(t_o)\rangle + (1/i\hbar)\int_{t_o}^{t} dt'[V_{qp}(t') + V_{ret-qa}(t')]_I \ |\psi_I(t')\rangle$$

where

$$[V_{qp}(t) + V_{ret-qa}(t)]_I$$

$$= U(t-t_o) [V_{qp}(t) + V_{ret-qa}(t)]_H \ U(t-t_o)^{-1}$$

$$= \exp[i(H_0)_S(t-t_o)] \ [(V_{qp})_S + (V_{ret-qa})_S] \ \exp[-i(H_0)_S(t-t_o)]$$

Since the above equations are also nonlocal in time, retarded, integro-differential equations then within them there exist characteristic light travel times $\Delta E/h < (t-t_o) < (t - R/c)$ in between the creation of the "quantum potentia" by the action of the $V_{qp}(t)$ and their conversion into "quantum actua" by the time irreversible action of $V_{ret-qa}(t)$.

In this context let us define the quantities $R_i$, $i = 1,2,..$ as representing the various values of the physical sizes and/or spatial separations associated with the physical aggregate of Measurement Color symmetric fermionic states, which contribute to the local-in-time bare observer-participant quantum states $|Eo\rangle$ into which the state vector $|\psi_I(t)\rangle$ is expanded.

Then in the context of the $|Eo\rangle$ states the expectation value of the relativistic, retarded effects of $V_{ret-qa}(t)_I$ will be negligible compared to $V_{qp}(t)_I$ for light travel time intervals - $(R_i/c) < t < (R_i/c)$, where $|(R_i/c)| \ggg |(h/\Delta E)|$. These light travel time intervals can be thought of as occurring "in-between" the "preparation of quantum states" by the action of $V_{ret-qa}(t)_I$ at $t \sim (-R_i/c)$ and the "measurement of quantum states" at $t \sim (R_i/c)$, (after which time the $V_{ret-qa}(t)_I$ then acts to convert the "quantum potentia" generated by $V_{qp}(t)_I$ into the "quantum actua" of observer-participant measurement events).

Hence for light travel time intervals $-(R_i/c) < t < (R_i/c)$, $|(R_i/c)| >>> |(h/\Delta E)|$, during which the operator $V_{qp}(t)_I$ dominates $V_{ret-qa}(t)_I$, the state vector in the Interaction Picture can be approximated by $|\psi_I(t)> \approx |\psi_I(t)>_{qp}$ where $|\psi_I(t)>_{qp}$ is the state vector which represents the "quantum potentia" state of the system and obeys the equation of motion

$$i\partial_t |\psi_I(t)>_{qp} = [V_{qp}(t)]_I \; |\psi_I(t)>_{qp}$$

Then during the light travel time intervals $-(R_i/c) < t < (R_i/c)$, which occur in between preparation and measurement, the above equation can be formally iterated using the Wick "time ordered product operator" $T$ to obtain the S-matrix approximation, associated with the "quantum potentia" created $V_{qp-in}(t)_I$ as

$$|\psi_I(t)>_{qp-out} = \left\{ T\left(\exp[(-i/h) \int_{-(R_i/c)}^{t<(R_i/c)} dt' V_{qp-in}(t')_I\right) \right\} |\psi_I(-R_i/c)>_{qp-in}$$

$$= S \; |\psi_I(-R_i/c)>_{qp-in}$$

where for $|(R_i/c)| >>> |(h/\Delta E)|$ the S-matrix in MC-QED is given by

$$S = \left\{ T\left(\exp[(-i/h) \int_{-(R_i/c)}^{t<(R_i/c)} dt' V_{qp-in}(t')_I\right) \right\}$$

The above expression represents the S-matrix approximation in MC-QED for the time evolution of the state vector in the Interaction Picture. (see Appendix V for a further more technical discussion of the structure of the S-matrix in the MC-QED formalism)

In the language of von Neumann this would be called PROCESS 2 evolution from which the probability of events associated with quantum potentia can be calculated.

We emphasize that the above described S-matrix approximation to MC-QED is valid only for the characteristic time intervals $-(R_i/c) < t < (R_i/c)$, $|(R_i/c)| >>> |(h/\Delta E)|$, which occur in between quantum state preparation and measurement. In this context the $R_i$ represent the various values of the physical sizes and/or spatial separations associated with the aggregate of Measurement Color symmetric fermionic states, which contribute to the local-in-time bare observer-participant quantum states $|Eo>$ into which the state vector $|\psi_I(t)>$ is expanded.

On the other hand for the characteristic time intervals $t > (R_i/c)$ the equation for $|\psi_I(t)>$ becomes

$$i\partial_t|\psi_I(t)> = [V_{qp}(t) + V_{ret-qa}(t)]_I |\psi_I(t)> \qquad t > (R_i/c)$$

which can be formally integrated as

$$|\psi_I(t)> = |\psi_I(R/c)>_{qp} + (1/i\hbar)\int_{(R_i/c)}^{t} dt'[V_{ret-qa}(t')]_I |\psi_I(t')>$$

For characteristic time intervals $t > (R_i/c)$, the above equation formally describes how the superposition of "quantum potentia" states in the MC-QED S-matrix associated with

$$|\psi_I(R/c)>_{qp} = \left\{T(\exp[(-i/\hbar)\int_{-(R_i/c)}^{(R_i/c)} dt' V_{qp-in}(t')_I )\right\} |\psi_I(-R_i/c)>_{qp}$$

are converted, by the time reversal violating quantum measurement interaction operator $[V_{ret-qa}(t)]_I$, into the "quantum actua" state of an observer-participant measurement event.

In the language of von Neumann this would be called PROCESS 1 evolution where the probability of events associated with the quantum potentia are time irreversibly converted into the quantum actua of
real events.

## IV. DYNAMIC TRANSITION FROM THE QUANTUM TO THE CLASSICAL LEVEL

In this section we will now show that for a sufficiently large aggregate of atomic systems, described by the by the bare state component of MC-QED Hamiltonian and assumed to exist in an "environment" associated with the retarded quantum measurement interaction component of the Hamiltonian, the net effect of the quantum measurement interaction in MC-QED will generate time reversal violating decoherence effects on the reduced density matrix in a manner which can give large aggregates of atomic systems apparently classical properties.

Hence, in contradistinction the Copenhagen Interpretation of QED with its strict form of "Macroscopic Realism", it follows that MC-QED obeys a <u>dynamic form of Macroscopic Realism</u> in which the classical level of physics emerges dynamically in the context of local <u>intrinsically time reversal violating quantum decoherence effects</u> which can project out individual states since they are generated by the time reversal violating quantum measurement interaction in the formalism.

This is in contrast to the time reversal symmetric case of QED where the <u>local quantum decoherence (Schlosshauer, M., 2007) effects only appear to be irreversible</u>. This occurs in the time symmetric description of decoherence in QED because a local observer does not have access to the entire wave function and, while interference effects appear to be eliminated, individual states have not been projected out.

Hence we conclude that the resolution of the problem of the asymmetry between microscopic quantum objects and macroscopic classical objects inherent in the laws of quantum physics can be found in the MC-QED formalism, because the intrinsically time reversal violating quantum decoherence effects inherent within it imply that <u>MC-QED does not require an independent external complementary classical level of physics obeying strict Macroscopic Realism in order to obtain a physical interpretation</u>.

In the Heisenberg picture the MC-QED Hamiltonian operator $H$ is

$$H = [H_0 + (V_{qp} + V_{ret-qa})] = [H_0 + (V)]$$

Now the successive state vector transformations on the Heisenberg Picture State vector $|\psi_H\rangle$, through the Schrodinger Picture state vector $|\psi_S\rangle$, that finally leads to the Interaction Picture state vector $|\psi_I\rangle$ can be formally represented by $|\psi_I(t)\rangle = U(t-t_o) |\psi_H\rangle$) where the unitary operator $U(t-t_o)$ is given by

$$U(t-t_o) = \exp[i(H_0)_S(t-t_o)] \exp[-iH(t-t_o)]$$

where the Schrodinger Hamiltonian operators $(H_0)_S$ and $H = H_S$ are constant in time. It then follows that the equation of motion of the state vector in the Interaction Picture is

$$i\partial_t|\psi_I(t)\rangle / dt = [V(t)]_I |\psi_I(t)\rangle$$

and that the density matrix operator $\rho_I(t) = |\psi_I(t)\rangle\langle\psi_I(t)|$ obeys the equation of motion

$$id\rho_I(t)/dt = i\partial_t\rho_I(t) = [V(t), \rho_I(t)]_I$$

where
$$[V(t)]_I = \exp[i(H_0)_S(t-t_o)] \; [V]_S \exp[-i(H_0)_S(t-t_o)]]_S$$

which is time-reversal violating because $[V]_I = [(V_{qp} + V_{ret-qa})]_I$ is not invariant under Wigner time reversal. However time reversal violating time evolution of $\rho_I(t)$ is unitary because conserves the total hamiltonian operator $H$. Because of this fact that the full density matrix satisfies the two conditions required for unitarity to hold, namely

$$Tr_{|Eo>}\{\rho_I(t)\} = <\psi_I(t)|\psi_I(t)> = 1 \qquad (\text{where } H|E_o> = E_o|E_o>)$$

and

$$\rho_I(t)^2 = \rho_I(t)$$

In standard QED the local Wigner time reversal invariant properties of the Hermetian Hamiltonian operator guarantees a unitary, time reversal invariant evolution of the state vector $|\psi_I(t)>$ of the quantum system. For this reason, in Copenhagen Interpretation of standard QED, macroscopic bodies associated with measuring instruments and observers are assumed to obey a strict form of "Macroscopic Realism" on a complementary classical level of physics external to the microscopic quantum electrodynamic system. Macroscopic bodies which satisfy the concept of classical "Macroscopic Realism" are assumed to have the property that they are at all times it is in one of their macroscopically distinct states. In addition it is also assumed that one can observationally determine that the macroscopic system is in a particular macroscopically distinguishable state without affecting its subsequent behavior.

However in contradistinction to standard QED we have that in MC-QED the nonlocal, time-reversal violating, observer-participant structure of MC-QED does not require an independent external complementary classical level of physics obeying "Macroscopic Realism" in order to obtain a physical interpretation. This is because the Hermetian Hamiltonian operator in MC-QED maintains unitary time evolution of the state vector $|\psi_I(t)>$ in the context of Wigner Time reversal violating, nonlocal, observer-participant interaction operators, associated with the quantum measurement interaction given by

$$Tr_{|Eo>}\{\rho_I(t)[V_{in-qp}(t) + V_{ret-qa}(t)]_I\} = <\psi_I(t)|\{V_{in-qp}(t) + V_{ret-qa}(t)\}_I|\psi_I(t)>$$

$$= <\psi_I(t)| \int dx^3 \sum_{(k)} \{J\mu^{(k)} A\mu^{(k)} obs\}_I |\psi_I(t)>$$

where

$$\int dx^3 \sum_{(k)} \{J\mu^{(k)} A\mu^{(k)} obs\}_I = \int dx^3 \sum_{(k)} \sum_{(j) \neq (k)} \{J\mu^{(k)}[A\mu^{(k)}(-) + A\mu^{(j)}(ret)]\}_I$$

Hence in MC-QED the action of the relativistic, time reversal violating property of the quantum measurement interaction $\langle \psi_I(t)| \,[V_{in\text{-}qp}(t) + V_{ret\text{-}qa}(t)]_I\, |\psi_I(t)\rangle$ dynamically causes the value of the quantity $\langle \psi_I(t)| \,[H_0(t)]_I\, |\psi_I(t)\rangle$ to change in a time reversal violating manner while still preserving the expectation value of the energy of the total hamiltonian $\langle \psi_I(t)|\,|[H(t)]_I\,|\psi_I(t)\rangle$ and hence the unitarity of the total state vector $|\psi_I(t)\rangle$.

Now inside of $\langle \psi_I(t)|\,|[H_0(t)]_I\,|\psi_I(t)\rangle$ let the $R_i$ represent the values of the various physical sizes and/or spatial separations associated with the aggregate of Measurement Color symmetric fermionic states, which contribute to the local-in-time bare observer-participant quantum states $|Eo\rangle$ into which the state vector $|\psi_I(t)\rangle$ is expanded.

Then the photons associated with the electromagnetic radiation emitted and absorbed within the context of these observer-participant states will occur:

a)  in a highly efficient manner in the "wave-zone" after a light travel time $\Delta t \sim R_i/c \gg (h/\Delta E)]$ if the characteristic spatial dimension of the observer-participant fermion states within $|Eo\rangle$ are $R_i \gg \lambda \sim hc/\Delta E$, and

b)  in a relatively inefficient manner in the "induction-zone" over time intervals $(h/\Delta E) < \Delta t \ll R_i/c$ if the characteristic spatial dimension of the observer-participant system is $R_i \sim \lambda \sim hc/\Delta E$,

Hence in $\langle \psi_I(t)|\,|[H_0(t)]_I\,|\psi_I(t)\rangle$ the relativistic, retarded effects of the operator $V_{ret\text{-}qa}(t)_I$ will be negligible compared to the effects of the operator $V_{in\text{-}qp}(t)$ for the light travel time intervals $(h/\Delta E) < t \ll (R_i/c)$, which occur in between the creation of the "quantum potentia" by the action of $V_{in\text{-}qp}(t)$ and their conversion into "quantum actua" by the action of $V_{ret\text{-}qa}(t)$ after the light travel time intervals $t \geq (R_i/c)$.

In this manner the action of operator $V_{ret\text{-}qa}(t)$ is responsible for the "preparation" of the quantum state which occurs at $t = t_o \sim -(R_i/c)$ as well as the "measurement process" which acts on the quantum state at $t = t_o \sim (R_i/c)$ and converts the quantum potentia (which exist during the intermediate time intervals $(h/\Delta E) < t \ll (R_i/c)$ within which the operator $V_{in\text{-}qp}(t)$ dominates the state vector equation of motion).

Hence in MC-QED it follows that macroscopic bodies do not obey a *strict form of Macroscopic Realism*, because in the context of this formalism they are considered to be fully quantum mechanical. However within the context of MC-QED it is possible for macroscopic bodies to obey a *"Dynamically Conditional Form of Macroscopic Realism"* in the following sense:

a)  If the physical dimension of the correlation length of the currents contained within the "object" is <u>larger</u> than the physical dimension of the wave-zone associated with its internal radiation fields, the physical effects of the time reversal violating "induced emission" interaction term will dominate the "spontaneous emission" interaction term inside of $<\sum_{(k)} \{J\mu^{(k)} A\mu^{(k)} obs\}>$. Then the resultant time reversal violating evolution of $<\psi_I(t)| \, |[H_0(t)]_I \, |\psi_I(t)>$ will lead to the generation of a rapid time reversal violating quantum de-coherence effect which will ultimately lead to $<\psi_I(t)| \, |[H_0(t)]_I \, |\psi_I(t)>$ being in a "pointer-basis defined" classical state.

**(e.g. if Ls ~ Ns(ao) >> λ ~ (c/ν) ~ $10^{-5}$ cm;   Ns >> $10^3$ atoms)**

b)  On the other hand if the physical dimension of the correlation length of the currents contained within the "object" is <u>smaller</u> than the physical dimension of the wave-zone associated with its internal radiation fields, the physical effects of the time reversal violating "induced emission" interaction term will be dominated by the "spontaneous emission" interaction term inside of $<\sum_{(k)} \{J\mu^{(k)} A\mu^{(k)} obs\}>$. Under these conditions the resultant time reversal violating unitary evolution of the density matrix <u>will not lead</u> to the generation a rapid quantum de-coherence effect on $<\psi_I(t)| \, |[H_0(t)]_I \, |\psi_I(t)>$. *Instead* $<\psi_I(t)| \, |[H_0(t)]_I \, |\psi_I(t)>$ will be in a quantum superposition of states with a lifetime associated with the spontaneous decay of its internal states if they are unstable.

**(e.g. if Ls ~ Ns(ao) << λ ~ (c/ν) ~ $10^{-5}$ cm;   Ns << $10^3$ atoms)**

These sizes associated with classical–quantum threshold of $10^{-5}$ cm and $10^3$ atoms are consistent those calculated on page 135, tbl 3.2, of the book by M. Schlosshauer "Decoherence And The Quantum To Classical Transition" as follows:

$$a_{(dust)} \sim 10^{-3} \text{ cm} > Ls \sim 10^{-5} \text{ cm} > a_{(Large\ molecule)} \sim 10^{-6} \text{ cm})$$

In this context the process of decoherence in MC-QED will always accompanied by the time reversal violating effects of dissipation. This phenomenon what produces such a profound effect on the physical nature of the transition from quantum to classical in the MC-QED formalism.

To see this more specifically we note that by virtue of its Measurement Color labeling structure, the total MC-QED Hamiltonian H is

$$H = \big(H_{0(sys)} + H_{0(env)}\big)_I + (V)_I$$

where $H_{0(sys)I} = \sum_{(k)} (H_f^{(k)})$ will be associated with Measurement Color fermion operators (k=1,2,...N ≥2) and the environment will be associated with the charge-field Hamiltonian $(H_{ph})_I$. In this context the bare state Hamiltonian operators $H_{0(sys)I}$ and $H_{0(env)I}$ can be used to define the bare states associated with the system and the environment respectively as

$$H_{f(sys)I} |E_f\rangle = E_f |E_f\rangle$$

$$H_{ph(env)I} |E_{ph}\rangle = E_{ph} |E_{ph}\rangle$$

Now choosing to = 0 and noting that $|E_{env}\rangle = |E_{ph}\rangle$ and $|E_{sys}\rangle = |E_f\rangle$, it follows that the reduced density matrix associated with the fermion "system" obtained by tracing over the "environment" is

$$\rho_I(t)_f = Tr_{|E_{ph}\rangle}\{\rho_I(t)\}$$

$$= Tr_{|E_{ph}\rangle}\{\exp[i(H_0)_S t] \rho_S(t) \exp[-iH_0 t]\}$$

$$= \exp[i(H_f)t] \; [Tr_{|E_{ph}\rangle}\{\exp[i(H_{ph})t] \rho_S(t) \exp[-i(H_{ph})t]\}] \; \exp[-i(H_f)t]$$

$$= \exp[i(H_f)t] \; [Tr_{|E_{ph}\rangle}\{\rho_S(t)\}] \; \exp[-i(H_f)t]$$

Hence for MC-QED we see that the reduced fermion density matrix of the system is

$$\rho_I(t)_f = Tr_{|E_{ph}\rangle}\{\rho_I(t)\} = \exp[i(H_f)t] \; [Tr_{|E_{ph}\rangle}\{\rho_S(t)\}] \; \exp[-i(H_f)t]$$

On the other hand while the reduced density matrix of the fermion system $\rho_I(t)_f = Tr_{|E_{ph}\rangle}\{\rho_I(t)\}$ obeys the time reversal violating time evolution equation given by

$$d\rho_I(t)_f / dt = -i \; Tr_{|E_{ph}\rangle}\{ [V(t), \rho_I(t)]_I \}$$

the time reversal violating evolution of the reduced density matrix is non-unitary.

This is because now the second of the two conditions for unitarity to hold is violated since now

$$Tr_{|E_{ph}\rangle}\{\rho_I(t)_f\} = Tr_{|E_o\rangle}\{\rho_I(t)\} = 1$$

and

$$\rho_I(t)_f^2 = (Tr_{|E_{ph}\rangle}\{\rho_I(t)\})^2 \neq \rho_I(t)_f = Tr_{|E_{ph}}\{\rho_I(t)^2\}.$$

It is important to note that the above result is formally the same as that shown in section 8.4 and Appendix I of the book by M. Schlosshauer "Decoherence And The Quantum To Classical Transition", except now for the case of MC-QED the operator $(V)_I = (V_{qp} + V_{ret\text{-}qa})_I$ is both non-local-in-time and time reversal violating.

## V. CONCLUSIONS

In order to describe the microscopic quantum electrodynamic measurement process in a relativistic, observer-participant manner, an Abelian operator symmetry of "microscopic observer-participation" called Measurement Color (MC) was incorporated into the field theoretic structure of the Quantum Electrodynamics (QED).

Within the multi-field-operator theoretic Measurement Color paradigm upon which MC-QED was based, a microscopic, causal, electrodynamic arrow of time was found to exist in the universe, independent of any additional external thermodynamic or cosmological assumptions. This occurred because the Measurement Color symmetry dynamically prohibited the free photon operator from the formalism. Instead the physical effects of photons were generated by the measurement color symmetric, negative time parity Total Coupled Radiation Charge-Field Photon operator in the MC-QED formalism.

In contradistinction to the standard local formulation of quantum electrodynamics, this caused the phenomenon of spontaneous symmetry breaking with respect to CPT invariance to occur the MC-QED formalism. This occurred because MC-QED was a non-local quantum field theory in which the photon carried the arrow of time. In this manner the physical requirement of a stable vacuum state spontaneously broke the CPT symmetry and led to operator solutions which were CP invariant but not T invariant.

In the context of the MC-QED formalism, the empirically observed invariance of CP in quantum electrodynamics does not imply T invariance. Hence for the MC-QED formalism the C, P, and CP symmetry will be preserved but the CPT symmetry will be spontaneously violated. This implies, within the context of MC-QED, that the CPT transformation cannot turn our universe into its "mirror image" because the photon carries the arrow of time. Hence MC-QED implies that the flow of time in the universe can run forward in a causal sense but cannot not backward in an acausal sense.

Since the microscopic observer-participant paradigm of Measurement Color with its dynamically generated microscopic dynamic arrow of time is a general concept, its application can be applied to quantum gauge field theories which are more general than Quantum Electrodynamics. Hence Measurement Color generalizations of higher symmetry quantum gauge particle field theories associated with the Standard Model and Grand Unified Models should be attainable, within which the gauge bosons as well as the photon would carry the Arrow of Time.

In this manner we see that the dynamic existence of the microscopic arrow of time in MC-QED represents a fundamentally quantum electrodynamic explanation for irreversible phenomena associated with the Second Law of Thermodynamics, which complements the one supplied by the well-known statistical arguments in phase space (Zeh, D., 2007). This occurs because MC-QED dynamically generates a causal radiation arrow in the universe which dynamically implies that the entropy, associated with spontaneous emission of a cloud of photons from a aggregate of fermions, will always increase.

Hence the dynamic radiation arrow of time, caused by the spontaneous CPT violation in the MC-QED formalism, can be used to derive the Second Law of Thermodynamics, in the fundamental form which states that the heat associated with radiation is an irreversible process which spontaneously flows from hot bodies to cold bodies and not the other way around.

Since the MC-QED formalism resolved the apparent asymmetry in the description of the microscopic and macroscopic "Arrows of Time" in the universe, this allowed us to use it to solve the problem of the asymmetry between microscopic quantum objects and macroscopic classical objects inherent in the laws of quantum physics. We began by first noting that the origin this problem lies within the nature of Copenhagen Interpretation of QED. This is because within QED macroscopic bodies, associated with macroscopic measuring instruments and macroscopic conscious observers, are assumed to obey a strict form of "Macroscopic Realism", on a complementary classical level of physics external to the microscopic quantum electrodynamic system.

Because its Measurement Color symmetry implied that the photon operator carries the arrow of time, this concept of strict Macroscopic Realism was shown to not be valid for the case of MC-QED. This was because the photon carrying the arrow of time was shown to have a profound effect on the nature of the time evolution of the state vector in the Schrodinger Picture of the MC-QED formalism.

In particular we found that this caused the total Hamiltonian operator in the Schrodinger Picture of MC-QED to contain a time reversal violating quantum evolution component as well as a time reversal violating quantum measurement interaction component. The time reversal violating quantum measurement interaction part of the Hamiltonian operator was shown to have components which contained causal retarded light travel times, connected to the values of the physical sizes and/or spatial separations associated with the physical aggregate of Measurement Color symmetric fermionic states into which the fermionic sector of the state vector was expanded.

For retarded light travel time intervals *in between the preparation and the measurement*, the expectation values of the time-reversal violating retarded quantum measurement interaction operator was found to be negligible compared to the expectation values of the time reversal violating quantum evolution operator, and the net effect generated the "quantum potentia" of what may occur in the form of the S-matrix approximation to the formalism.

On the other hand for the retarded light travel time intervals *corresponding to the preparation and/or the measurement,* the expectation values of the time-reversal violating retarded quantum measurement interaction operator was found to be dominant compared to the expectation values of the time reversal violating quantum evolution operator. and the net effect caused the "quantum potentia" of what may occur to be converted into the "quantum actua" of actual observer-participant measurement events.

In this context it was found that for a sufficiently large aggregate of atomic systems, described by the bare state component of total MC-QED Hamiltonian and assumed to exist in an "environment" associated with the remaining time reversal violating components of the total Hamiltonian, the net effect of the quantum measurement interaction in MC-QED generated time reversal violating decoherence-dissipation effects on the reduced density matrix in a manner which could dynamically give large aggregates of atomic systems apparently classical properties.

Hence, in contradistinction the Copenhagen Interpretation of QED with its strict form of "Macroscopic Realism", we found that MC-QED obeyed a <u>dynamic form of Macroscopic Realism</u> in which the classical level of physics emerged dynamically in the context of non-local <u>intrinsically time reversal violating quantum decoherence effects</u> which were able to project out individual states associated with specific diagonal elements of the density matrix. This result was in stark contrast to that of QED, where local quantum decoherence effects only appeared to be irreversible since a local observer did not have access to the entire wave function and, while interference effects appeared to be eliminated, individual states associated with specific diagonal elements of the density matrix had not been projected out (Schlosshauer, M., 2007).

In this manner we showed that the intrinsically time reversal violating quantum decoherence effects inherent within MC-QED implied that it did not require an independent external complementary classical level of physics obeying strict Macroscopic Realism in order to obtain a physical interpretation. In this way we have been led to the conclusion that an elegant resolution of the problem of the asymmetry between microscopic quantum objects and macroscopic classical objects inherent in the laws of quantum physics could be found within the context of the MC-QED formalism.

This result has broader implications since it leads to the possibility of finding a physical explanation of how living, macroscopic conscious observers emerge from the microscopic laws of quantum physics. This is because the observer-participant nature of MC-QED, with its intrinsic arrow of time dynamically generated by spontaneous CPT violation, opens up the possibility of two possible approaches to explain the apparently spontaneous emergence of macroscopic conscious minds in the universe from the microscopic laws of quantum physics. .

The first approach is a local one which can be found by extending the Measurement Color paradigm into the recently developed quantum field theoretic domain of consciousness research called Quantum Brain Dynamics QBD, (Jibu, M., and Yasue, K., 1995) , (Vitiello, G., 2001 ). Since MC-QED is a quantum field theoretic formalism which contains both the effects of quantization and dissipation, it may be possible that the ideas underlying QBD can be consistently generalized into a (MC-QBD) formalism. In this way it may be possible to find a <u>local cybernetic description</u> of how macroscopic conscious observer-participant entities emerge in a microscopic observer-participant universe.

The second approach is a global one which can be found by noting the fact that MC-QED describes the universe in terms of myriads of microscopic, time reversal violating, observer-participant quantum field theoretic interactions which span both the classical and the quantum world. On the other hand living, macroscopic conscious observers also appear to have physical properties which simultaneously span both the classical and the quantum world. Because of this similarity the MC-QED formalism has the capability of being able to explain how macroscopic conscious observer-participant entities emerge in a microscopic observer-participant universe. Since this would occur in a Measurement Color quantum field theoretic manner, a <u>global quantum holographic description of consciousness</u> may exist which connects
the "minds of conscious observers" to the "mind of the observer-participant universe" as a whole.

## APPENDIX I.  MEASUREMENT COLOR  QUANTUM ELECTRODYNAMICS

Measurement Color Quantum Electrodynamics (MC-QED) is constructed by imposing an Abelian operator gauge symmetry of <u>microscopic operator observer-participation</u> called Measurement Color onto the operator equations of Quantum Electrodynamics (QED) in the Heisenberg picture. We do this by defining an Abelian quantum field operator labeling symmetry associated with the integer indices k = 1,2, …, N (where in the limit we will let N --> ∞) and then imposing this integer labeling in an operational manner onto the quantum field structure of the standard QED formalism. In doing this we use the metric signature (1,-1, -1,-1), and (h/2π) = c = 1 units and the relativistic notation, operator sign conventions, and operator calculation techniques, used to generalize and extend the standard QED formalism into the MC-QED theory, are formally similar to those used in Chapters 8 and 9 of "Introduction to Relativistic Quantum Field Theory" by Sylvan. S. Schweber, Harper & Row 2$^{nd}$ Edition (1962).

The MC-QED formalism which emerges operationally describes the microscopic observer-participant quantum electrodynamic process, between the electron-positron quantum operator fields $\psi^{(k)}$ and the charge field photon quantum operator fields $A_\mu^{(j)}$ ($k \neq j$) which they interact with, in the Heisenberg Picture operator field equations. Since MC-QED is a theory of mutual quantum field theoretic observer-participation, its action principle must be constructed in a manner such that time-symmetric self-measurement interaction terms of the form $J_\mu^{(k)} A^{\mu(k)}$ (k=1,2,… , N→ -∞) are dynamically excluded from the formalism.

In the Heisenberg picture this is dynamically accomplished by means of the charge-conjugation invariant MC-QED action principle given by (k, j =1,2,… , N→ -∞)

$$I = \left\{ - \int dx^4 \left[ \sum_{(k)} \left( 1/4 [\psi^{(k)\dagger} \gamma^o, (-i\gamma^\mu \partial_\mu + m)\psi^{(k)}] + \text{hermetian conjugate} \right) \right. \right.$$
$$\left. \left. + \sum_{(k)} \sum_{(j \neq k)} (1/2 \partial_\mu A^{\nu(k)} \partial^\mu A_\nu^{(j)} + J_\mu^{(k)} A^{\mu(j)}) \right] \right\}$$

In the Heisenberg picture, following the standard second quantization methods taken in generating QED from CED to the above action for MC-QED, we find that the MC-QED Heisenberg operator equations of motion are given by

$$(-i\gamma^\mu \partial_\mu + m - e\gamma^\mu A_\mu^{(k)}{}_{(obs)})\psi^{(k)} = 0 \quad \text{(Heisenberg equation for } \psi^{(k)} \text{ fermion operator)}$$

$$A_\mu^{(k)}{}_{(obs)} = \sum_{(j \neq k)} A_\mu^{(j)} \quad \text{(electromagnetic operator field } A_\mu^{(k)}{}_{(obs)} \text{ observed by } \psi^{(k)} \text{)}$$

$$\Box^2 A_\mu^{(k)} = J_\mu^{(k)} = -e[\psi^{(k)\dagger} \gamma^o, \gamma_\mu \psi^{(k)}] \quad \text{(Heisenberg equation for the } A_\mu^{(k)} \text{ operator)}$$

where the Measurement Color labels on the operator fields $\psi^{(k)}$, and $A_\mu^{(k)}$ range over (k= 1,2, , N --> ∞). In the context of an indefinite metric Hilbert space, the Subsidiary Condition

$$\langle \psi | (\partial^\mu A_\mu^{(k)}) | \psi \rangle = 0 \qquad (k =1,2, … , N\text{-->} \infty))$$

must also be satisfied.

Then the expectation value of the Heisenberg Picture operator equations of MC-QED are will be invariant under the Abelian Measurement Color gauge transformation

$$\psi^{(k)'}(x) = \psi^{(k)}(x) \exp(ie\Lambda(x))$$

$$A_\mu^{(k)'}(x)_{obs)} = A_\mu^{(k)}(x)_{(obs)} + \partial_\mu \Lambda(x)$$

where $\Lambda(x)$ is a scalar field obeying $\Box^2 \Lambda(x) = 0$ (k =1,2, ... , N→ -∞). Hence the individual Measurement Color currents $J_\mu^{(k)}$ are conserved as $\partial^\mu J_\mu^{(k)} = 0$ (k =1,2, ... , N→ -∞) which implies that the individual Measurement Color charge operators $Q^{(k)} = \int dx^3 J_0^{(k)}$ (k =1,2, ... , N→ -∞) commute with the total Hamiltonian operator of the theory.

Following the standard procedures for the canonical quantization of fields applied to MC-CED leads to the canonical equal-time commutation and anti-commutation relations in the MC-QED formalism as

$$[A_\mu^{(k)}(x, t), \partial_t A_\nu^{(j)}{}_{(obs)}(x', t)] = i\eta_{\mu\nu} \delta^{kj} \delta^3(x' - x)$$

$$\{\psi^{(k)}(x, t), \psi^{(j)\dagger}(x', t)\} = \delta^{kj} \delta^3(x' - x) \qquad (k, j =1,2, ... , N→ \infty)$$

$$[A_\mu^{((k)}(x, t), \psi^{(j)}(x', t)] = [A_\mu^{((k)}(x, t), \psi^{(j)\dagger}(x', t)] = 0$$

where $\text{sig}(\eta_{\mu\nu}) = (1, -1, -1, -1$, with other equal-time commutators and anti-commutators vanishing respectively, (k, j =1,2, ... , N→ ∞).

In this context the structure of the MC-QED operator equations of motion and the equal-time commutation and anti-commutation relations dynamically enforces a form of mutual operator observer-participation which dynamically excludes time-symmetric Measurement Color self-interaction terms of the form $e\gamma^\mu A_\mu^{(k)} \psi^{(k)}$ (k =1,2, ... , N→ ∞) from the operator equations of motion.

The $N \geq 2$ Maxwell field operator equations must be solved for the charge-field operator solutions $A_\mu^{(k)}$ within the context of the multi-operator field theoretic Measurement Color paradigm upon which MC-CED is based. Hence the MC-QED paradigm excludes the local time-symmetric free radiation field operators $A_\mu^{(0)}$ from contributing to the $A_\mu^{(k)}$ charge-field operator solutions, because the $A_\mu^{(0)}$ field operators cannot be defined in terms of Measurement Color charge-field operator currents. This is in contrast to the case of QED where the local time-symmetric free radiation field operators $A_\mu^{(0)}$ cannot be excluded from $A_\mu$ since Measurement Color does not play a role in the Maxwell field operator structure in QED. Hence in solving the $N \geq 2$ Maxwell field operator equations for the charge-field operators $A_\mu^{(k)}$ the MC-CED paradigm implies that a universal time-symmetric boundary condition, which mathematically excludes local time reversal invariant free uncoupled radiation field operators $A_\mu^{(0)}$ from contributing to the $N \geq 2$ charge-fields $A_\mu^{(k)}$, has been imposed on <u>each</u> of the $A_\mu^{(k)}$ operator solutions to the $N \geq 2$ Maxwell operator field equations. Hence "free uncoupled radiation field operators" are excluded from MC-QED and in their place the physical effects of radiation are operationally described in a microscopic observer-participant manner by the "nonlocal, time anti-symmetric, total coupled radiation charge-field operator" $A_\mu^{(TCRF)}$

$$A_\mu^{(TCRF)} = \sum_{(j)} A_\mu^{(j)}(-) = \int dx^{4'} D_{(-)}(x-x') \sum_{(j)} J_\mu^{(j)}(x') \qquad (k=1,2,\ldots,N \to \infty))$$

where

$$D_{(-)}(x-x') = (D_{(ret)}(x-x') - D_{(adv)}(x-x'))/2$$

The $A_\mu^{(TCRF)}$ operator is a real, nonlocal operator which has a negative parity under Wigner time reversal $T_W$ (defined as the product of the Hermetian complex conjugate operator and the operator which takes t into –t) It has the unique property of being non-locally coupled to the sum of all of the current operators while still obeying the charge field photon operator equation $\Box^2 A_\mu^{(TCRF)} = 0$

Now within the operational observer-participant context of the MC-QED Heisenberg operator field equations, the electron-positron operator fields $\psi^{(k)}$ (k = 1,2, … N) "observe" the electromagnetic operator field $A_\mu^{(k)}{}_{(obs)}$ field, where $A_\mu^{(k)}{}_{(obs)}$ is given by the superposition of the "time-symmetric" electromagnetic field operators $A_\mu^{(j)}(+) = 1/2 \int dx^{4'} (D_{(ret)}(x-x') + D_{(adv)}(x-x')) J^{(k)}(x')$, ($j \neq k = 1,2, \ldots N$) and the "time-anti-symmetric" total coupled radiation field operator $A_\mu^{TCRF}$ as

$$A_\mu^{(k)}{}_{(obs)} = \sum_{(k \neq j)} A_\mu^{(j)} = \sum_{(j \neq k)} A_\mu^{(j)}(+) + (2p-1) A_\mu^{(TCRF)} \qquad (k = 1, 2, \ldots, N \to \infty))$$

where p is a c-number whose value, in the absence of "free radiation field operators", determines nature of the Arrow of Time in the MC-QED formalism independent of any external cosmological or thermodynamic assumptions.

To see this more explicitly we re-write the operator equations for $A_\mu^{(k)}{}_{(obs)}$ in the following form as

$$A_\mu^{(k)}{}_{(obs)} = pA^{(k)}{}_{(obs)(ret)} + (1-p)A^{(k)}{}_{(obs)(adv)} + A^{(k)}{}_{(obs)(in-p)}$$
$$= pA^{(k)}{}_{(obs)(adv)} + (1-p)A^{(k)}{}_{(obs)(ret)} + A^{(k)}{}_{(obs)(out-p)}$$

where the negative time parity coupled charge field photon "in and out" operators are

$$A^{(k)}{}_{(obs)(in-p)} = (2p-1) A^{(k)}{}_{(-)}$$

$$A^{(k)}{}_{(obs)(out-p)} = A^{(k)}{}_{(obs)(in-p)} + 2(2p-1) A^{(k)}{}_{(obs)(-)}$$

$$A^{(k)}{}_{(-)} = 1/2 \int dx^{4'} (D_{(ret)}(x-x') - D_{(adv)}(x-x'))J^{(k)}(x')$$

Now, in the <u>absence</u> of non-operational free radiation fields, the <u>presence</u> of the negative time parity Total Coupled Radiation Field operator $A_\mu^{(TCRF)}$ in MC-QED implies that the MC-QED operator equations violate <u>both</u> the $T_p$ <u>and</u> the $T_w$ symmetry operations defined as follows:

c) The "Radiation Flow Symmetry Operator" $T_p$, for which p → (1-p) occurs, is violated in the

   operator equations (3) since they have a negative parity under the $T_p$ operation

d) The Wigner Time Reversal operator symmetry $T_w$, for which Hermetian complex conjugation and
   t → -t occurs, is violated in the operator equations since by virtue of the presence of the Total Coupled Radiation Field operator $A_\mu^{(TCRF)}$ they have a negative parity under the $T_w$ operation

However, even though equations separately violate the $T_p$ and the $T_w$ symmetry, they are still invariant under the generalized Time Reversal operator $T = T_w \times T_p$ which is the product of the Wigner Time Reversal Operator $T_w$ and the Radiation Flow Symmetry Operator $T_p$. Since the operator field equations of motion of the MC-QED formalism in the Heisenberg Picture are also invariant under the respective action of the Charge Conjugation operator C, and the Parity operator P, then even though it violates the $T_w$ time reversal symmetry, we find that MC-QED is still CPT invariant where the T symmetry is generalized to become $T = T_w \times T_p$.

Now in the context of the Heisenberg Picture state vector $|\Psi\rangle$ one defines the "in-out" operator field solutions by imposing the "In-Asymptotic Condition" as $\psi^{(k)}(x, t \to -\infty) = \psi^{(k)}(in)$ (k =1, 2, ... , N$\to \infty$))

*(In-kinematic condition)*

$$\langle A_\mu^{(k)}{}_{(obs)}(x, t \to -\infty)\rangle = \langle(\int dx^3 (J_\mu^{(k)}{}_{(obs)}(x', t \to -\infty) / 4\pi |x-x'|) + A^{(k)}{}_{(obs)}(in)\rangle$$

*(in-dynamic stability condition)*

$$\langle \partial_t J_\mu^{(k)}(x, t \to -\infty)\rangle = 0$$

in the limit as $t \to -\infty$ of the operator equations (1) as

$$\langle(-i\gamma^\mu\partial_\mu + m - e\gamma^\mu A_\mu^{(k)}{}_{(obs)}(x, t \to -\infty))\psi^{(k)}(in)\rangle = 0$$
$$\langle \Box^2 A_\mu^{(k)}(in)\rangle = \langle J_\mu^{(k)}(in)\rangle = -e \langle[\psi^{(k)}(in)^\dagger \gamma^o, \gamma_\mu \psi^{(k)}(in)]\rangle$$

and also imposing "Out-Asymptotic Condition" as $\psi^{(k)}(x, t \to +\infty) = \psi^{(k)}(out)$ (k =1, 2, ... , N$\to \infty$))

*(out-kinematic condition)*

$$\langle A_\mu^{(k)}{}_{(obs)}(x, t \to +\infty)\rangle = \langle(\int dx^3 (J_\mu^{(k)}{}_{(obs)}(x', t \to +\infty) / 4\pi |x-x'|) + A^{(k)}{}_{(obs)}(out)\rangle$$

*(out-dynamic stability condition)*

$$\langle \partial_t J_\mu^{(k)}(x, t \to +\infty)\rangle = 0$$

in the limit as $t \to +\infty$ of the operator equations (1) as

$$\langle(-i\gamma^\mu\partial_\mu + m - e\gamma^\mu A_\mu^{(k)}{}_{(obs)}(x, t \to +\infty))\psi^{(k)}(out)\rangle = 0$$

$$\langle \Box^2 A_\mu^{(k)}(out)\rangle = \langle J_\mu^{(k)}(out)\rangle = -e \langle[\psi^{(k)}(out)^\dagger \gamma^o, \gamma_\mu \psi^{(k)}(out)]\rangle$$

Now by applying the Asymptotic Conditions to the MC-QED operator equations of motion it follows that a retarded quantum electrodynamic arrow of time emerges dynamically. This is because in the absence of "free uncoupled radiation field operators": To see this more specifically note that for the case of p =1 the Heisenberg Picture operator equations of motion have the form

$$\langle(-i\gamma^\mu\partial_\mu + m - e\gamma^\mu A_\mu^{(k)}{}_{(obs)})\psi^{(k)}\rangle = 0$$

$$\langle A_\mu^{(k)}{}_{(obs)}\rangle = \langle \sum_{(j \neq k)} A_\mu^{(j)}(ret) + A^{(k)}(-)\rangle \qquad (k, j = 1,2, ... , N \to \infty))$$

The expectation value of the above operator equations physically describe the situation where charge field photons are causally emitted and absorbed between the $\psi^{(k)}$ and $\psi^{(j)}$ $k \neq j$ fermion operators, while being spontaneously emitted into the vacuum by the $\psi^{(k}$ fermion operators, (k, j = 1,2, ... , N--> ∞)). For this reason these operator equations predict that electron-positron states can form bound states which spontaneously decay into charge field photons. Hence these operator equations will satisfy the dynamic stability component of the Asymptotic Condition because they predict that their expectation values imply that a stable vacuum state exists.

On the other hand for the case of p =0 the Heisenberg Picture operator equations of motion have the form

$$\langle(-i\gamma^\mu\partial_\mu + m - e\gamma^\mu A_\mu^{(k)}(obs))\psi^{(k)}\rangle = 0$$

$$\langle A_\mu^{(k)}(obs)\rangle = \langle \sum_{(j \neq k)} A_\mu^{(j)}(adv) - A^{(k)}(-)\rangle \qquad (k, j = 1,2, ... , N \to \infty))$$

On the other hand the expectation value of these operator equations physically describe the situation where charge field photons are causally absorbed and emitted between the $\psi^{(k)}$ and $\psi^{(j)}$ $k \neq j$ fermion operators, while being spontaneously absorbed from the vacuum by the $\psi^{(k}$ fermion operators, (k, j = 1,2, , N--> ∞)). For this reason these operator equations predict that electron-positron states will be spontaneously excited from the vacuum. Hence these operator equations cannot satisfy the dynamic stability component of the Asymptotic Condition because they predict that their expectation values imply that a stable vacuum state cannot exist.

In addition, because of the operational presence negative time parity Total Coupled Radiation Charge-Field $A_\mu^{(TCRF)}$ in the MC-QED formalism, the dynamic component of the time-symmetric Asymptotic Condition, which requires that a stable vacuum state must exist, dynamically determines a retarded quantum electrodynamic Physical Arrow of Time associated with p =1 independent of any Thermodynamic or Cosmological boundary conditions. Hence this implies that MC-QED has a negative parity under the $T_w$ and $T_p$ operations while still remaining Invariant under the CPT symmetry operation where T is generalized to become T = $T_w$ x $T_p$.

# APPENDIX II. OBSERVER-PARTICIPANT FORMALISM IN THE MC-QED INTERACTION PICTURE

The S-matrix approximation to MC-QED, which holds for time intervals $|(h / \Delta E)| < t \ll |(R/c|$, and its connection to the well-known Feynman diagrammatic explanations of quantum electrodynamic processes, will now be discussed in more detail in the this section where the observer-participant bare state structure of MC-QED will be developed.

In the Heisenberg picture the MC-QED Hamiltonian operator $H = H_S$ is

$$H = H^\dagger = [H_0 + V_{qp} + V_{ret\text{-}qa}]$$

where $H_0$ contains the bare fermion and bare components of $H$

$$H_0 = (H_f + H_{ph})$$

In $H_0$ the Measurement Color symmetric bare electron-positron Hamiltonian operator $H_f$ is given by ($k = 1, 2, \ldots, N \to \infty$)

$$H_f = \sum_{(k)} \{: \int dx^3 [\psi^{(k)\dagger}(\alpha \cdot p + \beta m - e\varphi^{(k)}_{(ext)}) \psi^{(k)} + J^{\mu(k)} A_\mu^{(k)}{}_{Breit}{}^{(obs)}] :\}$$

(where the Breit potential operator is given by

$$A_\mu^{(k)}{}_{Breit}{}^{(obs)} = \sum_{(j) \neq (k)} \int dx^3 J_\mu^{(j)}(x',t) / 4\pi |x-x'|$$

and an external potential $\varphi^{(k)}_{(ext)}$ has been included in order to represent the lowest order Coulombic effects of baryonic nuclei in the MC-QED) and the Measurement Color symmetric bare electromagnetic field Hamiltonian operator where $H_{ph}$ is given by ($k = 1, 2, \ldots, N \to \infty$)

$$H_{ph} = -1/2 \sum_{(k)} \{: \int dx^3 [(\partial_t A^{\mu(k)}{}_{(rad)} \partial_t A_\mu^{(k)}{}_{(rad)}{}^{(obs)} + \nabla A^{\mu(k)}{}_{(rad)} \cdot \nabla A_\mu^{(k)}{}_{(rad)}{}^{(obs)})] :\}$$

where

$$A_\mu^{(k)}{}_{(rad)} = (\alpha_\mu - A_\mu^{(k)}{}_{(-)})$$

$$\alpha_\mu = \sum_{(j)} A_\mu^{(j)}{}_{(-)} / (N-1)$$

$$A_\mu^{(k)}{}_{(rad)}{}^{(obs)} = \sum_{(j) \neq (k)} A_\mu^{(j)}{}_{(rad)} = A_\mu^{(k)}{}_{(-)}$$

and $A_\mu^{(k)}{}_{(-)}$ is the non-local, negative time parity, Heisenberg picture operator defined as

$$A_\mu^{(k)}{}_{(-)} = \int dx'^4 (D_{(-)}(x-x')) J_\mu^{(k)}(x') \qquad (k=1,2,\ldots,N\to\infty)$$

Note that the $A_\mu^{(k)}{}_{(rad)}$, $A_\mu^{(k)}{}_{(rad)}{}^{(obs)}$, $\alpha_\mu$ all have a negative parity under Wigner Time reversal since they are linear functions of $A_{\mu(-)}^{(k)}$.

The operators $V_{qp}$ and $V_{ret-qa}$ are given $(k=1,2,\ldots,N\to\infty)$ by

$$V_{qp} = \sum_{(k)} \{: \int dx^3 J_\mu^{(k)} A^{\mu(k)}{}_{(-)} :\}$$

and

$$(V_{ret-qa}) = \sum_{(k)} \{: \int dx^3 [J_\mu^{(k)} (A^{\mu(k)}{}_{(ret)}{}^{(obs)} - A_\mu^{(k)}{}_{Breit}{}^{(obs)})]$$
$$-1/2 [\partial_t A^\mu{}_{(ret)}^{(k)} \partial_t A_{\mu(ret)}^{(k)(obs)} + \partial_t A^\mu{}_{(ret)}^{(k)} \partial_t A_{\mu(rad)}^{(k)(obs)}$$
$$+ \partial_t A^\mu{}_{(rad)}^{(k)} \partial_t A_{\mu(ret)}^{(k)(obs)}] + \nabla A^\mu{}_{(ret)}^{(k)} \cdot \nabla A_\mu^{(k)}{}_{(ret)}^{(obs)}$$
$$+ \nabla A^\mu{}_{(ret)}^{(k)} \cdot \nabla A_\mu^{(k)}{}_{(rad)}^{(obs)} + \nabla A^\mu{}_{(rad)}^{(k)} \cdot \nabla A_{\mu(ret)}^{(k)(obs)}]:\}$$

where

$$A_\mu^{(k)}{}_{(ret)}^{(obs)} = \sum_{(j)\neq(k)} A_\mu^{(j)}{}_{(ret)}$$

where $J_\mu^{(k)} = -e[\psi^{(k)\dagger} \gamma^o, \gamma_\mu \psi^{(k)}]$ and the symbols $:$ $:$ indicate that normal ordering of operators has been taken. In the Heisenberg Picture the MC-QED *equal-time* commutation and anti-commutation relations for $(k,j=1,2,\ldots,N\to\infty)$ are given, $(k,j=1,2,\ldots,N\to\infty)$, $sig(\eta_{\mu\nu}) = (1,-1,-1,-1)$, by

$$[A_\mu^{(k)}(\mathbf{x},t), \partial_t A_\nu^{(j)}{}_{(obs)}(\mathbf{x}',t)] = -i\eta_{\mu\nu}\delta^{kj}\delta^3(\mathbf{x}'-\mathbf{x})$$
$$[A_\mu^{((k)}(\mathbf{x},t), \psi^{(j)}(\mathbf{x}',t)] = 0$$
$$[A_\mu^{((k)}(\mathbf{x},t), \psi^{(j)\dagger}(\mathbf{x}',t)] = 0$$
$$\{\psi^{(k)}(\mathbf{x},t), \psi^{(j)\dagger}(\mathbf{x}',t)\} = \delta^{kj}\delta^3(\mathbf{x}'-\mathbf{x})$$
$$\{\psi^{(k)}(\mathbf{x},t), \psi^{(j)}(\mathbf{x}',t)\} = 0$$
$$\{\psi^{(k)\dagger}(\mathbf{x},t), \psi^{(j)\dagger}(\mathbf{x}',t)\} = 0$$

All other equal-time commutators and anti-commutators vanishing respectively.

Now the successive state vector transformations on the Heisenberg Picture State vector $|\psi_H\rangle$, through the Schrodinger Picture state vector $|\psi_S\rangle$, that finally leads to the Interaction Picture state vector $|\psi_I\rangle$ can be formally represented by $|\psi_I(t)\rangle = U(t-t_o) |\psi_H\rangle)$ where the unitary operator $U(t-t_o)$ is

$$U(t-t_o) = \exp[i(H_0)_S(t-t_o)] \exp[-iH(t-t_o)]$$

and Schrodinger Hamiltonian operators $(H_0)_S$ and $H = H_S$ are constant in time. It then follows that the equation of motion of the state vector in the Interaction Picture is

$$i\partial_t|\psi_I(t)\rangle / dt = [V_{qp}(t) + V_{ret-qa}(t)]_I |\psi_I(t)\rangle$$

where Interaction Picture operators $O_I(x, t)_I$ are related to Heisenberg Picture operators $O_H(x, t)$ as

$$O_I(x,t)_I = U(t-t_o) O_I(x,t)_H U(t-t_o)^{-1} = U(t-t_o) O_I(x,t)_H U(t-t_o)^\dagger$$

The Interaction Picture the S-matrix approximation to the MC-QED is applicable during the time intervals for which $V_{qp}(t)_I$ dominates $V_{ret-qa}(t)_I$ in the state vector equation of motion. The S-matrix approximation is valid during the time intervals $-(R/c) < t < (R/c)$ in between *preparations* occurring at $t = -(R/c)$ and *measurements* occurring at $t = (R/c)$, (where $|(R/c)| \ggg (h/\Delta E)$ is the magnitude of the characteristic size and/or spatial separation of the physical components associated with observer-participant quantum states of the bare hamiltonian operator $H_0$). In this context it follows that the Interaction Picture Hamiltonian operator is given by $H_I = (H_0)_I + (V_{qp})_I$

For simplicity of notation, the subscript " $_I$ " will be dropped and understood to hold for all equations in what follows. In this context we have

$$H = H_0 + V_{qp}$$

where

$$H_0 = H_f + H_{ph}$$

and the bare fermion Interaction Picture Hamiltonian operator $H_{in-f}$ is ($k = 1, 2, \ldots, N \rightarrow$

$$H_{in-f} = \sum_{(k)} \{ :\int dx^3 [\psi^{(k)\dagger}(\alpha \cdot p + \beta m - e\varphi^{(k)}_{(ext)}) \psi^{(k)} + J^{\mu(k)} A_\mu^{(k)(obs)}_{Breit}] : \}$$

and the bare Interaction Picture Hamiltonian operator $H_{ph}$ is ($k = 1, 2, \ldots, N \rightarrow$

$$H_{ph} = -1/2 \sum_{(k)} \{: \int dx^3 [ (\partial_t A^{\mu(k)}{}_{(rad)} \partial_t A_\mu{}^{(k)}{}_{(rad)}{}^{(obs)}$$
$$+ \nabla A^{\mu(k)}{}_{(rad)} \cdot \nabla A_\mu{}^{(k)}{}_{(rad)}{}^{(obs)})]:\}$$

In $H_{ph}$ the operators $A_\mu{}^{(k)}{}_{(rad)}$ and $A_\mu{}^{(k)}{}_{(rad)}{}^{(obs)}$ are given by

$$A_\mu{}^{(k)}{}_{(rad)} = (\alpha_\mu - A_\mu{}^{(k)}{}_{(-)})$$

$$A_\mu{}^{(k)}{}_{(rad)}{}^{(obs)} = \sum_{(j) \neq (k)} A_\mu{}^{(j)}{}_{(rad)} = A_\mu{}^{(k)}{}_{(-)}$$

where $\alpha_\mu = \sum_{(j)} A_\mu{}^{(j)}{}_{(-)} /(N-1)$ and $A_\mu{}^{(k)}{}_{(-)}$ is given by

$$A_\mu{}^{(k)}{}_{(-)} = \left( \int dx'^4 D_{(-)}(x-x') U(t)U(t')^{-1} J^{\mu(k)}(x') U(t')U(t)^{-1} \right)$$

where

$$U(t)U(t')^{-1} = (\exp[i(H_0)_S t] \exp[-iH_S t]) (\exp[-i(H_0)_S t'] \exp[iH_S t'])$$

The "quantum potentia" Interaction Picture operator which couples $H_f$ to $H_{ph}$ is given by

$$V_{qp} = \sum_{(k)} \{: \int dx^3 J_\mu{}^{(k)} A^{\mu(k)}{}_{(-)} :\} = \sum_{(k)} V_{qp}{}^{(k)} \qquad (k=1,2,\ldots,N \longrightarrow \infty)$$

The Interaction Picture operator equations of motion for MC-QED are given by

$$(-i\gamma^\mu \partial_\mu + m - e\varphi^{(k)}{}_{(ext)} - e\gamma^\mu A_\mu{}^{(k)}{}_{(Breit-obs)})\psi^{(k)} = 0$$

$(k, j = 1, 2, \ldots, N \to \infty)$

$$\Box^2 A_\mu{}^{(k)}{}_{(-)} = 0$$

where

$$A_\mu{}^{(k)}{}_{(Breit-obs)} = \sum_{(j \neq k)} A_\mu{}^{(j)}{}_{(Breit)}$$

$$A_\mu{}^{(j)}{}_{(Breit)}(x,t) = \int dx'^3 J_\mu{}^{(j)}(x', t) / 4\pi |x - x'|$$

and the Coulombic effects of baryonic nuclei in the MC-QED formalism is represented to lowest order by the external potential $\varphi^{(k)}{}_{(ext)}$).

The *equal-time commutation and anti-commutation relations* in the Interaction Picture are (k, j =1,2, ... , N→ ∞), sig($\eta_{\mu\nu}$) = (1, -1, -1, -1) , (REF SCHWEBER PG 242)

$$[A_\mu{}^{(k)} (\mathbf{x}, t), \partial_t A_\nu{}^{(j)}{}_{(obs)} (\mathbf{x'}, t)] = -i\eta_{\mu\nu} \delta^{kj} \delta^3(\mathbf{x'} - \mathbf{x})$$
$$[A_\mu{}^{((k)} (\mathbf{x}, t), \psi^{(j)}(\mathbf{x'}, t)] = [A_\mu{}^{((k)} (\mathbf{x}, t), \psi^{(j)\dagger}(\mathbf{x'}, t)] = 0$$
$$\{\psi^{(k)} (\mathbf{x}, t), \psi^{(j)\dagger}(\mathbf{x'}, t)\} = \delta^{kj} \delta^3(x' - x)$$
$$\{\psi^{(k)} (\mathbf{x}, t), \psi^{(j)}(\mathbf{x'}, t)\} = 0$$
$$\{\psi^{(k)\dagger} (\mathbf{x}, t), \psi^{(j)\dagger}(\mathbf{x'}, t)\} = 0$$

All other equal-time commutators and anti-commutators vanishing respectively, (k, j =1,2, ... , N→ ∞).

Using the operator equations of motion the above equal time commutation and anti-commutation relations can be solved for the general commutation and anti-commutation relations in the Interaction Picture as

(k, j =1,2, ... , N→ ∞) , sig($\eta_{\mu\nu}$) = (1, -1, -1, -1)

$$[A_\mu{}^{(k)} (x), A_\nu{}^{(j)}{}_{(obs)} (x')] = \delta^{kj} [i\eta_{\mu\nu} D(x' - x)] = \delta^{kj} D_{\mu\nu}(x' - x)$$
$$[A_\mu{}^{((k)} (x), \psi^{(j)}(x')] = [A_\mu{}^{((k)} (x), \psi^{(j)\dagger}(x')] = 0$$
$$\{\psi^{(k)} (x), \psi^{(j)\dagger}(x')\} = \delta^{kj} S(x' - x)$$
$$\{\psi^{(k)} (x), \psi^{(j)}(x')\} = 0$$
$$\{\psi^{(k)\dagger} (x), \psi^{(j)\dagger}(x)\} = 0$$

where $D_{\mu\nu}(x' - x) = [i\eta_{\mu\nu} D(x' - x)]$ and $D(x' - x) = (1/2\pi) \varepsilon(x_0)\delta(x) = -D_{(-)}(x' - x)$, and all other equal-time commutators and anti-commutators vanishing respectively, (k, j =1,2, ... , N→ ∞).

When the relationships

$$A_\mu{}^{(k)}(x) = (\alpha_\mu(x) - A_\mu{}^{(k)}{}_{(-)}(x))$$
$$\alpha_\mu(x) = \sum_{(j)} A_\mu{}^{(j)}{}_{(-)}(x) /(N-1)$$
$$A_\mu{}^{(k)}{}_{(obs)}(x) = \sum_{(j) \neq (k)} A_\mu{}^{(j)}(x) = A_\mu{}^{(k)}{}_{(-)}(x)$$

are inserted, the commutation relations for $A_\mu{}^{(k)}{}_{(-)}(x)$, $A_\nu{}^{(k)}(x')$, and $\alpha_\mu(x)$

become (k, j =1,2, ... , N→ ∞).

$$[A_\mu^{(k)}{}_{(-)}(x), A_\nu^{(j)}{}_{(-)}(x')] = (1 - \delta^{kj})(D_{\mu\nu}(x'-x))$$
$$[\alpha_\mu(x), A_\nu^{(k)}{}_{(-)}(x')] = (D_{\mu\nu}(x'-x))$$
$$[\alpha_\mu(x), \alpha_\nu(x')] = (N/N-1)(D_{\mu\nu}(x'-x)) \longrightarrow D_{\mu\nu}(x'-x)$$
$$[\alpha_\mu(x), A_\nu^{(k)}(x')] = (1/N-1)(D_{\mu\nu}(x'-x)) \longrightarrow 0$$
$$[A_\mu^{((k))}{}_{(-)}(x), \psi^{(j)}(x')] = 0$$
$$[A_\mu^{((k))}{}_{(-)}(x), \psi^{(j)\dagger}(x')] = 0$$

where $D_{\mu\nu}(x'-x) = [i\eta_{\mu\nu}D(x'-x)]$ and $D(x'-x) = (1/2\pi)\varepsilon(x_0)\delta(x) = -D_{(-)}(x'-x)$, and all other equal-time commutators vanishing (k, j =1,2, ... , N→ ∞).

## APPENDIX III. PHOTON BARE STATE STRUCTURE IN THE MC-QED INTERACTION PICTURE

We now consider the quantum state structure associated with the MC-QED charge field photon Hamiltonian operator

$$H_{ph} = \sum_{(k)} \{:\int dx^3 [-1/2 (\partial_t A^{\mu(k)}{}_{(rad)} \partial_t A_\mu^{(k)}{}_{(rad)}{}^{(obs)} + \nabla A^{\mu(k)}{}_{(rad)} \cdot \nabla A_\mu^{(k)}{}_{(rad)}{}^{(obs)})]:\}$$

(k=1,2,...,N --- > ∞)

where

$$A_\mu^{(k)}{}_{(rad)} = (\alpha_\mu - A_\mu^{(k)}{}_{(-)})$$
$$A_\mu^{(k)}{}_{(rad)}{}^{(obs)} = \sum_{(j) \neq (k)} A_\mu^{(j)}{}_{(rad)} = A_\mu^{(k)}{}_{(-)}$$

and
$$\alpha_\mu = \sum_{(j)} A_\mu^{(j)}{}_{(-)} /(N-1)$$

are linear functions of the negative time parity operator $A_\mu^{(k)}{}_{(-)}(x)$ which obeys $\Box^2 A_\mu^{(k)}{}_{(-)}(x) = 0$ then

$$\Box^2 A_\mu^{(k)}{}_{(rad)}{}^{(obs)} = \Box^2 A_\mu^{(k)}{}_{(rad)} = \Box^2 \alpha_\mu(x) = 0$$

Hence the operators $A_\mu^{(k)}{}_{(rad)}$, $A_\mu^{(k)}{}_{(rad)}{}^{(obs)}$ and $\alpha_\mu$ (k =1,2, ... , N→ ∞)

can be respectively expanded as

$$A_\mu^{(k)}(x) = (\alpha_\mu(x) - A_\mu^{(k)}{}_{(-)}(x))$$
$$= \int d\lambda^3 / \sqrt{[2(2\pi)^3 \lambda^2]} \{ a_\mu^{(k)}(\lambda) e^{-i\lambda \cdot x} + a_\mu^{(k)}(\lambda)^\dagger e^{i\lambda \cdot x} \}$$

$$A_\mu^{(k)}{}_{(rad)}{}^{(obs)}(x) = A_\mu^{(k)}{}_{(-)}(x)$$
$$= \int d\lambda^3 / \sqrt{[2(2\pi)^3 \lambda^2]} \{ a_\mu^{(k)}{}_{(-)}(\lambda) e^{-i\lambda \cdot x} + a_\mu^{(k)}{}_{(-)}(\lambda)^\dagger e^{i\lambda \cdot x} \}$$

$$\alpha_\mu(x) = \sum_{(j)} A_\mu^{(j)}{}_{(-)}(x) / (N-1)$$
$$= \int d\lambda^3 / \sqrt{[2(2\pi)^3 \lambda^2]} \{ \alpha_\mu(\lambda) e^{-i\lambda \cdot x} + \alpha_\mu(\lambda)^\dagger e^{i\lambda \cdot x} \}$$

where in the above

$$a_\mu^{(k)}{}_{(-)}(\lambda) = a_\mu^{(k)}{}_{(rad)}{}^{(obs)}(\lambda), \quad \alpha_\mu(\lambda) = \sum_{(j)} a_\mu^{(j)}{}_{(-)}(\lambda) / (N-1) = \sum_{(j)} a_\mu^{(j)}(\lambda)$$
$$a_\mu^{(k)}(\lambda) = (\alpha_\mu(\lambda) - a_\mu^{(k)}{}_{(-)}(\lambda)) = (\sum_{(j)} a_\mu^{(j)}{}_{(-)}(\lambda) / (N-1) - a_\mu^{(k)}{}_{(-)}(\lambda))$$

*In the context of the above operator equations and their commutation relations, will now show that the time reversal violating Measurement Color symmetric operators $\alpha_\mu(\lambda)$ and $\alpha_\mu(\lambda)^\dagger$ act respectively as destruction and creation operators for Measurement Color symmetric charge field photon states in MC-QED which have a negative parity under the Wigner Time Reversal operator $T_w$.*

We begin by substituting the above representations of $A_\mu^{(k)}(x)$, $A_\mu^{(k)}{}_{(obs)}(x)$, and $\alpha_\mu(x)$ into the above MC-QED commutation relations to find (k, j =1,2, ... , N→ ∞) that

$$[a_\mu^{(k)}{}_{(-)}(\lambda), a_\nu^{(j)}{}_{(-)}{}^\dagger(\lambda')] = (1 - \delta^{kj})(-\eta_{\mu\nu}\lambda_0 \delta^3(\lambda - \lambda'))$$
$$[\alpha_\mu(\lambda), a_\nu^{(j)}{}_{(-)}{}^\dagger(\lambda')] = -\eta_{\mu\nu}\lambda_0 \delta^3(\lambda - \lambda')$$
$$[\alpha_\mu(\lambda), \alpha_\nu(\lambda')] = (-\eta_{\mu\nu}\lambda_0 \delta^3(\lambda - \lambda'))(N/(N-1))$$
$$[\alpha_\mu(\lambda), a_\nu^{(k)}(\lambda')^\dagger] = 0$$
$$[a_\mu^{(k)}{}_{(-)}{}^\dagger(\lambda), a_\nu^{(j)}{}_{(-)}{}^\dagger(\lambda')] = 0$$
$$[a_\mu^{(k)}{}_{(-)}(\lambda), a_\nu^{(j)}{}_{(-)}(\lambda')] = 0$$

where $\lambda_0 = \sqrt{(\lambda^2)} = \omega(\lambda)$ and all other commutators vanish. Next we substitute above representations of $A_\mu^{(k)}(x)$ and $A_\mu^{(k)}{}_{(obs)}(x)$ into the charge field photon hamiltonian $H_{ph}$ which gives the charge field photon hamiltonian as

$$H_{ph} = \sum_{(k)} \{: [-\int d\lambda^3/\lambda_0 \; (\omega(\lambda) a^{\dagger}{}_{(\lambda)\mu}^{(k)}(\lambda)) a^{\mu(k)}{}_{(-)}(\lambda)]:\}$$

and normal ordering of operators inside of the symbols $\{: \; :\}$ has been taken. $H_{ph}$ is Hermetian since by inserting $a_\mu^{(k)}(\lambda) = (\alpha_\mu(\lambda) - a_\mu^{(k)}{}_{(-)}(\lambda))$ and $\alpha_\mu(\lambda) = \sum_{(j)} a_\mu^{(j)}{}_{(-)}(\lambda)/(N-1)$ into the above expression we find that it can also be written as

$$H_{ph} = :[-\int d\lambda^3/\lambda_0 \; (\omega(\lambda) \{\sum_{(k)} \sum_{(j)} a_{(\lambda)(-)}{}^{\dagger\mu(j)} a_{(\lambda)(-)\mu}^{(k)}(\lambda)/(N-1)$$
$$- \sum_{(k)} a_{(\lambda)(-)}{}^{\dagger\mu(k)} a_{(\lambda)(-)\mu}^{(k)}(\lambda)\}:$$
$$= H_{ph}{}^{\dagger}$$

In this context if the bare MC-QED charge field photon vacuum state $|0_{ph}\rangle$ is defined by

$$a_\mu^{(k)}{}_{(-)}(\lambda) | 0_{ph} \rangle = 0 \qquad (k = 1, 2, \ldots, N \rightarrow \infty)$$

this implies that $H_{ph} | 0_{ph} \rangle = 0$ as required. Now since $\alpha_\mu(\lambda) = \sum_{(j)} a_\mu^{(j)}{}_{(-)}(\lambda)/(N-1)$ the above definition of $|0_{ph}\rangle$ also implies that the bare charge field photon vacuum state also obeys

$$\alpha_\mu(\lambda) | 0_{ph} \rangle = 0$$

In this context the bare single charge-field photon in MC-QED can be defined as

$$|\lambda 1_\mu\rangle = \alpha_\mu(\lambda 1)^{\dagger} |0\rangle = (1/(N-1)) \sum_{(j)} a_\mu^{(j)}{}_{(-)}{}^{\dagger}(\lambda 1)^{(j)} |0\rangle$$

This can be seen by calculating

$$H_{ph} |\lambda_1\rangle = - \sum_{(k)} \int d\lambda^3/\lambda_0 \; (\omega(\lambda) a_\nu^{(k)}(\lambda)^{\dagger} a^{\nu(k)}{}_{(-)}(\lambda) \alpha_\mu(\lambda 1)^{\dagger} |0\rangle$$

Then using the fact that $[\alpha_\mu(\lambda), a^{\nu(j)}{}_{(-)}{}^{\dagger}(\lambda')] = -\delta_{\mu\nu} \lambda_0 \delta^3(\lambda - \lambda')$ and $\alpha_\mu(\lambda 1) = \sum_{(J)} a_\mu^{(j)}(\lambda 1)$ in the above equation we have that

$$H_{ph} |\lambda_1\rangle = \sum_{(J)} \int d\lambda^3 \, (\omega(\lambda) a_\nu^{(j)}(\lambda)^\dagger)(-\delta_\mu^{\,\nu}) \delta^3(\lambda - \lambda_1) |0\rangle$$

$$= \omega(\lambda_1) \sum_{(J)} a_\mu^{(j)\dagger}(\lambda_1) |0\rangle$$

$$= \omega(\lambda_1) (\alpha_\mu(\lambda_1))^\dagger |0\rangle$$

$$= \omega(\lambda_1) |\lambda_1\rangle$$

as required.

Hence multiple bare charge-field photon states in MC-QED are defined as

$$|\lambda_{\alpha 1}, \lambda_{2\beta}, \lambda_{3\gamma}, \ldots \rangle = 1/\sqrt{(N_p!)} \, \alpha_\alpha(\lambda_1)^\dagger \alpha_\beta(\lambda_2)^\dagger \alpha_\gamma(\lambda_3)^\dagger \ldots |0\rangle$$

In a similar manner as that of the covariant form of QED, consistency with the expectation value of the operator form of Maxwell equations in the covariant form of MC-QED requires that an Indefinite Metric Hilbert space must be used. <u>Note that the time parity of the N-photon state $|\lambda_{\alpha 1}, \lambda_{2\beta}, \lambda_{3\gamma}, \ldots \rangle$ is $(-1)^N$ and that the coherent state defined by $\exp(\alpha_{\mu-}) |0\rangle$ is not symmetric under time reversal in</u>

In the context of an Indefinite Metric Hilbert space, the subset of physical bare charge field photon states in MC-QED contained within the above set of multiple charge field photon eigenstates of $H_{ph}$ are required to obey the Weak Subsidiary Condition

$$\lambda^\mu a_\mu^{(k)}(\lambda) |\psi\rangle = 0$$

where $a_\mu^{(k)}(\lambda) = (\alpha_\mu(\lambda) - a_\mu^{(k)}(obs)) = (\sum_{(j)} a_\mu^{(j)}{}_{(-)}/(N-1) - a_\mu^{(k)}{}_{(-)})$

which requires them to contain equal numbers of timelike and longitudinal charge field photons. Since the Indefinite Metric Hilbert space implies that charge field photon states with an odd number of time-like charge field photons have an additional negative sign associate with their inner product, the combination of the Weak Subsidiary Condition and the Indefinite Metric Hilbert space together imply that the physical bare charge field photon states have a positive semi-definite norm and energy momentum expectation values. Hence from the above analysis we conclude that the Measurement Color symmetric bare charge field photon state structure of MC-QED is similar to that of QED, with the key exception being that the Measure Color symmetric bare charge field photon creation and annihilation operators $\alpha_\mu(\lambda)^\dagger$ and $\alpha_\mu(\lambda)$ have a negative parity under Wigner Time reversal $T_w$.

## APPENDIX IV. FERMION BARE STATE STRUCTURE IN THE MC-QED INTERACTION PICTURE

We next discuss the bare state electron-positron structure associated with the fermion Hamiltonian operator in the context of the Furry Interaction Picture, (where an external potential $\varphi^{(k)}_{(ext)}$) has been included in order to represent the lowest order Coulombic effects of baryonic nuclei in the MC-QED)
given by

$$H_f = \; : \int dx^3 \sum_{(k)} [\psi^{(k)\dagger}(\alpha \cdot p + \beta m - e\varphi_{(ext)}) \psi^{(k)}] + J^{\mu(k)} A^{(k)}_{\mu\,Breit}{}^{(obs)} \; :$$

where $(k = 1, 2, \ldots, N \to \infty)$ and

$$A^{(k)}_{\mu\,Breit}{}^{(obs)} = \sum_{(j) \neq (k)} \int dx^3 J^{(j)}_\mu(x',t) / 4\pi |x-x'|$$

$(k, j = 1, 2, \ldots, N \to \infty)$

and the equal time anti-commutation relations in the Furry Interaction Picture are

$$\{\psi^{(k)}(\mathbf{x}, t), \psi^{(j)\dagger}(\mathbf{x}', t)\} = \delta^{kj} \delta^3(x' - x)$$
$$\{\psi^{(k)}(\mathbf{x}, t), \psi^{(j)}(\mathbf{x}', t)\} = 0$$
$$\{\psi^{(k)\dagger}(\mathbf{x}, t), \psi^{(j)\dagger}(\mathbf{x}', t)\} = 0$$

All other equal-time anti-commutators vanishing respectively, $(k, j = 1, 2, \ldots, N \to \infty)$.

Since the bare fermion Hamiltonian operator $H_f$ is summed over all of its internal Measurement Color indices $(k = 1, 2, \ldots, N \to \infty)$ it does not single out any particular Measurement Color label and hence it is a Measurement Color scalar. This implies that the multi-electron-positron eigenstates of $H_f$ must be $N_f \geq 2$ Measurement Color singlet states which are symmetric in their Measurement Color labels.

In the Furry Interaction Picture, the MC-QED fermion operator equations of motion, and their charge-conjugate equations of motion respectively have positive energy operator solutions (denoted by $\psi^{(k)\dagger}_{(+)}, \psi^{(k)}_{(+)}$ and $\psi^{c(k)\dagger}_{(+)}, \psi^{c(k)}_{(+)}$) which, when acting on the vacuum state $|0\rangle$ respectively annihilate and create electrons and positrons in a manner formally similar to that of QED. in this context the equal-time anti-commutation relations and the Measurement Color symmetry property of $H_f$ work together to generate Measurement Color symmetric electron-positron eigenstates.

Here we will start first by considering case of $N_f \geq 2$ electrons interacting with each other in the presence of an external field since $N_f = 1$ fermion states are ruled out by the requirement of Measurement Color symmetry. In this context it follows that the equal-time anti-commutation relations and the Measurement Color symmetry property of $H_f$ imply that the *Measurement Color symmetric* multi-electron eigenstates for $N_f = N_e \geq 2$ bare electrons in the Furry Interaction picture for MC-QED has the form

$$|E, (N_1),(N_2),..(N_e)\rangle = (1 / N_e!) \int dx_1^3 .. dx_{Ne}^3 \sum (s_1.. s_{Ne}) \chi_E(\mathbf{x}_1,s_1... \mathbf{x}_{Ne},s_{Ne})$$
$$\cdot \psi(\mathbf{x}_1,s_1)^{(1)}{}_{(+)}{}^\dagger ... \psi(\mathbf{x}_{Ne},s_{Ne})^{(Ne)}{}_{(+)}{}^\dagger |0\rangle$$

Note that in the $|E, (N_1),(N_2) ...(N_e)\rangle$ state the combination of the anti-commutation properties of the $\psi_{in}^{(k)}{}_{(+)}{}^\dagger$, with the requirement of measurement color symmetry of the $|E, (N_1),(N_2) ...(N_e)\rangle$ state imposed by $H_f$, automatically requires that the wave function for the Ne electrons given by $\chi_E(\mathbf{x}_1,s_1... \mathbf{x}_{Ne},s_{Ne})$ must be anti-symmetric in the configuration space and spin coordinates $(\mathbf{x}_1,s_1... \mathbf{x}_{Ne},s_{Ne})$ consistent with the Pauli Exclusion Principle.

The anti-symmetric Ne-electron wave function $\chi_E(\mathbf{x}_1,s_1... \mathbf{x}_{Ne},s_{Ne})$ is given by the positive energy eigenstate solution to the configuration space Hamiltonian for Ne electrons in an external field given by

$$H_{cs} \chi_E(\mathbf{x}_1,s_1... \mathbf{x}_{Ne},s_{Ne}) = E \chi_E(\mathbf{x}_1,s_1... \mathbf{x}_{Ne},s_{Ne})$$

where Ne $\geq 2$ and (k, j =1,2, ... , Ne)

$$H_{cs} = \sum_{(k)} \Lambda^k_+ (\alpha^k \cdot \mathbf{p}^k + \beta^k m - e\varphi^{(k)}{}_{(ext)}) \Lambda^k_+$$
$$+ (e^2 / 4\pi) \sum_{(k)} \sum_{(j \neq k)} [\Lambda^k_+ \Lambda^j_+ (1 - \alpha^k \cdot \alpha^j) \Lambda^k_+ \Lambda^j_+] / |\mathbf{x}^k - \mathbf{x}^j|$$

and the $\Lambda^k_+$ which appear in $H_{cs}$ are appropriately chosen positive energy projection operators for the kth electron interacting with the $j \neq k$ other electrons in the presence of an external field. For example in the simplest case of Ne = 2 electrons, depending on the choice of the external field representing the lowest order coulomb coupling of the nucleus, the measurement color symmetric |Ne =2 electron⟩ state could describe either the quantum states of either a Helium atom or the quantum states of two spatially separated Hydrogen atoms

In a similar manner Measurement Color symmetric $N_p \geq 2$ positron states involving operator products of $\psi^{c(k)}{}_{(+)}{}^\dagger$ acting on $|0\rangle$, and Measurement Color symmetric ($N_e + N_p$) $\geq 2$ electron-positron states, involving products of both $\psi^{(k)}{}_{(+)}{}^\dagger$ and $\psi^{c(k)}{}_{(+)}{}^\dagger$ acting together on $|0\rangle$, can be constructed in the MC-QED formalism.

# APPENDIX V.   ON THE CALCULATION OF THE S-MATRIX OF QUANTUM POTENTIA IN MC-QED

In the context of the previous discussion of the bare charge field photon and bare electron-positron state structure in the MC-QED Interaction Picture, we demonstrated that the predictive properties of the Measurement Color symmetric bare charge field photon and bare electron-positron state structure were similar to that of QED.

In this appendix we will show how these bare states can be used in calculating the S-matrix in the quantum potentia approximation to MC-QED. However in this context we will see that differences in the dynamic description of the source of radiative corrections will occur between QED and MC-QED.  This is because, instead of being defined by local in time free charge field photon operators as in the case of QED,  the observed bare in-field charge field photon operators in MC-QED are described by non-local in time operators with a negative parity under Wigner Time Reversal $T_W$ given by ($k = 1, 2, \ldots , N \to \infty$)

$$A_\mu^{(k)}{}_{(-)} = \left( \int dx'^4 \, D_{(-)}(x-x') \, U(t)U(t')^{-1} \, J^{\mu(k)}(x') \, U(t')U(t)^{-1} \right)$$

where

$$U(t)U(t')^{-1} = \left( \exp[i(H_0)_S \, t] \exp[-iH_S \, t] \right) \left( \exp[iH_S \, t'] \exp[-i(H_0)_S \, t'] \right)$$

Now recall that the successive state vector transformations on the Heisenberg Picture State vector $|\psi_H\rangle$, through the Schrodinger Picture state vector $|\psi_S\rangle$, that finally lead to the Interaction Picture state vector $|\psi_I\rangle$ can be formally represented by $|\psi_I(t)\rangle = U(t-t_o) |\psi_H\rangle$  where the unitary operator $U(t-t_o)$ is

$$U(t-t_o) = \exp[i(H_0)_S(t-t_o)] \exp[-iH_S(t-t_o)]$$

where the Schrodinger Hamiltonian operators $(H_0)_S$ and $H_S = H$ are constant in time. It then follows that the equation of motion of the state vector in the Interaction Picture is

$$i\partial_t |\psi_I(t)\rangle / dt = [V_{qp}(t) + V_{ret-qa}(t)]_I \, |\psi_I(t)\rangle$$

and the Interaction Picture operators $O_I(x, t)_I$ are related to Heisenberg Picture operators $O_H(x, t)$ as

$$O_I(x,t)_I = U(t-t_o) \, O_I(x,t)_H \, U(t-t_o)^{-1} = U(t-t_o) \, O_I(x,t)_H \, U(t-t_o)^\dagger$$

Now setting to = 0 for simplicity in what follows we have

$$J^{\mu(k)}(x)_I = U(t) J^{\mu(k)}(x)_H U(t)^{-1}$$

$$(A^{\mu(k)}(x)_{(-)})_I = U(t)(A^{\mu(k)}(x)_{(-)})_H U(t)^{-1}$$

Note that since

$$(A_\mu^{(k)}(x)_{(-)})_I = \int dx'^4 \, D_{(-)}(x-x') \, U(t)U(t')^{-1} \, J_\mu^{(k)}(x')_I \, U(t')U(t)^{-1}$$

is a nonlocal in time operator functional of the current operator then the commutation relations

$$[A_\mu^{((k))}{}_{(-)}(x), \psi^{(j)}(x')]_I = [A_\mu^{((k))}{}_{(-)}(x), \psi^{(j)\dagger}(x')]_I = 0$$

(required for the existence of the bare local-in-time fermion-charge field photon "in-states") creates a nonlocal in time operator constraint relationship, involving a spacetime integration over the Green functions $D_{(-)}(x'-x)$ and $S(x'-x)$ whose form is required to be consistent with the anti-commutation relations

$$\{\psi_{in}^{(k)}(x), \psi_{in}^{(j)\dagger}(x')\}_I = \delta^{kj} S(x'-x)$$

Hence in the Wick T-product decomposition of the S-matrix in the Interaction Picture, this leads to two different kinds of Wick contraction terms associated with the $J_\mu^{(k)}(x)$ and the $A_\mu^{(k)}{}_{(-)}(x')$ operators:

1) the first kind are local contractions between the $J_\mu^{(k)}(x)$ operators and the $A_\mu^{(k)}{}_{(-)}(x')$ operators as a whole which vanish as

$$(\overline{J_\mu^{(k)}(x) \, A_\mu^{(k)}{}_{(-)}(x')}) = 0$$

2) the second kind are contractions between the $J_\mu^{(k)}(x)$ operators and the nonlocal functional dependence of the $J_\mu^{(jk)}(x)$ operators which appear inside of the $A_\mu^{((k))}{}_{(-)}[J_\mu^{(k)}(x')]$ operators which generates two types of non-zero contraction terms

$$\overline{J_\mu^{(k)}(x) \, A_\mu^{((k))}{}_{(-)} [J_\mu^{(k)}(x')]} \neq 0$$

$$\overline{J_\mu^{(k)}(x) \, A_\mu^{((k))}{}_{(-)} [J_\mu^{(k)}(x')]} \neq 0$$

These contractions of the second kind, which are nonlocal since they involved spacetime integrations over products of $S_F(x'-x)$ and $D_{(-)}(x'-x) = (D_{(ret)}(x'-x) - D_{(adv)}(x'-x))/2$ are time reversal violating and generate time reversal violating radiative corrections to the bare fermion states which occur in the MC-QED S-matrix. The details associated with the calculation of the S-matrix in MC-QED will be presented elsewhere in future papers.

**APPENDIX VI. EFFECTS OF SPONTANEOUS CPT SYMMETRY BREAKING IN MC-QED**

From the above discussion we see that the Measurement Color symmetry in MC-QED automatically excluded time-symmetric free photon operators from the formalism. Instead the photon operator in MC-QED was described by a nonlocal Measurement Color Symmetric "Total Coupled Radiation" charge-field photon operator which carried a negative time parity under Wigner Time Reversal. In this context the physical requirement of a stable vacuum state dynamically required that the Heisenberg operator equations for fermions must contain a causal retarded quantum electrodynamic arrow of time, independent of any external thermodynamic or cosmological assumptions. Hence this dynamically implied that the photon carries the quantum electrodynamic arrow of time in the MC-QED formalism.

This result is better understood in a broader context by noting that, within the nonlocal quantum field theoretic structure of the MC-QED formalism, the physical requirement of a stable vacuum state generated a spontaneous symmetry breaking of both the T and the CPT symmetry. Spontaneous symmetry breaking of the T and the CPT symmetry occurred in MC-QED because the nonlocal photon operator acting within it has a negative parity under Wigner time reversal. In this manner the requirement of a stable vacuum state dynamically selected the operator solutions to the MC-QED formalism that contained a causal, retarded, quantum electrodynamic arrow of time, independent of any external thermodynamic or cosmological assumptions. In this manner the existence of the causal microscopic arrow of time in MC-QED represents a fundamentally quantum electrodynamic explanation for irreversible phenomena associated with the Second Law of Thermodynamics which complements the one supplied by the well-known statistical arguments in phase space [Zeh, 2007].

The fact that the Measurement Color symmetry in MC-QED implies that the photon operator carries the arrow of time has a profound effect on the nature of the time evolution of the combination of "systems + environment" in the Interaction Picture of the formalism (Leiter, 2010). This is because it causes the reduced density matrix of the "system" in the presence of its "environment" to contain both Von Neumann Type 2 (quantum potentia) time evolution of the state vector as well as Von Neumann Type 1 (quantum actua) time evolution.

This is because the reduced density matrix of the system takes the form of a differential-delay equation containing time reversal violating quantum evolution and quantum measurement interaction components. The time reversal violating quantum measurement interaction part of the quantum interaction has components that contain causal retarded light travel times, which are connected to the values of the physical sizes and/or spatial separations associated with the physical aggregate of Measurement Color symmetric fermionic states into which the fermionic sector of state vector is expanded.

For the retarded light travel time intervals in between the preparation and the measurement, the expectation values of the time-reversal violating retarded quantum measurement interaction operator will be negligible compared to the expectation values of the quantum evolution operator which generates the "quantum potentia" of what may occur. On the other hand for retarded light travel time intervals corresponding to the preparation and/or the measurement, the expectation values of the time-reversal violating retarded quantum measurement interaction operator will be dominant compared to the expectation values of the quantum evolution operator and this will cause the "quantum potentia" to be converted into the "quantum actua" of observer-participant measurement events.

Hence in this manner MC-QED contains its own time reversal violating microscopic observer-participant description of the quantum measurement process, independent of the Copenhagen Interpretation or the Everett "Many Worlds Interpretation". It is for this reason that the paradigm of MC-QED can be used to solve the problem of macroscopic quantum reality.

This is because Measurement Color Quantum Electrodynamics (MC-QED) has the form of a non-local quantum field theory which describes the quantum measurement process in terms of myriads of microscopic electron-positron quantum operator fields undergoing spontaneous CPT symmetry breaking time observer-participant quantum measurement interaction processes mediated by the charge-field photon quantum operator fields through which they interact.

In this context it has been shown (Leiter, 2010) for a sufficiently large aggregate of atomic systems, described by the bare state components of MC-QED Hamiltonian and interacting with each other through the effects of the time reversal violating quantum measurement interaction operator, that the effects of the CPT violating quantum measurement interaction will generate time reversal violating (Quantum Decoherence + Dissipation) effects on the reduced density matrix in a manner which will give these large aggregates of atomic systems apparently classical properties.

Since MC-QED obeys a dynamic form of Macroscopic Realism, the classical level of physics emerges in the context of local intrinsically time reversal violating quantum decoherence effects which project out individual states since they are generated by the time reversal violating quantum measurement interaction in the formalism. Hence MC-QED does not require an independent external complementary classical level of physics obeying strict Macroscopic Realism in order to obtain a physical interpretation.

Since it does not require an independent external complementary classical level of physics in order to obtain a physical interpretation of the quantum measurement process, the MC-QED formalism represents a more general observer-participant approach to quantum electrodynamics in which a consistent description of quantum electrodynamic measurement processes at both the microscopic and macroscopic levels can be obtained.

This is in contrast to the time reversal symmetric case of QED where the local quantum decoherence effects only have the appearance of being irreversible because a local observer does not have access to the entire wave function and, while interference effects appear to be eliminated, individual states have not been projected out.

The phenomenon of Quantum Decoherence in MC-QED is described in terms of quantum systems interacting with their environments, in a time irreversible manner which prevents different components in the [quantum superposition](#) of the [wave function](#) of the (system + environment) from [interfering](#) with each other. However MC-QED differs from QED in that the phenomenon of Quantum Decoherence occurs in the context of a microscopic, time irreversible, process generated by spontaneous CPT breaking in the MC-QED formalism.

Hence the phenomenon of Quantum Decoherence in MC-QED always includes the effects of Quantum Dissipation. This is due to the fact that spontaneous CPT symmetry breaking in the MC-QED formalism causes the photon to carry the arrow of time. The combination of (Quantum Decoherence + Quantum Dissipation), created by the spontaneous CPT symmetry breaking inherent in the MC-QED formalism, generates an overall time reversal violating process.

This causes the reduced density matrix of the system to become diagonalized by Quantum Decoherence effects over a "decoherence time period" after which the effects of Quantum Dissipation over a "relaxation time period" >> "decoherence time period" causes specific diagonal elements of the of the reduced density matrix to become equal to unity with all others equal to zero. In this manner the quantum field theoretic [dynamics](#) of the reduced density matrix of the system by itself will be both microscopically time [irreversible](#) as well as being non-unitary.

The combination of microscopically time reversal violating (Quantum Decoherence + Quantum Dissipation), generated by the spontaneous CPT symmetry breaking inherent in the MC-QED formalism, predicts both the probability of an outcome (i.e. a quantum potentia) as well as an actual outcome (i.e. a quantum actua). Because of this fact the well known paradox of the "problem of outcomes", associated with the process of Quantum Decoherence in QED [Schlosshauer, 2007], can resolved in the context of the MC-QED formalism